\documentclass[twocolumn,english,prl,superscriptaddress,nofootinbib,nobibnotes]{revtex4-1}
\usepackage[T1]{fontenc}
\usepackage[latin9]{inputenc}
\usepackage{color}
\usepackage{babel}
\usepackage{units}
\usepackage{amsmath}
\usepackage{amssymb}
\usepackage{graphicx}
\usepackage[unicode=true,pdfusetitle,
bookmarks=true,bookmarksnumbered=false,bookmarksopen=false,
breaklinks=false,pdfborder={0 0 0},pdfborderstyle={},backref=false,colorlinks=true]
{hyperref}
\hypersetup{
	urlcolor=orange,linkcolor=blue,citecolor=red}

\makeatletter
\usepackage{tikz}
\usepackage{color}
\usepackage{newtxtext}
\usepackage{newtxmath}
\usepackage{dsfont}

\DeclareMathOperator*{\argmax}{argmax}

\makeatletter
\newsavebox{\@brx}
\newcommand{\llangle}[1][]{\savebox{\@brx}{\(\m@th{#1\langle}\)}%
	\mathopen{\copy\@brx\kern-0.5\wd\@brx\usebox{\@brx}}}
\newcommand{\rrangle}[1][]{\savebox{\@brx}{\(\m@th{#1\rangle}\)}%
	\mathclose{\copy\@brx\kern-0.5\wd\@brx\usebox{\@brx}}}
\makeatother

\makeatother
\begin{document}
	\title{Typical and atypical solutions in non-convex neural networks with discrete and continuous weights}
	
	\author{Carlo Baldassi}
	\affiliation{Department of Computing Sciences, Bocconi University, 20136 Milano, Italy}
	\author{Enrico M. Malatesta}
	\email{enrico.malatesta@unibocconi.it}
	\affiliation{Department of Computing Sciences, Bocconi University, 20136 Milano, Italy}
	\author{Gabriele Perugini}
	\affiliation{Department of Computing Sciences, Bocconi University, 20136 Milano, Italy}
	\author{Riccardo Zecchina}
	\affiliation{Department of Computing Sciences, Bocconi University, 20136 Milano, Italy}

	
	\date{\today}
	\begin{abstract}
		We study the binary and continuous negative-margin perceptrons as simple non-convex neural network models learning random rules and associations. We analyze the geometry of the landscape of solutions in both models and find important similarities and differences. Both models exhibit subdominant minimizers which are extremely flat and wide. These minimizers coexist with a background of dominant solutions which are composed by an exponential number of algorithmically inaccessible small clusters for the binary case (the frozen 1-RSB phase) or a hierarchical structure of clusters of different sizes for the spherical case (the full RSB phase).
		
		In both cases, when a certain  threshold in constraint density is crossed, 
		the local entropy of the wide flat minima becomes non-monotonic, indicating a break-up of the space of robust solutions into disconnected components. 
		This has a strong impact on the behavior of algorithms in binary models, which cannot access  the  remaining isolated clusters. For the spherical case the behaviour is different, since even beyond the disappearance of the wide flat minima the remaining solutions are shown to always be surrounded by a large number of other solutions at any distance, up to capacity. Indeed, we exhibit numerical evidence that algorithms seem to find solutions up to the SAT/UNSAT transition, that we compute here using an 1RSB approximation. 
		For both models, the generalization performance as a learning device is shown to be greatly improved by the existence of wide flat minimizers even when trained in the highly underconstrained regime of very negative margins.
	\end{abstract}
	\maketitle
	
	
	\section{Introduction}
	
	One of the most important and open problems in machine learning is to characterize at the theoretical level the typical properties of the loss landscape of neural networks. Given the impressive results that machine learning has achieved, e.g. in computer vision, image and speech recognition and translation, it is of primary importance to understand what are the key ingredients in those high-dimensional loss landscapes that allow very simple algorithms to work so well.
	
	In the field of the physics of disordered systems~\cite{mezard1987spin} there have been, in the last few decades, many efforts to develop analytical and numerical techniques suitable for the study of nonconvex high-dimensional landscapes. Despite the many differences in the types of landscapes studied~\cite{CastellaniCavagna2005,baity2019comparing}, the main questions addressed by physicists are very similar to those of interest to the machine learning community. For example: can we predict, given an initial condition, in which part of the landscape a given algorithm will converge? What is the role played by local minima and saddles? 
	
	In the last years several empirical results have emerged on the general structure of the landscape. First, it has been found that there exists a wide and flat region in the bottom of the landscape: empirical analysis of the Hessian of configurations obtained by using simple variants of gradient descent algorithms~\cite{sagun2016eigenvalues,sagun2017empirical} showed that such regions are very attractive for simple learning dynamics. In~\cite{FengYuhai}	it was also shown that the most common algorithm used in machine learning, Stochastic Gradient Descent (SGD), is intrinsically biased by its noisy component towards flatter solutions. Low dimensional representations of the landscape have also shown~\cite{LiVisualizing2018} that several heuristic techniques used in machine learning are effective because they tend to smooth the landscape, increasing therefore the convergence abilities of gradient descent algorithms. 
	Other lines of research~\cite{Draxler,entezari2022the,pittorino2022deep}, studied the connectivity properties of solutions, showing that minimizers of the loss function obtained with different initializations actually lie in the same basin, i.e. can be connected by a zero training-error path. In~\cite{jiang2019fantastic} the authors tested a large number of complexity measures on several and diverse models concluding that the ones based on flatness are more predictive of generalization properties.
	
	
	In the statistical mechanics literature the role of wide and flat minima in the neural network landscape has emerged recently from the study of simple one-layer~\cite{baldassi2015subdominant,baldassi_local_2016} and two-layer models~\cite{relu_locent,baldassi2020shaping}. Those works, that we summarize for convenience in the next section, have been limited so far to non-convex neural networks with binary weights; the connections between the geometrical properties of the landscape and the behavior of algorithms has been investigated in several works. 
	In this paper we bridge this gap, trying to shed new light on the geometrical organization of solutions in simple non convex continuous neural network models. 
	
	Our main findings show that both discrete and continuous models are characterized by the existence of wide-flat minima which undergo a ``clustering'' phase transition below (but close) to capacity.
	For the continuous model the disappearance of wide-flat minima does not imply the onset of algorithmic hardness, as it happens for the discrete case. We argue that this is due to the underlying full-RSB background of ground states which is different from the disconnected structure of ground states of the discrete case.
	We show numerically that the algorithms typically used for learning tend to end-up in the wide-flat regions when they exist and that these regions show remarkable generalization capabilities.
	
	For more complex neural network architectures, such as deep multi layer networks, we expect that the differences between discrete and continuous models to become less evident. However, more analytical studies will be needed to address this point.
	
	The paper is organized as follows: in Section~\ref{sec::binary_models} we summarize the main phenomenology that has emerged in the context of binary weights models; in Section~\ref{sec::continuous_models} we introduce the non-convex models studied in this paper, namely the binary and spherical negative-margin perceptrons and we summarize their known properties. In Section~\ref{sec::1rsb_alphac} we present the 1-step replica symmetry breaking (1RSB) computation of the SAT/UNSAT transition of the negative-margin perceptron. In Section~\ref{sec::FP_main} we employ the Franz-Parisi technique to study the local landscape of a configuration extracted with a given probability measure from the set of solutions. In particular we study the flatness of the typical minima of a generic loss function and we develop a method to extract configurations maximizing the flatness. We then explore how clusters of flat solutions evolve as a function of the training set size, revealing a phase transition in the geometrical organization of the solutions in the landscape. We call this \emph{Local Entropy} (LE) transition.
	In Section~\ref{sec::numerical_experiments} we present further numerical experiments: firstly we show numerical evidence suggesting that at the Local Entropy transition the manifold of atypical solutions undergoes a profound change in structure; secondly we show in a continuous-weights model how the generalization performance is greatly improved by the existence of wide and flat minimizers. Section~\ref{sec::conclusions} contains our conclusions.

	\section{Overview of the phenomenology in binary models} \label{sec::binary_models}
	
	In this section we summarize the main results on the connection between algorithmic behaviour and geometrical properties of the landscape of solutions in binary models of neural networks. We focus for simplicity on the paradigmatic example of the binary perceptron problem, defined as follows.
	
	Given a training set composed by $P = \alpha N$ binary random patterns $\boldsymbol{\xi}^\mu \in \left\{ -1, 1 \right\}^N$ and labels $y^\mu \in \left\{ -1, 1 \right\}$ with $\mu = 1, \dots, P$, we aim to find a binary $N$-dimensional weight $\boldsymbol{w} \in \left\{ -1, 1 \right\}^N$ such that 
	\begin{equation}
		\label{eq::constraints_binary}
		y^\mu = \text{sign}\left( \boldsymbol{w} \cdot \boldsymbol{\xi}^\mu \right)
	\end{equation}
	for all $\mu$. The non-convexity of the problem is induced by the binary nature of the weights. In the following, we will simply call $\boldsymbol{w} \in \left\{ -1, 1\right\}^N$ a \emph{solution} if it satisfies the $P$ constraints in~\eqref{eq::constraints_binary}. From now on, we will also say that a solution is \emph{typical}, if it is extracted with uniform measure among the set of all the possible solutions, namely from the Gibbs measure at zero temperature. A typical solution can also be recognised by looking at macroscopic observables (e.g. the distribution of stabilities $\Delta^\mu \equiv \frac{1}{\sqrt{N}}\boldsymbol{w} \cdot \boldsymbol{\xi}^\mu$) which tend to concentrate on their most probable value in the large $N$ and $P$ limit with the ratio $\alpha$ fixed.	
	A solution that is not sampled from the Gibbs measure will instead be considered ``atypical''. 
	The name ``atypical'' should also suggest that the probability of the solution being sampled by the Gibbs measure goes to zero in the thermodynamic limit, that is, we are analyzing large deviations of the Gibbs measure.
	
	
	Since the late 1980s this model has been studied using statistical physics techniques by Gardner and Derrida~\cite{gardner1988The,Gardner_1989} and by Krauth and M\'ezard~\cite{krauth1989storage} in the limit $N, \, P \to \infty$ with fixed $\alpha$. They found that one can find solutions to the problem only up to a given critical value of $\alpha$ that separates a SAT from an UNSAT phase: for $\alpha < \alpha_c$ the problem admits an exponential number of solutions, whereas for $\alpha>\alpha_c$ no solution to the problem can be found.   Huang and Kabashima~\cite{huang2014origin} computed the entropy of solutions at a given distance from \emph{typical} solutions (the so-called \emph{Franz-Parisi} entropy) showing that for $\alpha<\alpha_c$ the space of solutions splits into well-separated clusters of vanishing entropy. More in detail, they found that picking a configuration $\tilde{\boldsymbol{w}}$ at random from the set of all solutions, the closest solution is found by flipping an extensive number of weights. This means that typical solutions are completely isolated or ``\emph{sharp}'', namely the training error is larger than zero even at very small (but extensive) distances around them. The geometrical organization of typical solutions was later confirmed rigorously in a related model, the symmetric binary perceptron~\cite{perkins2021frozen,abbe2021binary}. In several constraint satisfaction problems (CSPs), such point-like solutions are generally believed to be hard to find algorithmically. In particular in Boolean satisfiability problems like $k$-SAT~\cite{mezard2002analytic,marino2016backtracking,Marino_2021} and in graph coloring~\cite{Zdeborova2008} all the best known algorithms start to fail in finding solutions at the so-called \emph{clustering} transition, which occurs at the constraint density for which only such isolated solutions remain.
	
	The analytical picture of the landscape of the binary perceptron, and its relation with the behaviour of algorithms, however, was far from being completely understood. Indeed, if only those kind of point-like, sharp solutions existed, it should be really hard to find them. Nevertheless, not only there exist algorithms that solve efficiently the optimization problem~\cite{Braunstein2006,baldassi2007efficient,baldassi2015max}, but the solutions they find also have substantially different properties from the typical ones: we mention in particular the fact that one can numerically show that solutions found by algorithms usually lie in dense clusters of solutions~\cite{baldassi2020shaping} and observables like the stability distribution~\cite{relu_locent} and the generalization error (in teacher-student models)~\cite{baldassi2015subdominant} do not coincide with the theoretical calculations cited above.
	
	This apparent paradox has been resolved in~\cite{baldassi2015subdominant,baldassi_local_2016}, where it was shown analytically that there exist clustered (or also called ``flat'') solutions in the landscape that extend to very large scale (also called ``wide'') which are highly attractive for common learning algorithm dynamics. Those solutions are found analytically by sampling from probability measures that are not the Gibbs measure: one should give more statistical weight to those solutions that have a large \emph{local entropy}, i.e. those ones having an higher number of solutions surrounding them\footnote{the precise definition of local entropy will be given in Section~\ref{sec::FP_main}.}. By definition those solutions are atypical and they can be shown to be subdominant (even if still exponential) in number with respect to typical ones. 
	It was then shown~\cite{baldassi2021unveiling,baldassi2022learning} that the sharp and wide, flat solutions coexist in the landscape until a certain critical value $\alpha_{\text{LE}}$ is reached; this new transition has been named \emph{Local Entropy} (LE) transition: for $\alpha < \alpha_{\text{LE}}$, both sharp and wide and flat minima can be found whereas for $\alpha>\alpha_{\text{LE}}$ only sharp and flat (but not wide) minima exist. This has a drastic impact on learning: the most efficient algorithms are not able to overcome this threshold, since beyond it only isolated solutions exists. $\alpha_{\text{LE}}$ can be therefore thought as a ``clustering transition'' as was studied earlier in various CSPs, but whose nature is different, being the result of disappearance of the most clustered solutions. 
	Moreover, this analytical picture was exploited to create new algorithms, that have been explicitly designed to target wide and flat minima: we mention in particular focusing Belief Propagation (fBP); for other types of ``replicated algorithms'' see~\cite{baldassi2015subdominant}. Those algorithms in turn are the ones that succeed in getting closer to the local entropy transition. Other algorithms like Reinforced Belief Propagation (rBP)~\cite{Braunstein2006}, reinforced max-sum~\cite{baldassi2015max} and Stochastic BP-inspired (SBPI)~\cite{baldassi2007efficient} also induce a bias towards clustered solutions.
	
	The phenomenology described above has been shown to hold qualitatively also on other binary models like, e.g. one-hidden layer tree committee machines and teacher student models. Those lines of research also showed that the heuristic techniques and architectural choices used in machine learning, such as the use of the cross-entropy loss~\cite{baldassi2020shaping}, or ReLU activation function~\cite{relu_locent}, or regularization~\cite{baldassi2020wide} tend to affect the learning landscape and to induce wider and flatter minima. In addition, a similar analysis of the role of overparameterization~\cite{baldassi2022learning} suggests that it allows  the emergence of atypical clusterized solutions at the local entropy transition. The appearance of those solutions makes the dynamics change from glassy to non-glassy~\cite{baity2019comparing}.
	
	\begin{figure*}
		\begin{centering}
			\includegraphics[width=1.8\columnwidth]{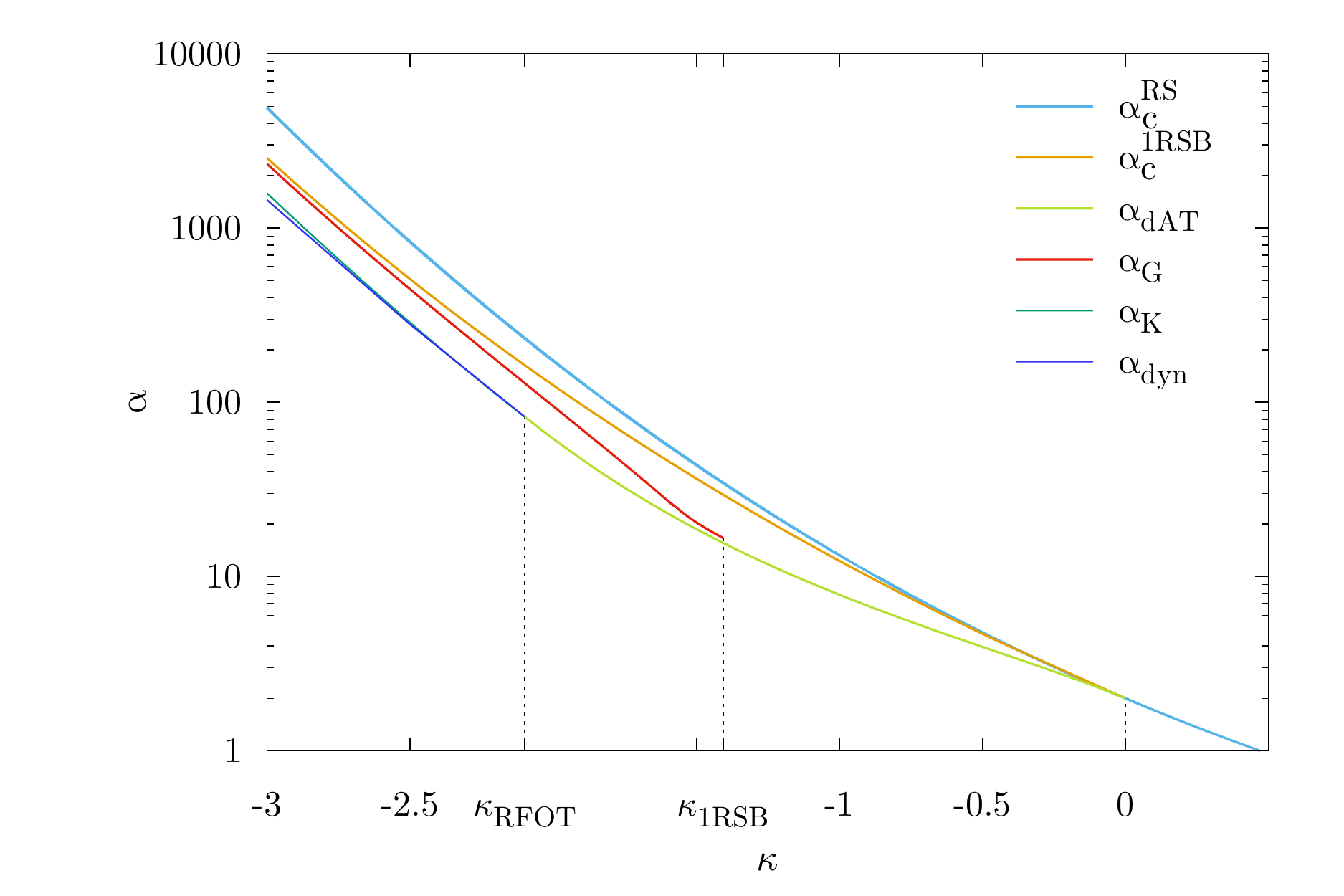}
		\end{centering}	
		\caption{Phase diagram of the negative spherical perceptron. The light blue line represents the SAT/UNSAT transition computed in the Replica-Symmetric approximation as presented originally in Gardner-Derrida's work~\cite{gardner1988optimal}. This line is correct only in the region $\kappa \ge 0$, where the model is convex. A better approximation to the SAT/UNSAT transition is the orange line, which is computed here using the 1RSB approximation. The other lines were computed in~\cite{gardner1988optimal} and~\cite{franz2017}. The green line represents the De Almeida-Thouless instability of the RS ansatz: when one crosses $\alpha_{\mathrm{dAT}}$ for $\kappa > \kappa_{\mathrm{1RSB}}$ one goes from an RS to a fRSB phase; when  $\kappa_{\mathrm{RFOT}} < \kappa < \kappa_{\mathrm{1RSB}}$ one instead goes from RS to a 1RSB stable phase. For a value of $\kappa < \kappa_{\mathrm{RFOT}}$ one encounters several transitions when $\alpha$ is increased: the \emph{dynamical} or \emph{clustering} transition $\alpha_{\mathrm{dyn}}$ (blue line), the \emph{Kauzmann transition} (green line) and the \emph{Gardner transition} (in red). For $\alpha < \alpha_{\mathrm{dyn}}$ the system is RS. For $\alpha_{\mathrm{dyn}} < \alpha < \alpha_{\mathrm{K}}$ the system is in a dynamical 1RSB phase, where $q_1>q_0$ and the Parisi block parameter $m=1$. For $\alpha_{\mathrm{K}} < \alpha< \alpha_{\mathrm{G}}$ the Gibbs measure is dominated by a 1RSB solution having $m<1$. Above the Gardner transition $\alpha > \alpha_{\mathrm{G}}$ the system is fRSB. Note that the dynamical transition presented in~\cite{franz2017} is slightly below the one presented here~\cite{franz2017_error}.}
		\label{Fig::phase_diagram} 
	\end{figure*}
	
	\section{The model} \label{sec::continuous_models}

	The scenario described in the previous section holds for binary weights neural network models. On the other hand, in non-convex continuous models less is known so far. In~\cite{baldassi2020shaping,relu_locent} the existence of wide and flat solutions was shown in one-hidden-layer continuous weights models. However, there has been no detailed study of the change in the structure of these regions as a function of overparameterization, nor of their algorithmic implications as in the case of binary models. 
	Here we study these questions in an even simpler model, the negative-margin perceptron. It is defined as follows.

	Given a set of $P = \alpha N$ normally distributed random patterns $\boldsymbol{\xi}^\mu$, $\mu = 1,\dots, P$ we want to find a vector of $N$ continuous weights $\boldsymbol{w}$ normalized on $\mathcal{S}_N$, the sphere of radius $\sqrt{N}$
	\begin{equation}
		\label{eq::normalization}
		\boldsymbol{w}\cdot \boldsymbol{w} = \sum_{i=1}^N w_i^2 = N
	\end{equation}
	that satisfy the set of $P$ constraints
	\begin{equation}
		\label{eq::constraints}
		\Delta^\mu(\boldsymbol{w}; \kappa) = \frac{1}{\sqrt{N}}\boldsymbol{w} \cdot \boldsymbol{\xi}^\mu - \kappa \ge 0\,, \quad \forall\mu = 1, \dots, P \,.
	\end{equation}
	$\kappa$ is fixed and it is called \emph{margin}. For $\kappa \ge 0$ the problem is convex and the space of solutions is always connected up to $\alpha_c$, therefore the model does not exhibit the phenomenology discussed in the previous section. Remarkably, however, it turns out that for $\kappa<0$ the problem becomes non-convex and exhibits a considerably richer phase diagram; owing to this peculiarity, it has recently attracted some attention (see below), earning its own name as a model, the ``negative spherical perceptron problem''~\cite{montanari2021tractability}. We will call a configuration $\tilde{w}$ satisfying~\eqref{eq::normalization} and~\eqref{eq::constraints} simply as a \emph{solution} to the problem. In the following we will adopt the same nomenclature for ``typical'' and ``atypical'' solutions as introduced in the previous section.
	
	In the following, in order to draw a fair comparison regarding the impact of the nature of the weights, we will also be interested in the binary analog of the problem ($w_i = \pm 1$). The geometrical organization of the solutions in the binary weight case have already been studied in the $\kappa = 0$ case in~\cite{baldassi2021unveiling}. In this paper we present results valid also for $\kappa < 0$, that we call the ``negative binary perceptron problem''. Even though the negative margin case has not been analyzed before, all the relevant computations can be found in~\cite{baldassi2021unveiling} for generic $\kappa$ and for this reason we do not report them in the Appendices. The results that we present here are in this sense novel, but the phenomenologies that we will describe in the following sections are similar to the $\kappa=0$ case analyzed in~\cite{baldassi2021unveiling}. Indeed, differently from the continuous-weight model, the model is always non-convex no matter what the value of the margin is. As we shall see, the continuous nature of the weights, together with the non-convexity of the landscape, will in some respects change the picture that emerged from the binary case.

	\subsection{Related work}
	
	The negative spherical perceptron has been studied in the statistical mechanics community since the 1980s in the seminal work of Gardner and Derrida~\cite{gardner1988optimal}. In particular they studied, for a fixed value of $\kappa$, the so called SAT/UNSAT transition $\alpha_c$: this is the maximum value of patterns per number of parameters that can be perfectly classified by the network, in the limit of large $N$. They computed it by using the replica method in the Replica-Symmetric (RS) approximation. They also computed the De Almeida-Thouless (dAT) line that marks the onset of the instability of the RS ansatz, showing that the SAT/UNSAT transition is correctly computed in the RS approximation only for $\kappa \ge 0$. We plot $\alpha_c(\kappa)$ in the RS approximation and $\alpha_{\text{dAT}}(\kappa)$ for $\kappa < 0$ in~Fig.~\ref{Fig::phase_diagram}.  
	
	More recently, the negative spherical perceptron has been studied as the ``simplest model of jamming''~\cite{franz2016}: the patterns $\boldsymbol{\xi}^\mu$ can be interpreted as point obstacles in fixed random positions on $\mathcal{S}_N$ and the constraints~\eqref{eq::constraints} correspond to imposing that a particle at position $\boldsymbol{w}$ on $\mathcal{S}_N$ is at a Euclidean distance larger than $\sigma = \sqrt{2N+2\kappa}$ from the point obstacles\footnote{For this reason the variable $\Delta^\mu$ has been named \emph{gap variable}. In the statistical mechanics literature it has been also called \emph{stability} of pattern $\mu$.}. The jamming point corresponds, for a fixed $\alpha$, to the maximization of the margin $\kappa$ (that we will denote from now on as $\kappa_{\text{max}}(\alpha)$\footnote{$\kappa_{\text{max}}(\alpha)$ is simply the inverse function of the SAT/UNSAT transition $\alpha_c(\kappa)$ mentioned before.}), i.e. to the maximization of the distance from the point obtacles. Because of isomorphism with the problem of packing of spheres, the negative perceptron problem has been studied using the replica method in~\cite{franz2017}, where the whole phase diagram of typical solutions has been derived. In particular the authors showed that for low enough margin the model exhibits the classical Random First Order Transition (RFOT) phenomenology: increasing $\alpha$ for a fixed $\kappa$ one first finds a \emph{clustering} transition, then a~\emph{Kautzmann} transition and finally a \emph{Gardner} transition. For clarity, we have plotted those lines in Fig.~\ref{Fig::phase_diagram}; we refer to the caption of the figure and the paper~\cite{franz2017} for their precise definitions. In the same paper, the critical exponents were calculated on the jamming line and they appeared to be equal to the ones found in the jamming of hard spheres in infinite dimensions~\cite{charbonneau2014fractal}. 
	
	In another work~\cite{montanari2021tractability}, Montanari and coworkers studied the performance of several algorithms by characterizing their algorithmic threshold and compared it with rigorous upper and lower bounds to the critical capacity $\alpha_c$. In particular they showed that there exists a gap between the interpolation threshold of Linear Programming (LP) algorithm and the lower bound to $\alpha_c$. They also showed numerically that other algorithms, such as Gradient Descent (GD) and Stochastic GD (SGD) on the cross-entropy loss function, behave much better since they have a much higher algorithmic threshold with respect to the one of LP. This raised the question of the existence of a fundamental computational barrier (such as is common in binary CSPs), whereby no algorithm may able to reach the SAT/UNSAT $\alpha_c$ transition. 
	
	In a previous work~\cite{elAlaoui2022algorithmic}, the authors presented an algorithm based on the Incremental Approximate Message Passing (IAMP), developed in~\cite{montanari2021optimization, elAlaoui2021optimization} for the Sherrington-Kirkpatrick (SK) and the mixed $p-$spin models, which is provably guaranteed to succeed arbitrarily close to the satisfiability threshold $\alpha_c$, provided that the so-called Overlap Gap Property (OGP) is absent. The OGP~\cite{gamarnik2021overlap} is intuitively the property that \emph{any} two near-optimal solution should either be close or far from each other, namely their overlap distribution should display a gap. A formal proof of OGP in the negative perceptron problem is still lacking.



	
	\section{1RSB critical capacity in the negative spherical perceptron}~\label{sec::1rsb_alphac}
	\begin{figure*}	
		\begin{centering}
			\includegraphics[width=0.49\textwidth]{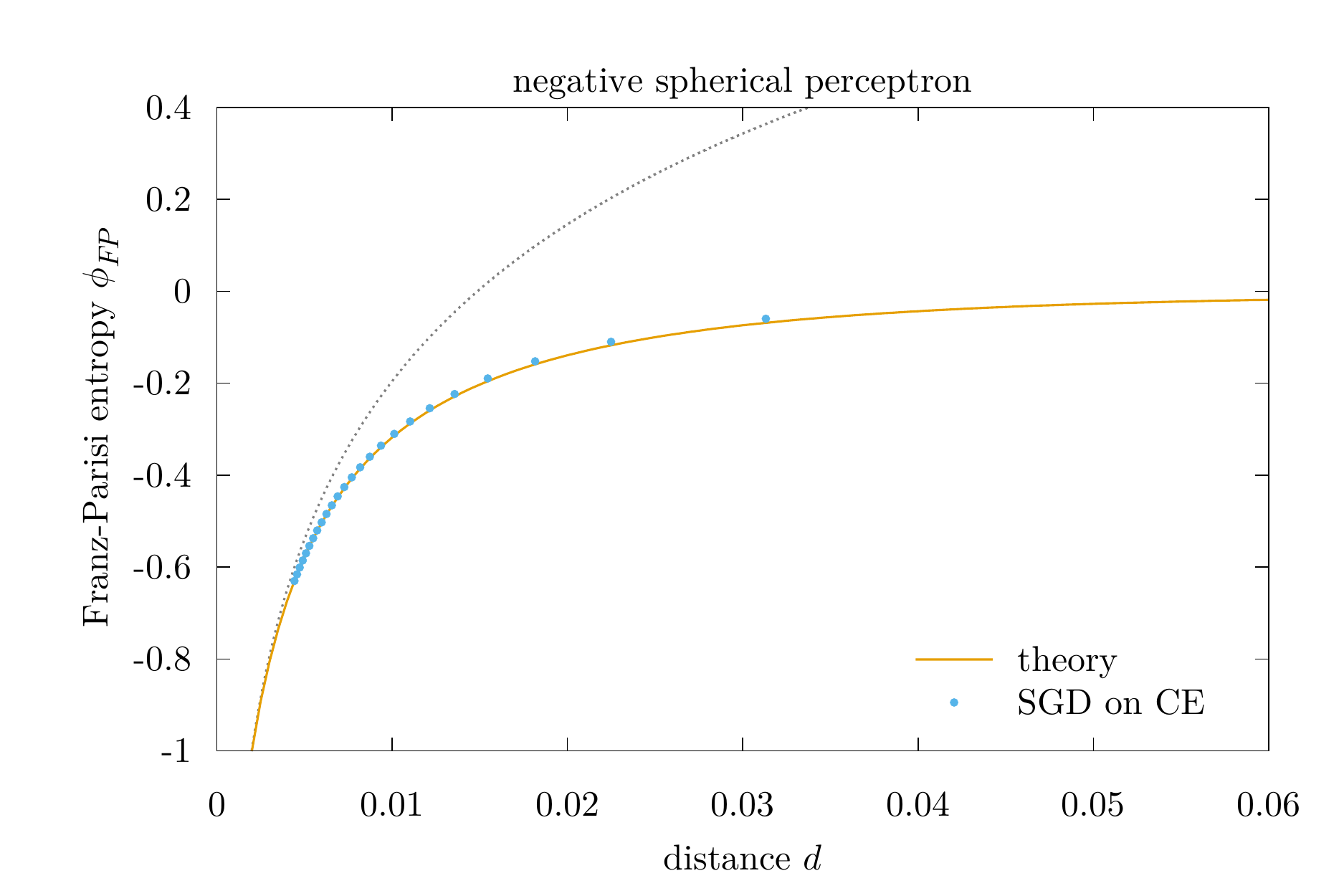}
			\includegraphics[width=0.49\textwidth]{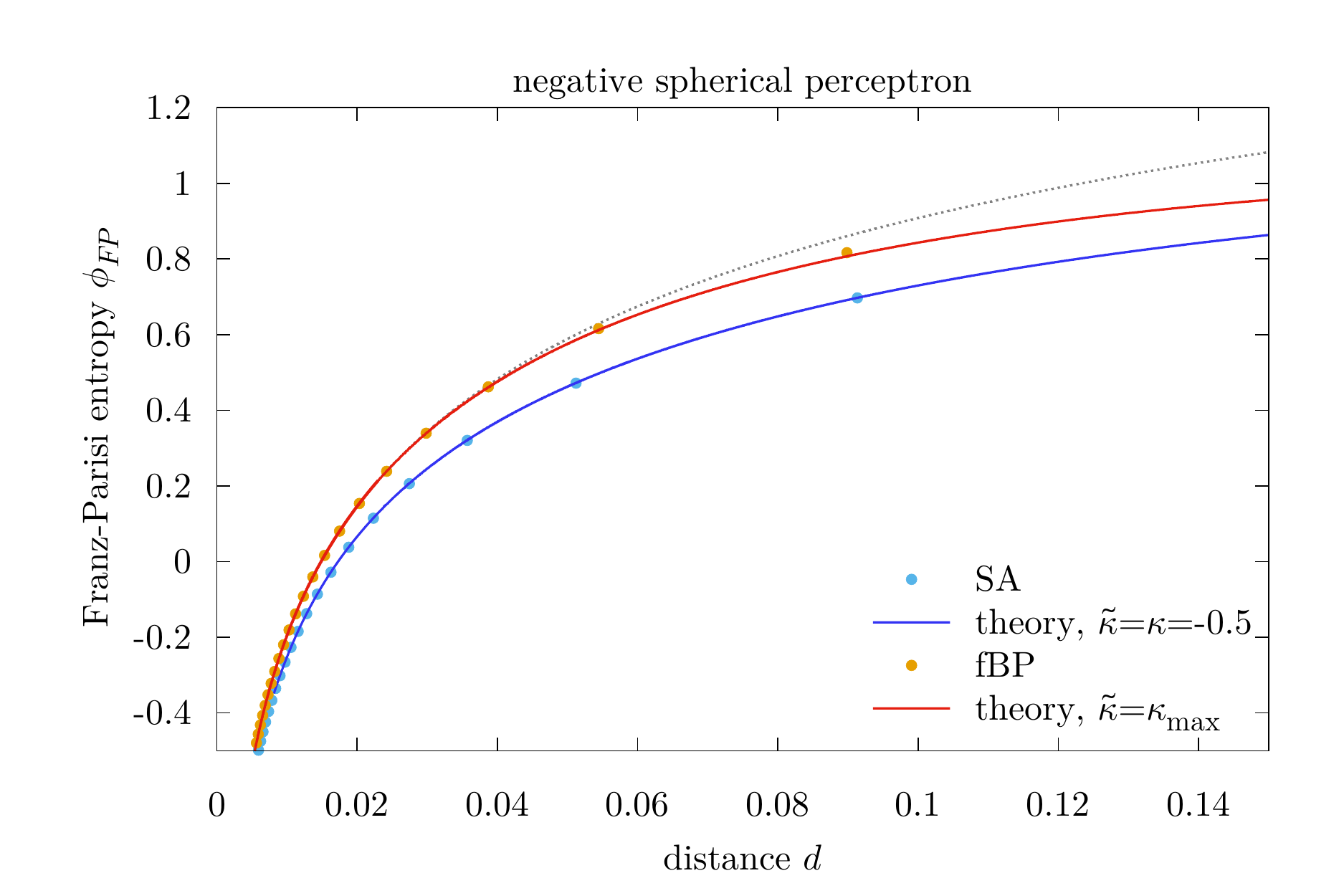}
		\end{centering}
		\caption{ 
			Left panel: average local entropy of a minimizer of the cross-entropy loss function defined in~\eqref{eq::cross_entropy}. The light blue dots are obtained by optimizing the same loss using SGD ($N=1000$). Here we have chosen $\alpha=3.0$,  $\kappa = -0.5$ and $\gamma = 5$.
			Right panel: comparison of the average local entropy of typical and atypical solutions in the spherical negative perceptron. Franz-Parisi entropy for $\kappa=-0.5$ and $\alpha=1$ for a reference extracted from the flat measure over solutions having margin $\tilde{\kappa}$ as in~\eqref{eq::flat_measure_kappat}. The blue line corresponds to the case $\tilde{\kappa} = \kappa$ (i.e. a typical solution); the red curve is instead the curve for the maximum margin $\tilde{\kappa} = \kappa_\text{max}$. Since $\alpha<2$ in this case the RS estimation of $\kappa_\text{max}$ is correct. The dashed line corresponds to the bound in~\eqref{eq::upper_bound_FP_entropy}. The points correspond to the average local entropy estimated using BP on solutions found by SA on the number of error loss (light blue point) and fBP (orange ones). The agreement between theory and simulations is remarkable.
		}
		\label{fig::FP_entropy}
	\end{figure*}
	Following the seminal work of Gardner~\cite{gardner1988The,gardner1988optimal,Gardner_1989} the partition function of the negative perceptron models is 
	\begin{equation}
		Z = \int \!d \mu(\boldsymbol{w}) \, \mathbb{X}_{\boldsymbol{\xi}} (\boldsymbol{w}; \kappa)
	\end{equation}
	where, denoting by $\Theta(\cdot)$ the Heaviside theta function, we have denoted with 
	\begin{equation}
		\mathbb{X}_{\boldsymbol{\xi}} (\boldsymbol{w}; \kappa) \equiv \prod_{\mu = 1}^{\alpha N} \Theta\left( \Delta^\mu(\boldsymbol{w}; \kappa) \right) 
		\label{eq::Gibbs_measure}
	\end{equation}
	the indicator that selects the solutions to the optimization problem~\eqref{eq::constraints}.
	$d \mu(\boldsymbol{w})$ is a measure over the weights that we have introduced to treat both the spherical and binary cases:
	\begin{subequations}
		\begin{align}
			d\mu_{\text{sph}}(\boldsymbol{w}) &= \delta\left( N - \sum_i w_i^2 \right) \prod_{i=1}^{N} dw_i\\
			d\mu_{\text{bin}}(\boldsymbol{w}) &= \prod_{i=1}^{N} dw_i \, \left[ \frac{1}{2}\delta\left( w_i - 1 \right) + \frac{1}{2} \delta\left( w_i + 1 \right)\right] \,.
		\end{align}
	\end{subequations}
	Notice that $d\mu(\boldsymbol{w}) \, \mathbb{X}_{\boldsymbol{\xi}}(\boldsymbol{w}; \kappa)$ being the Gibbs distribution at zero temperature, it is a flat measure over all possible solutions, i.e. it selects typical solutions of the problem according to the definitions we have introduced in Section~\ref{sec::binary_models}. 
	\begin{figure*}	
		\begin{centering}
			\includegraphics[width=0.49\textwidth]{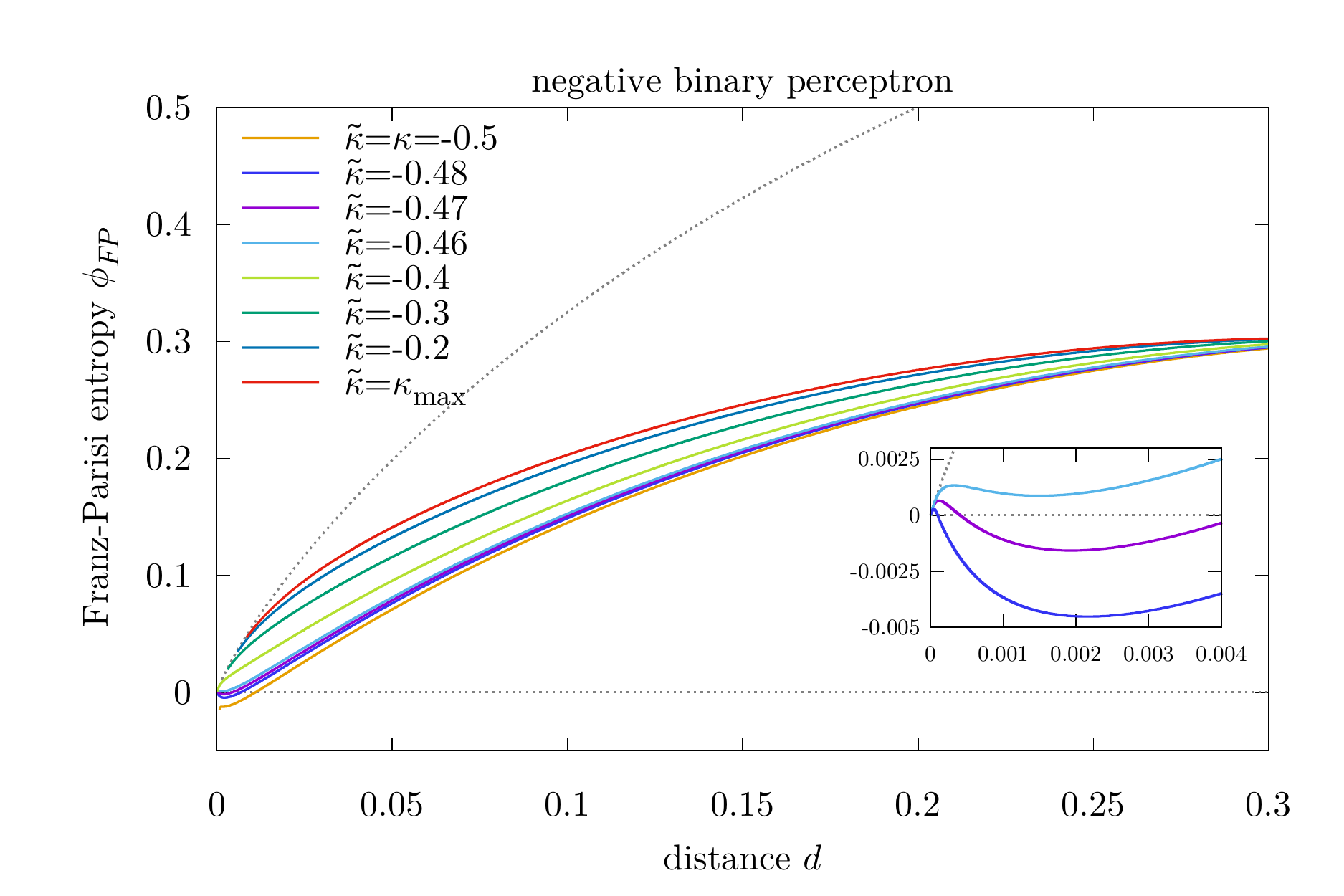}
			\includegraphics[width=0.49\textwidth]{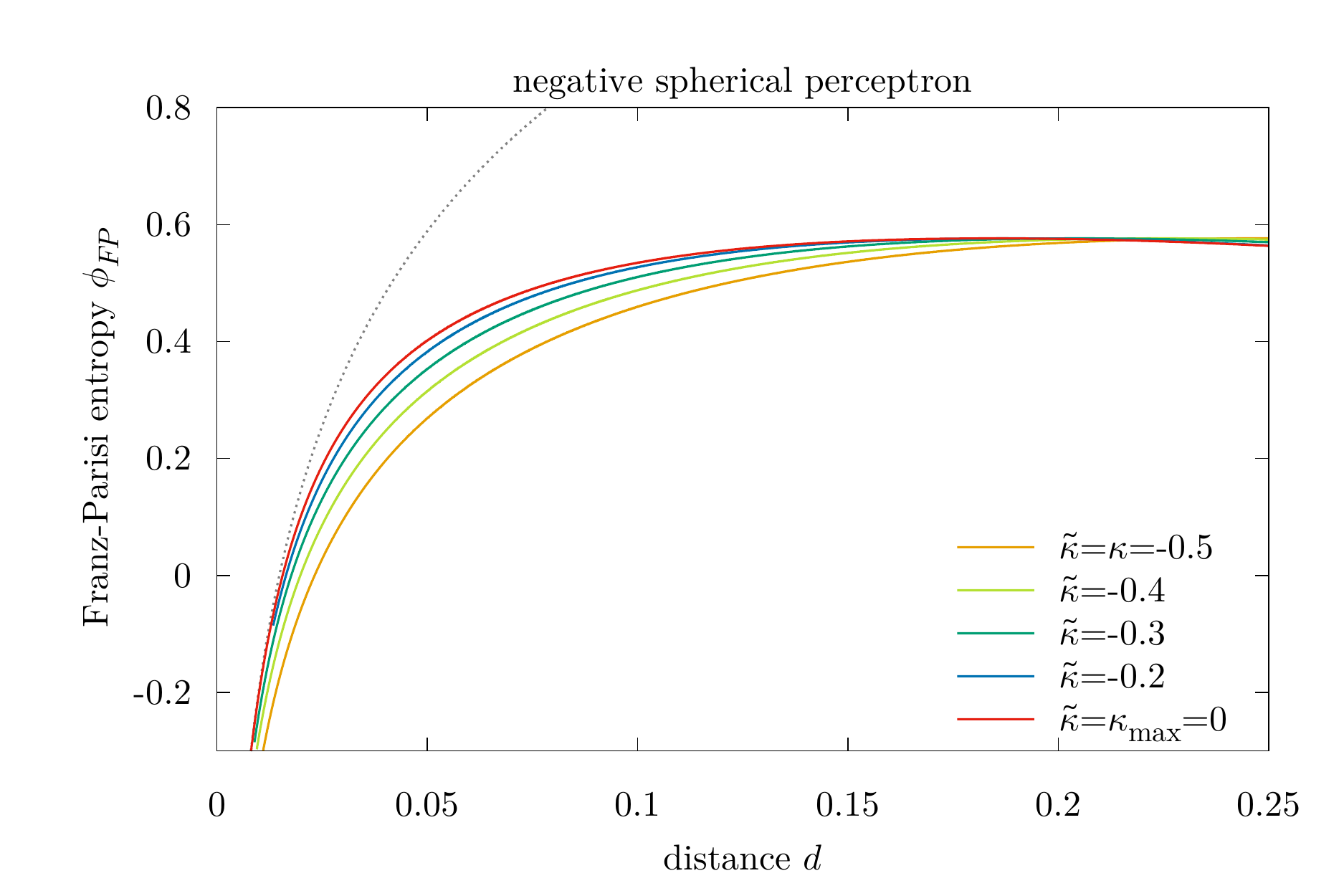}
		\end{centering}
		\caption{
			Average local entropy $\phi_{\mathrm{FP}}$ of typical (i.e. $\tilde{\kappa} = \kappa$) and atypical (i.e. $\tilde{\kappa}>\kappa$) solutions in the binary (left panel, with $\alpha = 1$, $\kappa = -0.5$) and continuous weight models (right panel, with $\alpha = 2$, $\kappa = -0.5$) as a function of the distance. In both cases $\phi_{\mathrm{FP}}$ is computed within the RS ansatz and for several values of the margin $\tilde{\kappa}$ of the reference solution. The dashed grey lines correspond to the upper bound to the entropy as in equation~\eqref{eq::upper_bound_FP_entropy_binary} and~\eqref{eq::upper_bound_FP_entropy} respectively for the binary and spherical cases. 
			The distance at which the curves attain the maximum correspond to the typical distance between solution having margin $\tilde{\kappa}$ and $\kappa$~\cite{Huang_2013}. In the inset plot in the left panel, we show a zoom at small distances of the main plot, to show that for low values of the reference margin the RS local entropy goes below zero; this is not possible in binary models and we interpret it as a lack of solutions at those distances (i.e., low-margin reference solutions are surrounded by thick shells devoid of other solutions). No such phenomenon is observed in continuous models, where negative entropies are valid.
		}
		\label{Fig::typ_vs_atyp}
	\end{figure*}
	One then is interested in computing the free entropy of the system in the thermodynamic limit
	\begin{equation}
		\phi \equiv \lim\limits_{N\to \infty}\frac{1}{N} \left\langle \ln Z \right\rangle_{\boldsymbol{\xi}} \,,
	\end{equation}
	where $\left\langle \cdot \right\rangle_{\boldsymbol{\xi}}$ is the average over all the random patterns $\left\{ \boldsymbol{\xi}^\mu \right\}_{\mu=1}^{\alpha N}$. The free entropy can be computed by using the replica trick~\cite{mezard1987spin}
	\begin{equation}
		\left\langle \ln Z \right\rangle_{\boldsymbol{\xi}} = \lim\limits_{n\to 0} \frac{1}{n} \ln \left\langle Z^n \right\rangle_{\boldsymbol{\xi}}
	\end{equation}
	The entropy can be fully characterized by a $n\times n$ order parameter matrix $q^{ab}$ which physically represents the most probable overlap between two replicas extracted from the Gibbs measure~\eqref{eq::Gibbs_measure}, i.e.
	\begin{equation}
		\label{eq::qab}
		q^{ab} = \frac{1}{N} \sum_{i=1}^{N} \left \langle w_i^a w_i^b \right \rangle \,.
	\end{equation}
	where we have indicated by $\left \langle \cdot \right \rangle$ the average over the Gibbs measure in~\eqref{eq::Gibbs_measure}. 
	We review in appendix~\ref{sec::Typical_case_scenario} the analytical calculations of the entropy in the case in which the structure of the overlap matrix $q^{ab}$ is Replica-Symmetric (RS) 
	\begin{equation}
		q_{ab} = \delta_{ab} + (1-\delta_{ab}) q
	\end{equation}
	or broken at 1-step Replica Symmetry Breaking level (1RSB)
	\begin{equation}
		q_{ab} = q_0 + (q_1-q_0) I_{ab}^{m} + (1-q_1) \delta_{ab}\\
	\end{equation}
	where $I_{ab}^m$ is the $n\times n$ matrix having elements equal to 1 inside the blocks of size $m$ located around the diagonal and 0 otherwise. 
	
	
	In both spherical and binary negative perceptron problems the free entropy is a decreasing function of $\alpha$, meaning that the solution space shrinks when constraints are added. Increasing $\alpha$ one then crosses a critical value $\alpha_c$ (the SAT/UNSAT transition) such that for larger constraint densities there is no solution to the problem. In binary models $\alpha_c$ can be computed easily by evaluating the value of $\alpha$ for which the Replica-Symmetric (RS) free entropy goes to 0~\cite{engel-vandenbroek,krauth1989storage}. In this model the most probable overlap between solutions does not go to 1 at the SAT/UNSAT transition. 
	
	In continuous models the estimation of $\alpha_c$ is instead much harder and requires the use of the full Replica Symmetry Breaking Ansatz (fRSB)~\cite{Parisi1980}. We present in appendix~\ref{sec::Typical_case_scenario} the computation of $\alpha_c$ (or, equivalently of the \emph{maximum margin} $\kappa_{\text{max}}$ at a fixed value of $\alpha$) in the spherical negative perceptron in the RS and 1RSB approximations. In the RS case, this requires to study the limit $q\to 1$. In the 1RSB case, we must consider instead the limit $q_1 \to 1$ with $m \to 0$ and $\tilde{m} \equiv m/(1-q_1)$ finite~\cite{engel1992storage}.
	We plot those two approximations in Fig.~\ref{Fig::phase_diagram}; the 1RSB ansatz shows a substantial change on the estimation of the critical capacity. It can be regarded as a good upper bound to the true SAT/UNSAT transition. For another upper bound (which is slightly less stringent than the one presented here) and a lower bound to the true value of $\alpha_c$ see~\cite{montanari2021tractability}.

	\section{Probing the local entropy landscape using the Franz-Parisi method}~\label{sec::FP_main}

	In this section we discuss how the landscape of solutions of the negative perceptron problem is composed by solutions that can be completely different in nature. In particular, among the various observables that we can compute analytically, we are interested in the so called ``\emph{local entropy}'' of a given solution. Given a configuration $\tilde{\boldsymbol{w}}$, normalized as in eq.~\eqref{eq::normalization} and that solves the set of constraints~\eqref{eq::constraints}, we define its local entropy $\mathcal{S}_\xi(\tilde{\boldsymbol{w}}, S; \kappa)$ as the log of the volume (or number in the binary case) of solutions at a given (normalized) distance $d$ from $\tilde{\boldsymbol{w}}$; namely
	\begin{equation}
		\begin{aligned}
			\mathcal{S}_\xi(\tilde{\boldsymbol{w}}, d; \kappa) &= \frac{1}{N}\ln\mathcal{N}_\xi(\tilde{\boldsymbol{w}}, d; \kappa) \\
			&\equiv \frac{1}{N}\ln\!\int\!\!d\mu(\boldsymbol{w}) \, \mathbb{X}_{\boldsymbol{\xi}} \left(\boldsymbol{w}; \kappa\right) \delta\!\left( N(1-2d)- \boldsymbol{w} \cdot \tilde{\boldsymbol{w}} \right)
		\end{aligned}
	\end{equation}
	Noticing that in the previous definition we have enforced the constraint over the distance by expressing it in terms of the overlap between $\boldsymbol{w}$ and $\tilde{\boldsymbol{w}}$, indeed the two quantities are connected as $d = (1-\boldsymbol{w} \cdot \tilde{\boldsymbol{w}})/2$.
	For any distance $d$ the local entropy is bounded from above by the value it attains for $\alpha = 0$, where $\mathbb{X}_{\xi}(\boldsymbol{w}; \kappa)$ = 1. We call this quantity $\mathcal{S}_\mathrm{max}$. In this case the previous equation simply measures the log of the \emph{total} volume (or the total number in the binary case) of configurations that are at a certain distance by $d$ from $\tilde{\boldsymbol{w}}$. Because of the homogeneity of space, $\mathcal{S}_\mathrm{max}$ cannot depend on $\tilde{\boldsymbol{w}}$, so that we can safely choose $\tilde{w}_i=1$ for every $i = 1, \dots, N$. In the spherical case one gets (see appendix~\ref{sec::upper_bound_franz_parisi})
	\begin{equation}
		\label{eq::upper_bound_FP_entropy}
		\begin{split}
			\mathcal{S}_\mathrm{max} (d) &\equiv \frac{1}{N}\ln\int d\mu_{\mathrm{sph}}(\boldsymbol{w}) \, \delta\left( N (1-2d) - \sum_i w_i \right) \\
			&\overset{N\to \infty}{=} \frac{1}{2} \left[1 + \ln(2\pi) + \ln\left( 1-(1-2d)^2 \right)\right] \,.
		\end{split}
	\end{equation}
	whereas in the binary case we have
	\begin{equation}
		\label{eq::upper_bound_FP_entropy_binary}
		\begin{split}
			\mathcal{S}_\mathrm{max}^{\mathrm{bin}}(d) &\equiv \frac{1}{N}\ln\int 	d\mu_{\mathrm{bin}}(\boldsymbol{w}) \, \delta\left( N (1-2d) - \sum_i w_i \right) \\
			&\overset{N\to \infty}{=} - d \ln(d) - (1-d) \ln(1-d) \,.
		\end{split}
	\end{equation}
	Given a probability distribution $P_\xi(\tilde{\boldsymbol{w}})$ over the set of configurations satisfying the constraints~\eqref{eq::constraints}, we are interested in computing the local entropy of a configuration $\tilde{\boldsymbol{w}}$ sampled from $P_\xi(\tilde{\boldsymbol{w}})$ and averaged over all the possible realizations of the $\alpha N$ patterns:
	\begin{equation}
		\phi_{\text{FP}}(d; \kappa) = \left\langle \int \! d\mu(\tilde{\boldsymbol{w}}) P_\xi(\tilde{\boldsymbol{w}})
		\mathcal{S}_\xi(\tilde{\boldsymbol{w}}, d; \kappa) \right\rangle_{\xi} \,.
	\end{equation}
	This ``averaged local entropy'' was introduced in the context of mean field spin glasses by Franz and Parisi~\cite{franz1995recipes} and for this reason we will call it \emph{Franz-Parisi entropy}. Intuitively, sampling from a distribution $P^1_\xi(\tilde{\boldsymbol{w}})$ that has a larger Franz-Parisi entropy with respect to another one $P^2_\xi(\tilde{\boldsymbol{w}})$ for all distances within some radius will result in solutions that lie in wider and flatter minima. Moreover, we expect a larger number of flat directions around a given solution as its local entropy curve approaches the upper bound $\mathcal{S}_{\text{max}}$ at small distances. The solution will be located in a wider region if the local entropy is monotonic and it tends to saturate the bound for a larger range of distances near the solution itself.
	
	It is interesting therefore to study how extracting the reference configuration $\tilde{\boldsymbol{w}}$ by using a different class of probability distributions $P_\xi(\tilde{\boldsymbol{w}})$ one obtains different types of local entropy curves. Those analytical results on the flatness of a particular class of solutions can then be compared with the ones actually found by several algorithms.
	
	In the next two subsections we explore two different ways of choosing $P_\xi(\tilde{\boldsymbol{w}})$. 
	
	\subsection{Extracting the reference from the minima of a loss function}
	
	\begin{figure*}	
		\begin{centering}
			\includegraphics[width=0.49\textwidth]{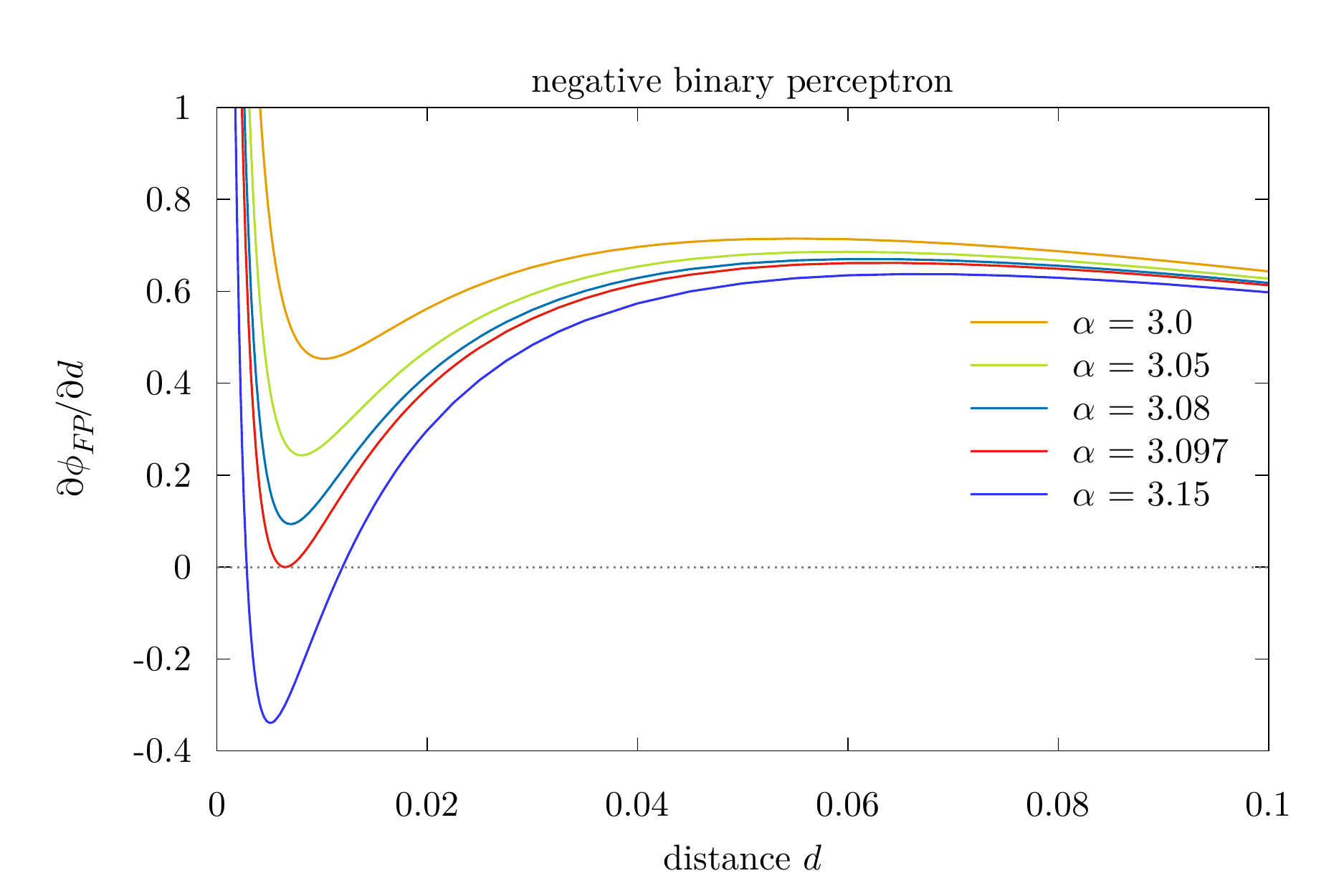}
			\includegraphics[width=0.49\textwidth]{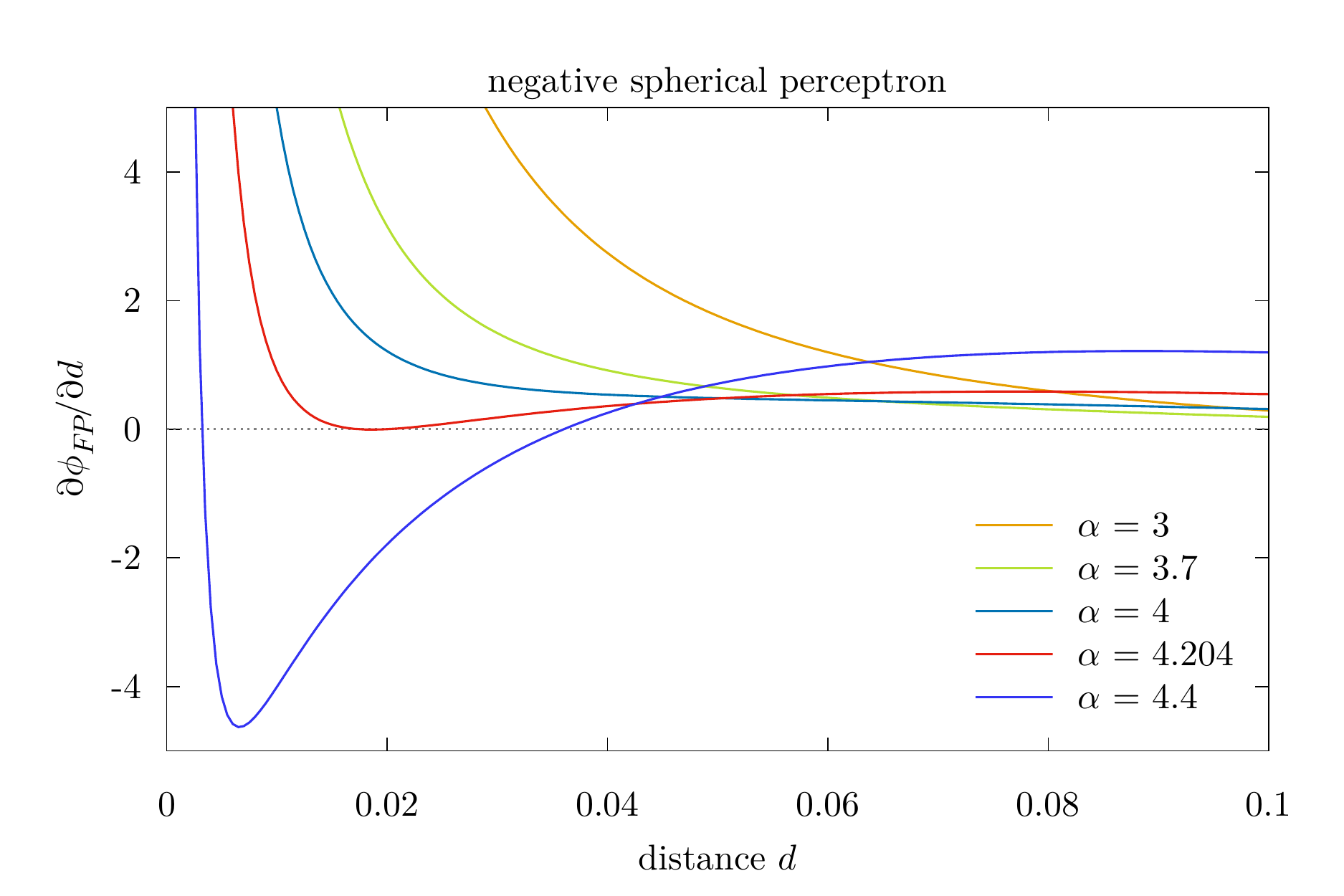}
		\end{centering}
		\caption{
			Derivative of the Franz-Parisi entropy $\partial \phi_{\text{FP}}/ \partial d$ for the maximum margin solutions as a function of $d$ for several values of $\alpha$. The left panel is for the binary case ($\kappa = -1$), whereas the right panel is for the spherical case ($\kappa = -0.5$). Qualitatively the two plots are similar: for small values of $\alpha$ the $\phi_{\text{FP}}$ is monotonic (the derivative is always positive). At $\alpha_{\text{LE}}$, $\partial \phi_{\text{FP}}/ \partial d$ develops a zero at small distances. For $\alpha > \alpha_{\text{LE}}$ the Franz-Parisi entropy is not monotonic. 
		}
		\label{Fig::LE_transition}
	\end{figure*}
	What one usually does to find a solution to~\eqref{eq::constraints} is to introduce a loss function $\mathcal{L}$
	\begin{equation}
		\label{eq:loss}
		\mathcal{L}(\tilde{\boldsymbol{w}}) = \sum_{\mu=1}^{\alpha N} \ell(\Delta^\mu(\tilde{\boldsymbol{w}}; \kappa))
	\end{equation}
	where $\ell(\cdot)$ is a loss per pattern. In first-order algorithms such as gradient descent (GD) and Stochastic GD (SGD) the loss $\mathcal{L}$ needs to be differentiable. 
	A very common loss that is used extensively in machine learning practice is the cross-entropy, which for a binary classification problem has the form\footnote{In standard deep learning practice the ``pseudo-inverse-temperature'' parameter $\gamma$ is not normally used since in the exponent it is redundant with the norm of the last layer; in our models the norm is fixed so we add it explicitly (and keep it fixed). The renormalization by $\gamma^{-1}$ keeps under control the limits of small or large $\gamma$, which could be also achieved by re-parameterizing the SGD learning rate.}
	\begin{equation}
		\label{eq::cross_entropy}
		\ell(x) = \frac{1}{2 \gamma} \log\left( 1 + e^{-2\gamma x}\right) 
	\end{equation}
	Notice that by introducing a loss function~\eqref{eq:loss} can dramatically change the non-convex nature of the original problem specified by~\eqref{eq::constraints}; indeed if the loss is convex (as in the cross-entropy case), since the constraints are linear in the weights, the problem of minimization of~\eqref{eq:loss} subject to the normalization of the weights~\eqref{eq::normalization} becomes convex as well. Nevertheless, the relationship between the minima of the loss and the solutions of the original problem is nontrivial at all~\cite{baldassi2020shaping}. 
	
	In order to study the local entropy landscape around the minimizer found by such algorithms we use as reference configurations $\tilde{\boldsymbol{w}}$ the ones extracted from the measure
	\begin{equation}
		\label{eq::generic_measure}
		P_\xi(\tilde{\boldsymbol{w}}) = \frac{e^{-\beta \sum_{\mu=1}^{\alpha N} \ell(\Delta^\mu(\tilde{\boldsymbol{w}}; \kappa))}}{\int d \mu(\tilde{\boldsymbol{w}}^\prime) \, e^{-\beta \sum_{\mu=1}^{\alpha N} \ell(\Delta^\mu(\tilde{\boldsymbol{w}}^\prime; \kappa))}}
	\end{equation}
	In order to focus on the minima of the loss function $\mathcal{L}$ we take the large $\beta$ limit.
	
	We plot in the left panel of Fig.~\ref{fig::FP_entropy} the RS Franz-Parisi entropy of a minimum of the cross entropy loss function. We show in the same plot that the theoretical computation is in striking agreement with numerical simulations done by optimizing the same loss with the SGD algorithm. The local entropy of a configuration of weights has been computed using Belief Propagation (BP), see~\cite{baldassi_local_2016} for the details of the implementation.
	
	\subsection{Flat measure over solutions having margin $\tilde{\kappa}$}
	
	Secondly, we studied the average local entropy of solutions sampled using a flat measure over all configurations having margin $\tilde{\kappa} \ge \kappa$, that is
	\begin{equation}
		\label{eq::flat_measure_kappat}
		P_\xi(\tilde{\boldsymbol{w}}) = 	\frac{\mathbb{X}_\xi(\tilde{\boldsymbol{w}}; \tilde{\kappa})}{\int d \mu(\tilde{\boldsymbol{w}}^\prime) \, \mathbb{X}_\xi(\tilde{\boldsymbol{w}}^\prime; \tilde{\kappa})}
	\end{equation}
	This measure can be obtained from~\eqref{eq::generic_measure} by using as loss function the ``error counting loss''
	\begin{equation}
		\mathcal{L}(\tilde{\boldsymbol{w}}) = \sum_{\mu=1}^{\alpha N} \Theta(- \Delta^\mu(\tilde{\boldsymbol{w}}; \tilde{\kappa}))
	\end{equation}
	which counts the number of violated constraints in the dataset, and then sending $\beta$ to infinity. In Appendix~\ref{sec::FP} we have therefore sketched the computations for the generic measure~\eqref{eq::generic_measure} and then we have specialized in Appendix~\ref{app::FP_error_counting_loss} to the error counting loss case.

	In the case $\tilde{\kappa} = \kappa$ we are sampling typical solutions to the problem since we are sampling among all solutions with flat measure. We studied how the Franz-Parisi curve changes as we vary the margin $\tilde{\kappa}$ in the RS ansatz (see Appendix~\ref{sec::FP} and Fig.~\ref{Fig::FP_1RSB}). Notice that if one samples using a flat measure over solutions having margin $\tilde{\kappa}> \kappa$, the measure~\eqref{eq::flat_measure_kappat} is not flat in the space of solutions of margin $\kappa$, i.e. they are ``atypical''. Moreover, as shown in~\cite{baldassi2021unveiling} the entropy of solutions with margin $\tilde{\kappa}>\kappa$ is exponentially lower than those ones having margin $\kappa$, making them to be ``subdominant''.
	In addition, since a solution having a margin $\tilde\kappa > \kappa$ solves a more constrained problem than~\eqref{eq::constraints}, we intuitively expect that its Franz-Parisi entropy will be higher than that for $\tilde{\kappa} = \kappa$.
	
	The Franz-Parisi entropy for the distribution~\eqref{eq::flat_measure_kappat} has been already analytically derived in~\cite{baldassi2021unveiling}, but only the problem with $\kappa = 0$ was analyzed. Here to make a fair comparison on the impact on the nature of the weights we have extended those reults to the case $\kappa < 0$ obtaining a similar phenomenology to that described in~\cite{baldassi2021unveiling}, see the left panel of Fig.~\ref{Fig::typ_vs_atyp}. 
	
	Firstly, if one samples a typical solution to the problem (i.e. with $\tilde
	\kappa = \kappa$) one will find that there exists, for any (arbitrarily small) value of $\alpha$, a neighborhood of $d=0$ where the Franz-Parisi entropy is negative~\cite{baldassi2021unveiling}. This implies that, within a certain distance from the reference, only a sub-extensive number of solutions can be found. One therefore says that typical solutions are \emph{isolated}~\cite{huang2014origin,abbe2021proof,perkins2021frozen}.  
	Secondly, if one samples solutions having larger margin with respect to the one of the problem $\tilde{\kappa} > \kappa$ one finds that there always exists a neighborhood around $d=0$ having positive average local entropy. Therefore solutions having larger margin are always surrounded by an exponential number of other solutions within a small but extensive distance. Moreover as one decreases the distance from the reference further, the local entropy curve becomes nearly indistinguishable from the total number of configurations at that distance, implying that the cluster where the reference is located is very dense. If $\alpha$ is sufficiently small, the local entropy becomes monotonic as the margin $\tilde{\kappa}$ is increased.
	
	Differently from the binary weights case, in the spherical case we found that no typical solution is actually isolated, there is always a non-vanishing volume of solutions at a given distance from it, see Fig.~\ref{Fig::typ_vs_atyp}, right panel. Moreover if $\alpha$ is low enough even a typical solution has a monotonic local entropy.
	Apart for those two differences, the general picture is similar: as the margin $\tilde{\kappa}$ is increased, the local entropy gets larger in a given range of distances from the reference, meaning that those (atypical) solutions are located inside a denser cluster. As in the binary case, the solution having the largest local entropy at small distances is the one sampled with the maximum margin $\tilde{\kappa} = \kappa_{\text{max}}$. We refer to appendix~\ref{sec::app_FP_1RSB} for a discussion of the RSB effects that play an important role in the large $\alpha$ regime.
	
	
	In the right panel of Fig.~\ref{fig::FP_entropy}, we also show the comparison between the average local entropy of typical and atypical solutions in the negative spherical perceptron and the agreement with the one found by two algorithms: Simulated Annealing on the number-of-errors loss and fBP. The latter algorithm was explicitly designed to target flatter solutions, see~\cite{baldassi2015subdominant} for the details of its implementation; we found that fBP finds solutions whose local entropy is comparable to (even slightly larger than) the theoretical one found imposing the maximum possible margin for the reference, as was previously observed in binary models~\cite{baldassi2021unveiling}.
	In Appendix~\ref{sec::APP_num_exp} we show that even for larger value of $\alpha$, where the RS ansatz on the order parameters for the reference configuration is wrong (i.e. $\alpha > \alpha_\text{dAT}(\tilde{\kappa})$) the agreement with numerical simulations is still rather good.
	
	\begin{figure*}	
		\begin{centering}
			\includegraphics[width=0.49\textwidth]{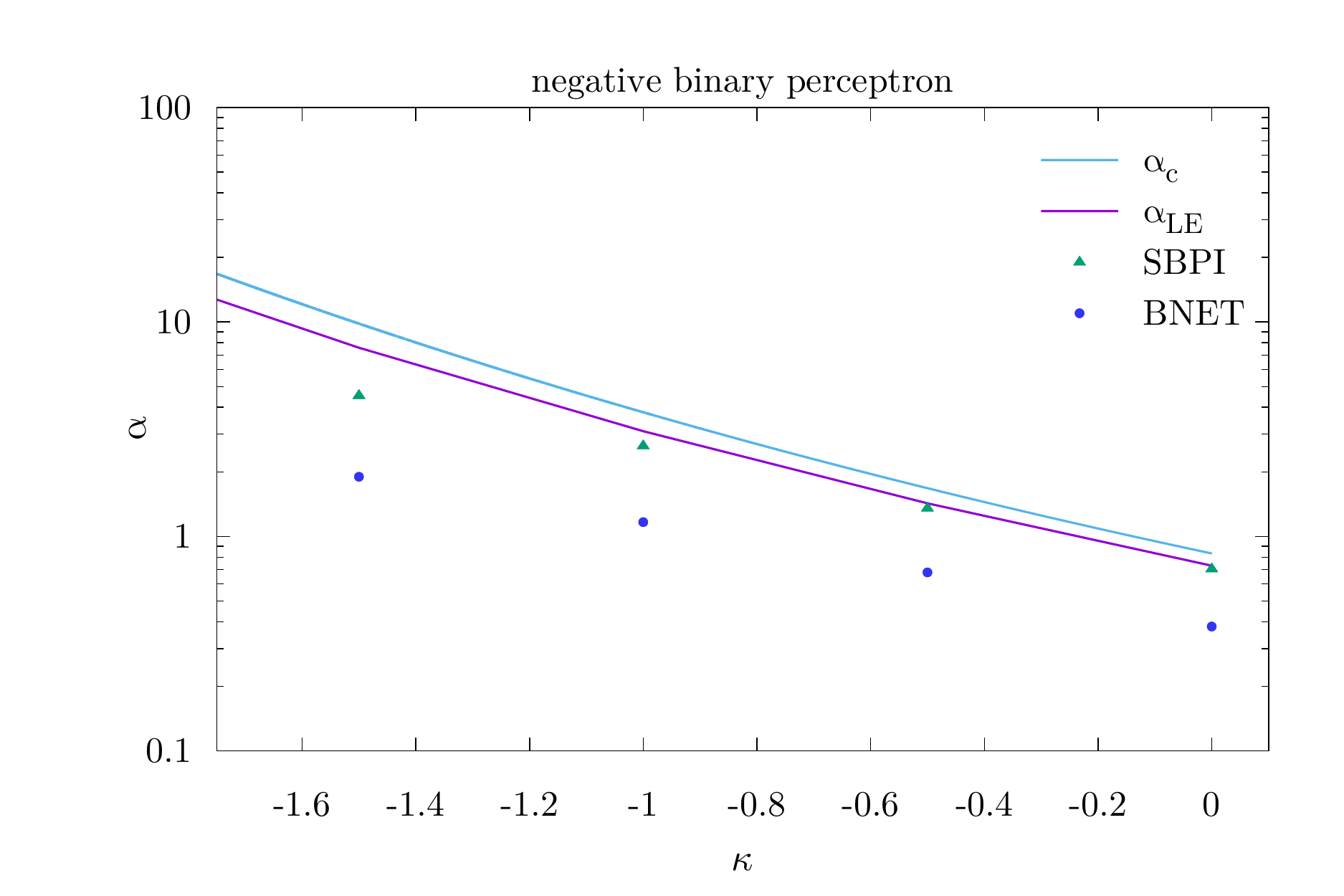}
			\includegraphics[width=0.49\textwidth]{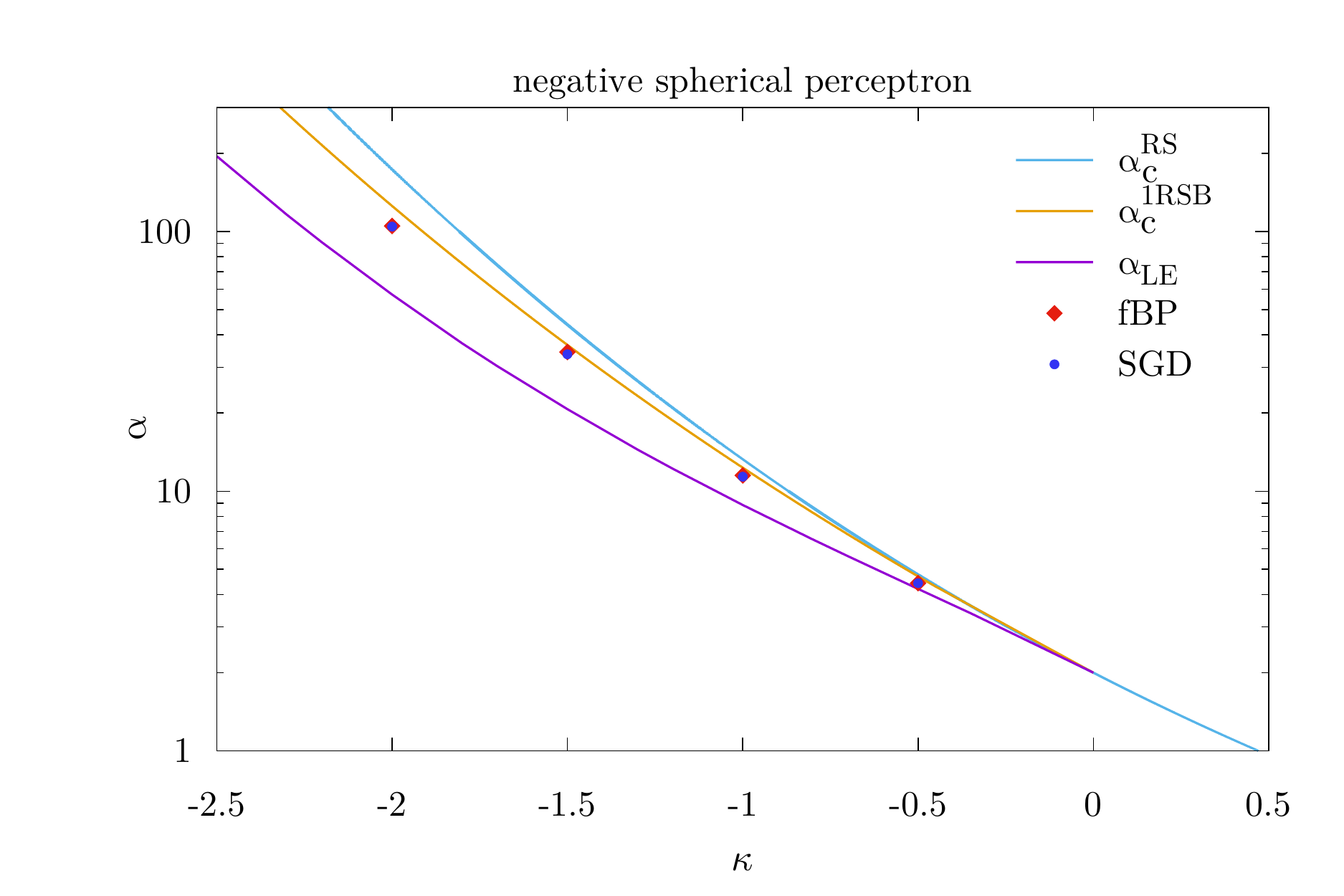}
		\end{centering}
		\caption{			
			SAT/UNSAT and local entropy transitions as a function of $\kappa$ for the binary (left) and spherical (right) cases. The points represent the highest value of $\alpha$ that we were able to reach at $N=1000$ with several algorithms: fBP, SBPI~\cite{baldassi2007efficient} and Binary NET (BNET)~\cite{hubara2016binarized} in the binary case; fBP and SGD for the continuous case. In the spherical case we plot the critical capacity both in the RS and 1RSB approximations. Notice also that in this case the local entropy transition coincides with the SAT/UNSAT transition for $\kappa \ge 0$, since the solution space is connected and convex.
		}
		\label{Fig::alphaLE}
	\end{figure*}
	\subsection{Phase transition in the geometrical organization of solutions: local entropy transition}
	Next, we investigate what happens to the widest and flattest minima as we increase the constraint density $\alpha$. We therefore study the local entropy profile of the maximum-margin solutions for several increasing values of $\alpha$, for a fixed value of $\kappa$. 
	
	
	In Fig.~\ref{Fig::LE_transition} we plot $\frac{\partial \phi_{\mathrm{FP}}}{\partial d}$ as a function of the distance $d$. The phenomenology is quite similar in both the binary and the spherical cases. If $\alpha$ is lower than a critical threshold $\alpha_{\mathrm{LE}}(\kappa)$, the local entropy profile exhibits only one maximum (not shown in the figure), located at the typical distance $d_{\mathrm{typ}}(\alpha)$ between solutions with margin $\tilde{\kappa} = \kappa_{\mathrm{max}}$ and $\kappa$; for $d < d_{\mathrm{typ}}$ the local entropy is monotonic with positive derivative. This means that the reference is located in a wide and flat region that extends to very large scales~\cite{baldassi2015subdominant,baldassi2021unveiling}. For $\alpha = \alpha_{\mathrm{LE}}$, there appears at small distances another point where the derivative $\frac{\partial \phi_{\mathrm{FP}}}{\partial d}$ vanishes.
	For $\alpha > \alpha_{\mathrm{LE}}$ the local entropy is non-monotonic and it has a local maximum at a distance $d_\star < d_{\mathrm{typ}}(\alpha)$: this suggests that the most robust solutions are no longer located in regions that extend to arbitrary large distances, but that have a typical size $d_\star$ instead. This \emph{Local Entropy transition}~\cite{baldassi2015subdominant, baldassi2021unveiling} occurring at $\alpha_{\mathrm{LE}}(\kappa)$ can therefore be interpreted as the point at which the cluster of atypical robust solutions fractures in many pieces. 
	
	In~\cite{baldassi2021unveiling} the local entropy transition has been computed for the binary perceptron model for $\kappa = 0$ and it has been shown that it gives similar results to the more precise method of finding the reference that maximizes the local entropy at every distance~\cite{baldassi2015subdominant}. In the same works, moreover, it has been shown that this change in the geometry of atypical solutions strongly affects the behaviour of algorithms: no known algorithm is able to find solutions for $\alpha > \alpha_{\text{LE}}$.
	We plot in Fig.~\ref{Fig::alphaLE} the local entropy transition as a function of $\kappa$ for the binary (left panel) and for the spherical case (right panel). In the same plots we show the algorithmic threshold of several algorithms. In the same plots we show the SAT/UNSAT transition, which was computed by using the zero entropy criterium in the binary case~\cite{krauth1989storage} and by using the RS and 1RSB approximations in the spherical case. In the left panel of Fig.~\ref{Fig::alphaLE} we can see that in the binary case no algorithm is able to cross the local entropy transition; in addition fBP which is an algorithm designed to target maximally entropic regions appears to stop working \emph{exactly} at the local entropy transition. In the spherical case, this is not the case: even if the atypical states fracture in many pieces, algorithms are still able to overcome the threshold and find solutions. Indeed the landscape of solutions is very different in the two models: in the spherical case even typical solutions are surrounded by an exponential number of solutions up to capacity. The algorithmic thresholds plotted in the right panel of Fig.~\ref{Fig::alphaLE} seem to suggest that algorithms are able to reach the SAT/UNSAT transition of the model, especially knowing that taking into account higher order RSB corrections can considerably lower the estimate of $\alpha_c$. Binary and spherical models are thus significantly different from the optimization point of view.
	
	Notice that the computation of $\alpha_{\mathrm{LE}}$ could still be very imprecise in the spherical case because of the presence of large RSB effects. However, when $\kappa$ is near zero, we expect the RSB corrections to play a minor role and our computation to be reliable; indeed the RS estimate of the maximum margin configuration is expected to approximate quite well the true value. On the other hand, when the margin is very negative, the RS estimate of $\kappa_{\mathrm{max}}$ is very imprecise (cf. Fig.~\ref{Fig::phase_diagram}); therefore we expect our estimate of $\alpha_{\mathrm{LE}}$ to be imprecise as well. In the appendix we describe a method to estimate the local entropy transition that does not rely on the ability to sample a replica in deep RSB phase; the method is therefore expected to give much more precise estimates when $\kappa$ is large in modulus.

	


	\section{Numerical experiments}	\label{sec::numerical_experiments}
	
	\subsection{Numerical justification of the local entropy transition}
	\begin{figure}	
		\begin{centering}
			\includegraphics[width=0.48\textwidth]{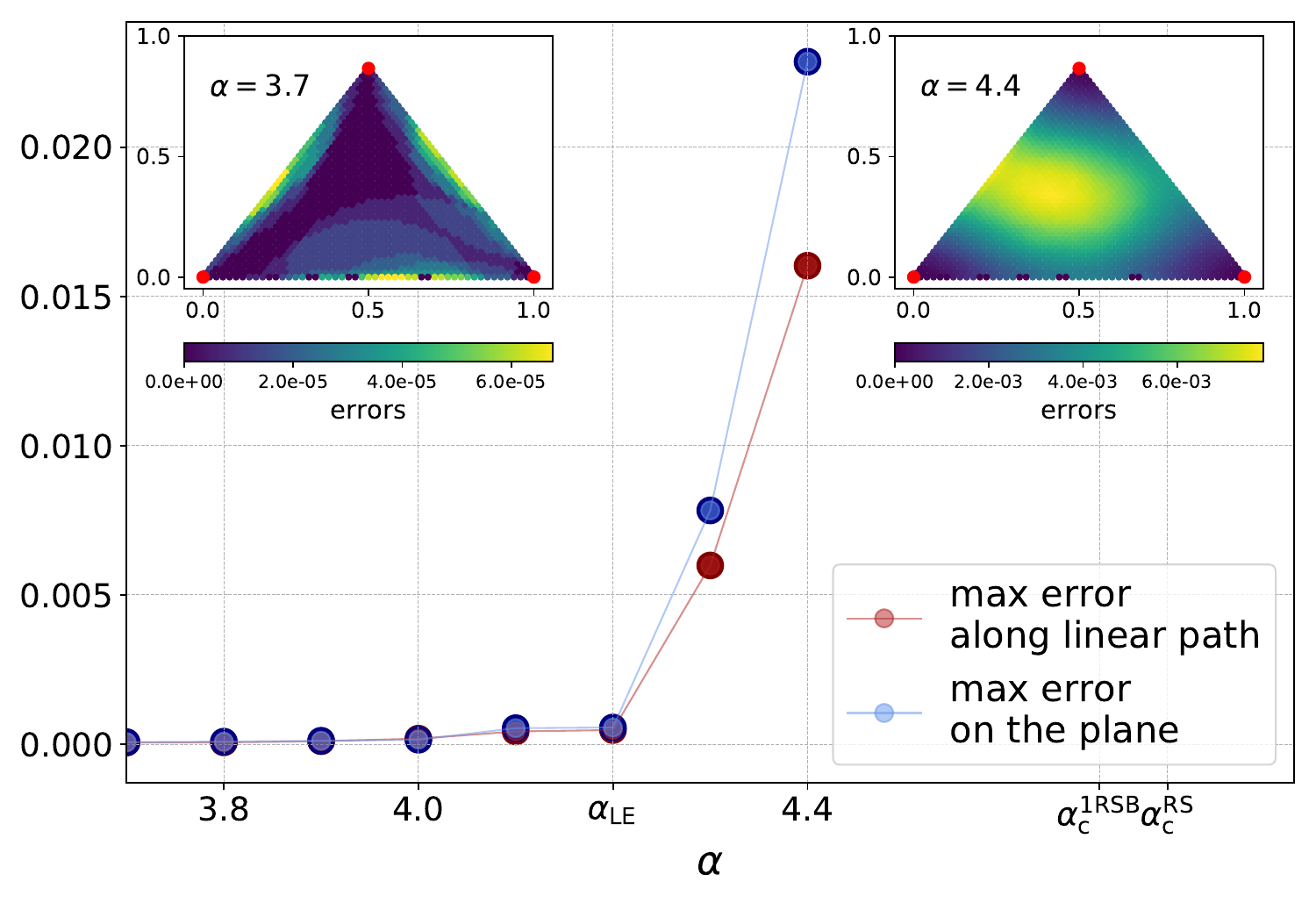}
		\end{centering}
		\caption{ 
			Average maximum error fraction along the geodesic path connecting two solutions (red points) and on the $2D$ manifold (``plane'') spanned by three solutions on the $N$-dimensional hypersphere surface (blue points).
			A high error barrier between solutions appears before the algorithmic threshold $\alpha_c$, and its onset is compatible with the local entropy transition (in this case $\alpha_{\mathrm{LE}}(\kappa=-0.5) \simeq 4.2$). 
			The two inset plots show the error on the plane spanned by three solutions (represented by red dots at the vertices of the triangle): 
			for $\alpha<\alpha_{\mathrm{LE}}$, the error along the linear paths (edges of the triangles) is tiny, and it is even smaller at the barycenter of the plane;
			for $\alpha>\alpha_{\mathrm{LE}}$ a high error peak appears in the barycenter. Configurations were obtained by optimizing the hinge loss with margin $\kappa=-0.5$ using SGD. Points are averages over 20 realizations of the patterns and 5 different runs for each dataset. The value of $N$ is 2000.  
		}
		\label{Fig::paths}
	\end{figure}

	While we expect to observe a clear difference in algorithmic behaviour  between the discrete and continuous versions of the model at the local entropy transition, we still expect to observe in both cases a  structural geometrical change. Beyond the local entropy transition the discrete model displays a disconnected 1-RSB structure of solutions also at the out-of-equilibrium level. 
	On the contrary, the continuous version displays a full-RSB structure which is expected to be accessible (though not particularly flat). For the discrete case several works have already clarified the  phenomenon both analytically and numerically, see~\cite{Braunstein2006,baldassi2007efficient,baldassi2022learning,baldassi2021unveiling}. Here we focus on the continuous case. 
	
	We measure numerically the error of a weight vector $\boldsymbol{w}_{\boldsymbol{\gamma}}$ obtained as a convex combination of $y$ solutions $\boldsymbol{w}^a$ with $a = 1\,, \dots\,, y$ and normalized on the sphere in $N$ dimension of radius $\sqrt{N}$, namely
	\begin{equation}
		\boldsymbol{w}_{\boldsymbol{\gamma}} \equiv \frac{\sqrt{N} \sum_{a=1}^y \gamma_a \boldsymbol{w}^a}{\lVert\sum_{a=1}^y \gamma_a \boldsymbol{w}^a \rVert}\,, \qquad \sum_{a=1}^y \gamma_a = 1\,, \quad \forall a:\,\gamma_a\ge 0
	\end{equation}
	The study of the training error around geodesic paths connecting the same or different classes of solutions is actually an interesting problem in its own right~\cite{annesi2023star}. 
	Here we provide some preliminary numerical results on the simple cases $y=2$ or $3$ in which the solutions $\boldsymbol{w}^a$ are obtained with SGD with the hinge loss $\ell\left(x\right)=\max\left(0,-x\right)$ (the margin is included in $x$, see eq.~\eqref{eq:loss}).
	In particular, the case $y=2$ amounts at computing the barrier along a ``linear'' (geodesic) path connecting two given solutions. 
	This topic has been widely studied in deep learning literature~\cite{Draxler,entezari2022the,pittorino2022deep} as it is believed to be a good proxy to probe the error landscape around solutions.
	Based on the phenomenology exhibited by deep networks, we expect two robust solutions to be connected by an almost zero error path~\cite{pittorino2022deep}.
	This is in fact what we observe in the overparameterized regime (low $\alpha$), see Fig.~\ref{Fig::paths}.
	However, as the constraint density is increased, a barrier in the linear path connecting solutions appears, in a region close to the RS estimate of the local entropy transition.
	Moreover, if we study the error landscape on the ``plane'' (2$D$ manifold) spanned by $y=3$ solutions, we see that, for $\alpha > \alpha_{\mathrm{LE}}$, a high error region appears in the barycenter, signaling that right above the local entropy transition SGD starts to find solutions that are likely to be located in different basins. 
	
	\subsection{Connections with generalization}
\begin{figure}	
		\begin{centering}
			\includegraphics[width=0.49\textwidth]{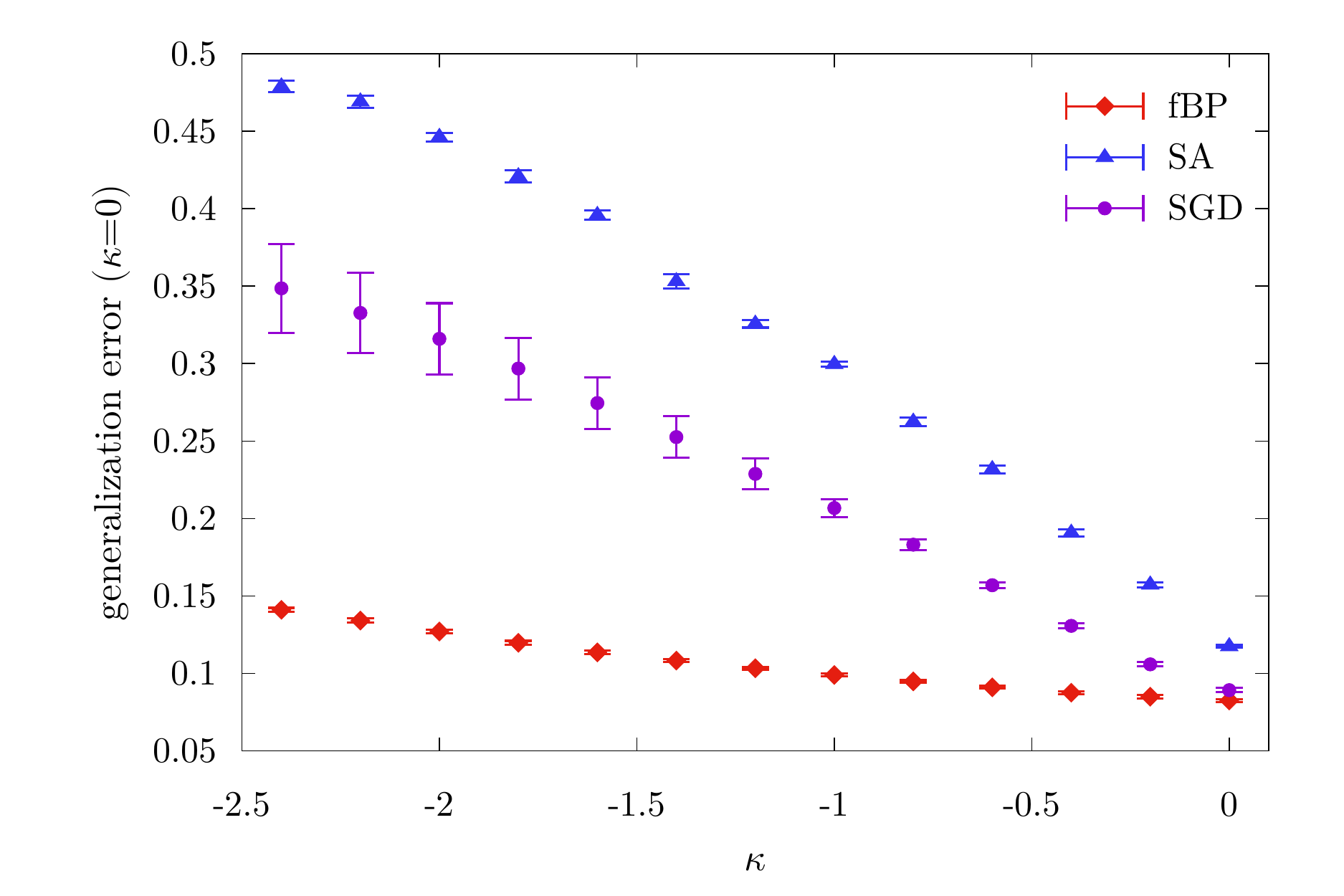}
		\end{centering}
		\caption{
			Generalization error in the teacher student setting as a function of the margin. The teacher assigns labels with zero margin, while the students solve the problem at fixed negative margin $\kappa$. The generalization error is then  computed using zero margin (see text). Results are averaged over $10$ independent teacher realizations and 10 random restarts for each dataset.
		}
		\label{Fig::generalization}
	\end{figure}

	In order to probe numerically the computational advantages of wide flat minima and to create a natural link to future studies on multilayered models, we have analyzed the generalization properties in a teacher-student setting.
	Specifically we generate data with a random teacher perceptron ($\kappa=0$) and train a student perceptron with negative $\kappa$. Once learning has completed, we test the generalization performance of the student with zero margin.
	Remarkably, we find that -- provided we converge into wide flat minima -- even learning with very negative values of $\kappa$ (a very under-constrained learning problem, with very little signal coming for the training set) leads to good generalization performance, see Fig.~\ref{Fig::generalization} for the continuous case. Learning with fBP leads to minimizers which are well inside the flat region and as such are effectively robust, even though the robustness condition coming from the learning constraint is very weak (the negative $\kappa$). Other algorithms display different degree of robustness depending  on the details, such as the effective temperature $\gamma$ of the cross-entropy loss minimized by SGD, or the cooling schedule for Simulated Annealing. Similar behaviours are found in the binary case, which we refer to the appendix. 

	\section{Conclusions}~\label{sec::conclusions}
	
	In this paper we have studied the binary and spherical negative-margin perceptrons, i.e. two simple non-convex neural networks learning a random rule with binary and spherical weights. We have analyzed in both models the geometry of the landscape of solutions, showing important similarities but also differences. First, we have pointed out how the typical solutions of the models are substantially different: in the binary case for any $\alpha>0$ the landscape is composed by an extensive number of clusters with vanishing entropy; in the spherical case, instead, typical solutions are always surrounded by an exponential number of other solutions, i.e. they are not isolated. This is the first result that shows that the low-energy landscapes of non-convex neural networks having binary and spherical weights are completely different in nature. Secondly, we have studied highly robust (i.e. high-margin) but exponentially rare solutions of both problems, and showed that in both cases those configurations have a larger local entropy compared to typical solutions. In binary models the configurations having the largest local entropy corresponds to the one having the largest margin; in continuous weights models the same is true, modulo RSB effects (see Appendix).
	
	Finally, we analyzed the solutions with the largest local entropy as a function of the constraint density $\alpha$ in both binary and spherical models, unveiling a phase transition in the geometry of the widest and flattest states: the local entropy transition $\alpha_{\mathrm{LE}}$. This transition can be computed by finding the largest value of $\alpha$ for which the maximum margin solutions (i.e. ones having the largest local entropy), have a local entropy that is monotonic. This transition signals a break-up of the space of robust solutions into disconnected components. Indeed, we have verified numerically in the spherical case, that for $\alpha < \alpha_{\mathrm{LE}}$ algorithms find solutions lying in the same basin, whereas for $\alpha > \alpha_{\mathrm{LE}}$ we observe a sudden rise of energy barriers along the geodesic path between pairs of solutions and in between triplets of solutions. Even if we cannot rule out at this time the existence of non-geodesic zero-energy paths connecting these solutions, this is already an indication of a profound change in the structure of the manifold of solutions. 
	
	In binary models we have verified that the transition has a very strong impact on the behavior of algorithms. In spherical models, even if we have a strong indication that the geometry of the space of solutions is undergoing a radical change in structure near the transition, it does not appear to have an impact on algorithmic hardness. As we have shown, efficient algorithms are probably able to reach the SAT/UNSAT transition, which we have computed here in the 1RSB ansatz.
	
	We have verified that similar analytical findings are found also in models, the so called \emph{tree-committee} machine~\cite{relu_locent}, where the non-convexity in the problem is not induced by the negative margin but by the presence of an additional layer and a generic non-linearity. Interesting future research directions involve the extension of those results to models presenting the notion of generalization and finally the analytical investigation of the connectivity properties of solutions in neural network models~\cite{annesi2023star}.

	\section*{Acknowledgements} 
	E.M.M. wishes to thank R. D\'iaz Hern\'andez Rojas, S. Franz, P. Urbani and F. Zamponi for several interesting discussions.

	\noindent \bibliographystyle{unsrturl}
	\bibliography{references}
	
	\onecolumngrid
	\appendix
	
	\tableofcontents
	
	\section{The model and its partition function (Gardner volume)}
	
	The partition function of a perceptron learning a set of patterns $\boldsymbol{\xi}^\mu$ ($\mu = 1, \dots, P=\alpha N$) having label $y^\mu = \pm 1$ with margin $\kappa$ optimizing a generic loss $\ell$ is
	\begin{equation}
		Z = \int \!d \mu(\boldsymbol{w}) \, e^{-\beta \sum_{\mu=1}^{P} \ell(\Delta^\mu(\boldsymbol{w}; \kappa))}
	\end{equation}
	where we have denoted by
	\begin{equation}
		\Delta^\mu(\boldsymbol{w}; \kappa) \equiv \frac{y^\mu}{\sqrt{N}} \sum_{i=1}^{N} w_i \xi_i^\mu - \kappa
	\end{equation}
	the stability of pattern $\boldsymbol{\xi}^\mu$ for the given set of weights $\boldsymbol{w}$. $d \mu(\boldsymbol{w})$ is a generic measure over the weights. In the paper it has been considered the case of spherical (i.e. $\sum_{i=1}^{N} w_i^2 = N$) and binary (i.e. $w_i = \pm 1$) weights. 
	To lighten the discussion, we will present in detail the computations for the spherical case only. For the corresponding derivations of the binary case we refer to~\cite{relu_locent,baldassi2021unveiling} where the same computations have already been presented. Notice that in~\cite{baldassi2021unveiling} only the organization of solutions of the problem with margin $\kappa = 0$ has been analyzed. In this paper we present results valid also for $\kappa < 0$, and we get phenomenologies similar to the $\kappa=0$ case. Indeed, differently from the continuous-weight model, the model remains always non-convex even when the margin is positive.
	
	We consider the case of random i.i.d. Gaussian patterns and $\pm 1$ labels with equal probability. In this case all the labels $y^\mu$ can be considered all to be 1, since they can reabsorbed in the patterns using the Gauge transformation $\xi^\mu \to y^\mu \xi^\mu$. 
	
	We will also consider a generic loss per pattern $\ell$; but in some sections we will specialize to the error counting loss $\ell(x) = \Theta(-x)$, where $\Theta(\cdot)$ is the Heaviside's theta function.

	\section{Replica computation for a generic loss function}~\label{sec::Typical_case_scenario}
	We want here to compute the free entropy of the model averaged over the training set disorder (which will be denoted by $\left\langle \cdot \right\rangle_{\boldsymbol{\xi}}$) in the large $N \to \infty$:
	\begin{equation}
		\phi = \lim\limits_{N\to \infty} \frac{1}{N} \left\langle\ln Z \right\rangle_{\boldsymbol{\xi}} \,.
	\end{equation}
	We resort to a standard tool used in the theory of disordered systems, namely the replica method: $\left\langle\ln Z \right\rangle_{\boldsymbol{\xi}} = \lim\limits_{n\to0} \frac{\left\langle Z^n \right\rangle_{\boldsymbol{\xi}}}{n}$. The averaged replicated partition function can be written as 
	\begin{equation}
		\left\langle Z^n \right\rangle_{\boldsymbol{\xi}} = \int \prod_{a\ne b} d q_{ab} \, \prod_{a b} d \hat q_{ab} \ e^{N 	\left[G_S(q_{ab}, \hat{q}_{ab}) + \alpha G_E(q_{ab})\right]}
	\end{equation}
	where we have defined the overlap $q_{ab}$ between two replicas as
	\begin{equation}
		q_{ab} = \frac{1}{N} \sum_{i=1}^{N} w_i^a w_i^b \,.
	\end{equation}
	Notice that, because of the spherical constraint, $q_{aa} = 1$. The ``conjugated order parameter'' $\hat q_{ab}$ was introduce in order to enforce the definition of $q_{ab}$ via a delta function, as usual in replica computations.
	Moreover we have defined the so called ``entropic'' and ``energetic'' terms, which are respectively given by
	\begin{subequations}
		\begin{align}
			G_{S} &=-\frac{1}{2}\sum_{a,b}q_{ab}\hat{q}_{ab}+\log\int\prod_{a}\mathrm{d}w^{a}\ e^{\frac{1}{2}\sum_{a,b}\hat{q}_{ab}w^{a}w^{b}} \\
			G_{E} &= \log\int\prod_{a}\frac{\mathrm{d}u^{a}\mathrm{d}\hat{u}^{a}}{2\pi} e^{-\beta \sum_a\ell(u^a - \kappa) -i\sum_{a}\hat{u}^{a}u^{a}-\frac{1}{2}\sum_{ab}q_{ab}\hat{u}^{a}\hat{u}^{b}}
		\end{align}
	\end{subequations}
	In the large $N$ limit we can use a saddle point approximation in order to evaluate the free entropy $\phi$. The saddle point equations are
	\begin{subequations}
		\begin{align}
			q_{ab} &= \frac{\int\prod_{c}\mathrm{d}w^{c}\ e^{\frac{1}{2}\sum_{c,d}\hat{q}_{cd}w^{c}w^{d}} w^a w^b}{\int\prod_{c}\mathrm{d}w^{c}\ e^{\frac{1}{2}\sum_{c,d}\hat{q}_{cd}w^{c}w^{d}}} \equiv \langle w^a w^b \rangle_{S}\\
			\hat q_{ab} &= -\frac{\int\prod_{a}\frac{\mathrm{d}u^{a}\mathrm{d}\hat{u}^{a}}{2\pi} e^{-\beta \sum_a\ell(u^a - \kappa) -i\sum_{a}\hat{u}^{a}u^{a}-\frac{1}{2}\sum_{ab}q_{ab}\hat{u}^{a}\hat{u}^{b}}}{\int\prod_{a}\frac{\mathrm{d}u^{a}\mathrm{d}\hat{u}^{a}}{2\pi} e^{-\beta \sum_a\ell(u^a - \kappa) -i\sum_{a}\hat{u}^{a}u^{a}-\frac{1}{2}\sum_{ab}q_{ab}\hat{u}^{a}\hat{u}^{b}}} \equiv - \alpha \langle \hat u^a \hat u^b\rangle_{E}
		\end{align}
	\end{subequations}
	In order to solve the saddle point equations one searches solutions restricting the space of the matrices to particular subspaces. Here we will adopt a so called Replica-Symmetric ansatz in section~\ref{app::RS_generic_loss} and a one-step replica Symmetry Breaking ansatz in section~\ref{sec::1RSB_ansatz}.

	\subsection{Replica-Symmetric ansatz} \label{app::RS_generic_loss}
	We now impose a Replica-Symmetric (RS) ansatz: 
	\begin{subequations}
		\begin{align}
			\label{eq::RS_reference}
			q_{ab} &= \delta_{ab} + (1-\delta_{ab}) q \\
			\hat{q}_{ab} &= -\hat{Q} \, \delta_{ab} + (1-\delta_{ab}) \hat q		
		\end{align}
	\end{subequations}
	We therefore have that the free entropy is
	\begin{equation}
		\phi = \mathcal{G}_{S} + \alpha \mathcal{G}_{E}
	\end{equation}
	where the new entropic and energetic terms are
	\begin{subequations}
		\label{eq::GS_GE_generic_loss}
		\begin{align}	
			\mathcal{G}_{S}  &\equiv \lim\limits_{n\to 0} \frac{G_S}{n} = \frac{1}{2} \hat Q + \frac{q \hat q}{2} + \frac{1}{2} \ln \frac{2\pi}{\hat Q + \hat q } + \frac{1}{2} \frac{\hat q}{\hat Q + \hat q } \\
			\mathcal{G}_E &\equiv \lim\limits_{n\to 0} \frac{G_E}{n} = \int Dz_0 \ln \int Dz_1 \, e^{-\beta \ell\left(\sqrt{q} \, z_0  + \sqrt{1-q} \, z_1 - \kappa \right)} \,.
		\end{align}
	\end{subequations}	
	The saddle point equations are
	\begin{subequations}
		\label{eq::SPeq_RS}
		\begin{align}
			q & = \frac{\hat q}{(\hat Q + \hat q)^2}\\
			1 & = \frac{\hat Q + 2 \hat q}{(\hat Q + \hat q)^2} \,,\\
			\hat q & = - 2 \alpha \frac{\partial \mathcal{G}_E}{\partial q} = \alpha \int Dz_0 \left[\frac{\int Dz_1 \, e^{-\beta \ell\left(\sqrt{q} \, z_0  + \sqrt{1-q} \, z_1 - \kappa \right)} \beta \ell'\left(\sqrt{q} \, z_0  + \sqrt{1-q} \, z_1 - \kappa \right) }{\int Dz_1 \, e^{-\beta \ell\left(\sqrt{q} \, z_0  + \sqrt{1-q} \, z_1 - \kappa \right)}}\right]^2 \label{eq::SP_loss_generic}
		\end{align}
	\end{subequations}
	where $\ell'$ is the derivative of the loss $\ell$. 
	The saddle point equations can be solved for a generical value of the margin which has to be intended as an external parameter of the problem. The order parameters therefore depend \emph{implicitly} on the margin via the the solve saddle point equations they solve.
	
	Notice that the first two saddle point equations can be explicitly inverted, which give an expression of conjugated parameters $\hat q$, $\hat Q$ in terms of $q$
	\begin{subequations}
		\label{eq::sol_eqRS_hat}
		\begin{align}
			\hat q & = \frac{q}{(1-q)^2}\\
			\hat Q & = \frac{1-2q}{(1-q)^2} \,.
		\end{align}
	\end{subequations}
	The spherical term therefore can be written as
	\begin{equation}
		\mathcal{G}_{S}  = \frac{1}{2(1-q)} + \frac{1}{2} \ln\left( 1-q \right)
	\end{equation}
	From the solution of the saddle point equations~\eqref{eq::SPeq_RS} one can compute all the quantities of interest. For example, the training loss is
	\begin{equation}
		E = \frac{1}{N} \langle \mathcal{L}(\boldsymbol{w})\rangle = - \frac{\partial \phi}{\partial \beta} = - \alpha \frac{\partial \mathcal{G}_E}{\partial \beta} = \alpha \int Dz_0 \, \frac{ \int Dz_1 \, e^{-\beta \ell\left(\sqrt{q} z_0  + \sqrt{1-q} z_1 - \kappa \right)}  \ell\left(\sqrt{q} z_0  + \sqrt{1-q} z_1 - \kappa\right)}{\int Dz_1 \, e^{-\beta \ell\left(\sqrt{q} z_0  + \sqrt{1-q} z_1 - \kappa\right)}}
	\end{equation}
	where we have indicated by $\langle \cdot \rangle$ the average over the Gibbs measure
	\begin{equation}
		\langle \bullet \rangle \equiv \mathbb{E}_{\boldsymbol{\xi}}
		\frac{\int \!d \mu(\boldsymbol{w}) \, e^{-\beta \sum_{\mu=1}^{P} \ell(\Delta^\mu(\boldsymbol{w}; \kappa))} \, \bullet}{Z}
	\end{equation}
	Taking the average over the Gibbs measure of the loss $ \ell(x) = \Theta(-x)$ where, $\Theta(\cdot)$ is the Heaviside Theta function, we get the training error
	\begin{equation}
		\label{eq::train_error}
		\begin{split}
			\epsilon_t &= \int Dz_0 \, \frac{ \int Dz_1 \, e^{-\beta \ell\left(\sqrt{q} z_0  + \sqrt{1-q} z_1 - \kappa \right)}  \Theta\left( \kappa -\sqrt{q} z_0 - \sqrt{1-q} z_1\right)}{\int Dz_1 \, e^{-\beta \ell\left(\sqrt{q} z_0  + \sqrt{1-q} z_1 - \kappa\right)}} = \int Dz_0 \, \frac{ \int_{-\infty}^{\frac{\kappa - \sqrt{q} z_0}{\sqrt{1 - q}}} Dz_1 \, e^{-\beta \ell\left(\sqrt{q} z_0  + \sqrt{1-q} z_1 - \kappa \right)} }{\int Dz_1 \, e^{-\beta \ell\left(\sqrt{q} z_0  + \sqrt{1-q} z_1 - \kappa\right)}}
		\end{split}
	\end{equation}
	The entropy can be computed using the thermodynamic relation $S = \phi + \beta E$.

	\subsection{The case of a convex loss: the infinite $\beta$ limit} \label{app::infinite_beta_limit_reference}
	
	Here we consider the case in which the loss $\ell(x)$ is convex and we will consider it to have a unique minimizer. In this case the large $\beta$ limit makes $q$ (the overlap between two different replica configuration) tend to 1. In the infinite $\beta$ limit one therefore expects the scaling 
	\begin{equation}
		\label{eq::q_large_beta}
		q \simeq 1 - \frac{\delta q}{\beta}	
	\end{equation}
	This implies two scalings for $\hat q$ and $\hat Q$
	\begin{subequations}
		\begin{align}
			\hat q &= \beta^2 \delta \hat q - \beta \delta \hat Q\\
			\hat Q&= - \beta^2 \delta \hat q + 2\beta \delta \hat Q 
		\end{align}
	\end{subequations}
	so that
	\begin{subequations}
		\begin{align}
			\hat q + \hat Q &= \beta \delta \hat Q\\
			\hat q + \frac{\hat Q}{2}&= \frac{\beta^2}{2} \delta \hat q
		\end{align}
	\end{subequations}
	so that the first two saddle point equations become
	\begin{subequations}
		\begin{align}
			\delta q & = \frac{1}{\delta \hat Q} \\
			1 & = \frac{\delta \hat q}{\delta \hat Q^2} \,.
		\end{align}
	\end{subequations}
	In the large $\beta$ limit we have
	\begin{equation}
		\begin{split}
			-f = \lim\limits_{\beta \to \infty} \frac{\phi}{\beta} & = \frac{\delta \hat Q}{2} -\frac{\delta \hat q \delta q}{2} + \frac{\delta \hat q}{2 \delta \hat Q} + \alpha \int Dz_0  \, A(z_0) =  \frac{1}{2 \delta q} +  \alpha \int Dz_0 \, A(z_0)  \,.
		\end{split}
	\end{equation}
	where we have defined
	\begin{equation}
		A(z_0) \equiv \max_{z_1}\left[-\frac{z_1^2}{2}- \ell\left(z_0  + \sqrt{\delta q} \, z_1 - \kappa\right) \right]
	\end{equation}
	Notice that doing the large $\beta$ limit for a convex loss is equivalent of doing in spirit the maximum margin limit for the theta loss (see later). Let us call 
	\begin{equation}
		\label{eq::z1star}
		z_1^\star(z_0; \kappa) \equiv \argmax_{z_1}\left[-\frac{z_1^2}{2} - \ell\left(z_0  + \sqrt{\delta q} \, z_1 - \kappa\right) \right]
	\end{equation}
	which satisfies the equation
	\begin{equation}
		z_1^\star = - \sqrt{\delta q} \, \ell'\left(z_0  + \sqrt{\delta q} \, z_1^\star - \kappa\right)
	\end{equation}
	We have removed for simplicity the explicit dependence of $z_1^\star$ on $z_0$ and $\kappa$. The saddle point equation for $\delta q$ then reads
	\begin{equation}
		\frac{1}{\delta q^2} = \alpha \int D z_0 \frac{d A(z_0)}{d \delta q} = - \alpha \int D z_0 \,  \ell'\left( z_0  + \sqrt{\delta q} \, z_1^\star - \kappa \right) \frac{z_1^\star}{2 \sqrt{\delta q}}  =  \frac{\alpha}{\delta q} \int Dz_0 \, \left( z_1^\star \right)^2
	\end{equation}
	or
	\begin{equation}
		\delta q = \left[ \alpha \int Dz_0 \, \left( z_1^\star \right)^2 \right]^{-1} \,.
	\end{equation}
	The training loss and error become
	\begin{subequations}
		\begin{align}
			E &= \alpha \int Dz_0 \, \ell\left(z_0  + \sqrt{\delta q} \, z_1^\star - \kappa \right) \,, \\
			\epsilon_t &= \int Dz_0 \, \Theta\left(\kappa - z_0  - \sqrt{\delta q} \, z_1^\star \right) \,.
		\end{align}
	\end{subequations}

	\subsection{The error-counting loss case}
	Starting from the RS expressions in Appendix~\ref{app::RS_generic_loss}, we specialize to the error counting loss case, that is we choose $\ell$ to be 
	\begin{equation}
		\ell(x) = \Theta(-x)
	\end{equation}
	The expression of the energetic term in~\eqref{eq::GS_GE_generic_loss} can be simplified. For clarity we report the full expression of the free entropy (the entropic term $\mathcal{G}_S$ does not depend at all on the loss function)
	\begin{subequations}
		\begin{align}
			\phi &= \mathcal{G}_{S}  + \alpha \mathcal{G}_{E}  \\
			\mathcal{G}_{S}  & =\frac{1}{2} \hat Q + \frac{q \hat q}{2} + \frac{1}{2} \ln \frac{2\pi}{\hat Q + \hat q } + \frac{1}{2} \frac{\hat q}{\hat Q + \hat q } = \frac{1}{2(1-q)} + \frac{1}{2} \ln\left( 1-q \right)\\
			\mathcal{G}_{E} & =\int Dz_{0}\,\ln H_\beta\left(\frac{\kappa + \sqrt{q} \, z_{0}}{\sqrt{1-q}}\right)
		\end{align}
	\end{subequations}
	where we have defined $H_\beta(x) = e^{-\beta} + (1-e^{-\beta}) H(x)$ and $H(x) = \frac{1}{2} \text{Erfc}\left(\frac{x}{\sqrt{2}} \right)$. The infinite $\beta$ limit of the previous expressions is trivial to perform, since we only need to set $H_{\beta} \to H$ in the log term of the energetic term $\mathcal{G}_E$. The saddle point equation~\eqref{eq::SP_loss_generic} is substituted by
	\begin{equation}
		\hat q = - 2 \alpha \frac{\partial \mathcal{G}_E}{\partial q} = \frac{\alpha}{1-q}  \int Dz_0 \left[\frac{(1-e^{-\beta}) G\left(\frac{\kappa + \sqrt{q} z_0}{\sqrt{1-q}}\right)}{H_\beta\left(\frac{\kappa + \sqrt{q} z_0}{\sqrt{1-q}}\right)}\right]^2 
	\end{equation}
	with $G(x)$ being a standard normal distribution. In particular, in the infinite $\beta$ limit, this reduces to
	\begin{equation}
		\label{eq::SP_err_beta_inf}
		\hat q = \frac{q}{(1-q)^2}=\frac{\alpha}{1-q}  \int Dz_0 \, \left[GH\left(\frac{\kappa + \sqrt{q} z_0}{\sqrt{1-q}}\right) \right]^2 
	\end{equation}
	where we have defined the function $GH(x) \equiv \frac{G(x)}{H(x)}$. 
	The training error in equation~\eqref{eq::train_error} becomes
	\begin{equation}
		\label{eq::train_error_theta}
		\epsilon_t = e^{-\beta} \int Dz \, \frac{H\left( - \frac{\kappa + \sqrt{q} \, z }{\sqrt{1-q}} \right)}{H_\beta\left(\frac{\kappa + \sqrt{q} \, z }{\sqrt{1-q}} \right)}
	\end{equation}
	which vanishes in the infinite $\beta$ limit. In the following we specialize on the infinite $\beta$ limit, which, as said before is trivial to perform (it does not require any scaling of $q$ with $\beta$ as in~\eqref{eq::q_large_beta}). 
	
	\subsubsection{The $\kappa_\text{max}$ limit} \label{sec::kmax_limit}
	To find the maximum possible margin for a fixed value of $\alpha$ we should do the $q\to 1$ limit. Here we follow the seminal work of Gardner~\cite{gardner1988The,gardner1988optimal}. Using the fact that $\ln H(x) \simeq - \frac{1}{2} \ln(2\pi) - \ln x - \frac{x^2}{2}$ as $x \to \infty$, retaining only the diverging terms we get
	\begin{equation}
		\begin{split}
			&\int Dz_0 \ln H\left(\frac{\kappa + \sqrt{q} z_{0}}{\sqrt{1-q}}\right) \simeq \int_{-\kappa}^{+\infty} Dz_0 \left[ \frac{1}{2} \ln \delta q - \frac{\left( \kappa + z_0 \right)^2}{2 \delta q} \right]=\frac{1}{2} \ln (\delta q) \, H\left( -\kappa \right) - \frac{B(\kappa)}{2 \delta q}
		\end{split}
	\end{equation}
	where we have indicated, with a slight abuse of notation $\delta q \equiv 1-q$ (it is indeed \emph{not} the same quantity that appears in scaling~\eqref{eq::q_large_beta} which was introduced to deal with the large $\beta$ limit of a generic convex loss) and
	\begin{equation}
		B(\kappa) = \int_{-\kappa}^{\infty} Dz_0 \, (\kappa + z_0)^2 = \kappa G\left( \kappa \right) + \left( \kappa^2 + 1 \right) H\left( -\kappa \right) \,.
	\end{equation}
	The free entropy is therefore
	\begin{equation}
		\label{eq::phi}
		\phi =  \frac{1}{2 \delta q} + \frac{1}{2} \ln \delta q + \frac{\alpha}{2} \left( \ln (\delta q) \, H\left( -\kappa \right) - \frac{B(\kappa)}{\delta q} \right)
	\end{equation}
	The derivative with respect to $\delta q$ is 
	\begin{equation}
		2 \frac{\partial \phi}{\partial \delta q} = \frac{1}{\delta q} - \frac{1}{\delta q^2} + \alpha \left( \frac{H\left( -\kappa \right)}{\delta q} + \frac{B(\kappa)}{\delta q^2}\right) = 0 \,.
	\end{equation}
	Now we can exploit the fact that when $q \to 1$ for fixed $\alpha$ we are near the maximum value of the margin we can impose $\kappa = \kappa_\text{max} - \delta \kappa$, i.e. $\delta q \simeq C' \delta \kappa$. However performing this expansion in the previous formula is a little bit tricky. It is convenient to do the critical capacity limit, with $\kappa$ fixed, i.e. $\alpha = \alpha_c - \delta \alpha$ and $\delta q = C \delta \alpha$. The two approaches of course are equivalent and are connected simply by $C' = C \frac{\partial \alpha}{\partial k}$. We get
	\begin{equation}
		\begin{split}
			2 \frac{\partial \phi}{\partial \delta q} &= \frac{1}{C \delta\alpha} - \frac{1}{C^2 \delta\alpha^2} + \left( \alpha_c - \delta \alpha \right) \left[ \frac{H\left(-\kappa \right)}{C \delta\alpha} + \frac{B(\kappa)}{C^2 \delta\alpha^2} \right] \\
			&= \frac{1}{C \delta\alpha} \left[ 1 + \alpha_c H\left( -\kappa \right) - \frac{B(\kappa)}{C}\right] + \frac{1}{C^2 \delta\alpha^2} \left[\alpha_c B(\kappa) - 1 \right] = 0 \,.
		\end{split}
	\end{equation}
	The first term gives the scaling of $\delta q$, the second gives us the critical capacity in terms of the margin~\cite{gardner1988The,gardner1988optimal}. 
	\begin{equation}
		\alpha_c = \frac{1}{B(\kappa)}
	\end{equation}
	or the maximum margin applicable $\kappa_\text{max}$ in terms of $\alpha$
	\begin{equation}
		B(\kappa_\text{max}) = \frac{1}{\alpha} \,.
	\end{equation}
	Notice that $\alpha_c = \frac{1}{B(\kappa)}$ is equivalent to imposing that the divergence $1/\delta q$ in the free entropy~\eqref{eq::phi} is eliminated at the critical capacity (so that it correctly goes to $-\infty$ in that limit).

	\subsubsection{Stability of the Replica-Symmetric Ansatz}
	As shown in~\cite{franz2017}, the RS ansatz is not always correct for any value of $\alpha$ in the SAT phase. Indeed, analogously to the de-Almeida-Thouless line in the Sherrington-Kirkpatrick model~\cite{AlmeidaThouless1978}, for any value of $\alpha$ it exists a value of $\kappa$ that we will call $\kappa_{\text{dAT}}(\alpha)$, above which the RS ansatz is unstable. This line can be found by finding the point at which one of the eigenvalues of the Hessian matrix evaluated on the RS saddle point vanishes. As shown in~\cite{franz2017}, $\kappa_{\text{dAT}}(\alpha)$ is obtained by solving the following equation
	\begin{equation}
		\begin{split}
			\label{eq::dAT}
			\frac{1}{(1-q)^2} &= \alpha \int D z \left[ \frac{\partial^2}{\partial z^2} \ln \left.H\left( \frac{z}{\sqrt{1-q}}\right) \right|_{z=\kappa + \sqrt{q} h}\right]^2 = \frac{\alpha}{(1-q)^2} \int D z \; \mathcal{W}^2 \left(\frac{\kappa + \sqrt{q} h}{\sqrt{1-q}}\right) 
		\end{split}
	\end{equation}
	together with the saddle point equation for $q$ in~\eqref{eq::SP_err_beta_inf}. In the previous equation we have also defined the function $\mathcal{W}$ that is
	\begin{equation}
		\label{eq::w(x)}
		\mathcal{W}(x) \equiv \frac{\partial^2}{\partial x^2} \ln H(x) = GH\left(x\right) \left(x - GH\left(x\right) \right)
	\end{equation}
	The dAT line is shown in Figure~\ref{Fig::phase_diagram} (orange line).
	
	\subsection{One-step Replica Symmetry Breaking Ansatz for the error-counting loss case}\label{sec::1RSB_ansatz}
	It is therefore useful to study the impact of Replica Symmetry Breaking (RSB) in the negative perceptron problem. In this section we impose a 1-step RSB ansatz (1RSB) on the order parameters
	\begin{subequations} 
		\begin{align}
			\label{eq::1rsb_ansatz}
			q_{ab} &= q_0 + (q_1-q_0) I_{ab}^{m} + (1-q_1) \delta_{ab}\\
			\hat q_{ab} &= \hat q_0 + (\hat q_1 - \hat q_0) I_{ab}^{m} - (\hat Q+\hat q_1) \delta_{ab}
		\end{align}
	\end{subequations}
	where $I_{ab}^m$ is the $n\times n$ matrix having elements equal to 1 inside the blocks of size $m$ located around the diagonal and 0 otherwise. 
	
	In this ansatz the free entropy is	
	\begin{subequations}
		\begin{align}
			\phi &= \mathcal{G}_{S}  + \alpha \mathcal{G}_{E} \\
			\mathcal{G}_{S}  & =\frac{1}{2}\hat{Q}+\frac{1}{2}\hat{q}_{1}q_{1}+\frac{m}{2}\left(q_{0}\hat{q}_{0}-q_{1}\hat{q}_{1}\right)+\frac{1}{2}\left[\ln\frac{2\pi}{\hat{Q}+\hat{q}_{1}}+\frac{1}{m}\ln\left(\frac{\hat{Q}+\hat{q}_{1}}{\hat{Q}+\hat{q}_{1}-m\left(\hat{q}_{1}-\hat{q}_{0}\right)}\right)+\frac{\hat{q}_{0}}{\hat{Q}+\hat{q}_{1}-m\left(\hat{q}_{1}-\hat{q}_{0}\right)}\right] \label{eq:Gs1rsb_sph} \\
			\mathcal{G}_{E} & =\frac{1}{m}\int Dz_{0}\,\ln\int Dz_{1} \, H^{m}\left(\frac{\kappa + \sqrt{q_0} \, z_{0}+\sqrt{q_1-q_0}z_{1}}{\sqrt{1-q_1}}\right) \label{eq:Ge1rsb}
		\end{align}
	\end{subequations}
	The saddle point equations for $\hat q_0$, $\hat q_1$ and $\hat Q$ can be explicitly expressed in terms of $q_1$, $q_0$ and $m$ as
	\begin{subequations}
		\label{eq::sol_eq1RSB_hat}
		\begin{align}
			\hat q_0 & = \frac{q_0}{(1 - q_1 + m (q_1 - q_0))^2}\\
			\hat q_1 & = \frac{q_0}{(1 - q_1 + m (q_1 - q_0))^2} + \frac{q_1-q_0}{(1-q_1)(1 - q_1 + m (q_1 - q_0))}= \frac{(1 - q_1) q_1 + m (q_1 - q_0)^2}{(1 - q_1) (1 - q_1 + m (q_1 - q_0))^2}\\
			\hat Q & = \frac{(1-2 q_1)(1-q_1) + m (2 + q_0 - 3 q_1) (q_1 - q_0) + m^2 (q_1 - q_0)^2}{(1 - q_1) (1 - q_1+ m (q_1 - q_0))^2}\,.
		\end{align}
	\end{subequations}
	The entropic term simplifies as
	\begin{equation}
		\mathcal{G}_{S}  = \frac{1}{2} \left(1 + 
		\ln(2\pi) + \frac{q_0}{1 - q_1 + m (q_1 - q_0)} + \frac{m-1}{m} \ln\left(1-q_1\right) + \frac{1}{m} 
		\ln\left(1 -q_1 + m (q_1-q_0)\right) \right)
	\end{equation}
	The saddle point equations for $q_0$ and $q_1$ read
	\begin{subequations}
		\begin{align}
			\hat{q}_0 &= - \frac{2 \alpha}{m} \frac{\partial \mathcal{G}_E}{\partial q_0} = \frac{\alpha}{1-q_1} \int D z_0 \left[\frac{\int Dz_{1} \, H^{m}\left(\frac{\kappa + \sqrt{q_0} \, z_{0}+\sqrt{q_1-q_0}z_{1}}{\sqrt{1-q_1}}\right) GH\left(\frac{\kappa + \sqrt{q_0} \, z_{0}+\sqrt{q_1-q_0}z_{1}}{\sqrt{1-q_1}}\right) }{\int Dz_{1} \, H^{m}\left(\frac{\kappa + \sqrt{q_0} \, z_{0}+\sqrt{q_1-q_0}z_{1}}{\sqrt{1-q_1}}\right)} \right]^2\\
			\hat{q}_1 &= - \frac{2 \alpha}{1-m} \frac{\partial \mathcal{G}_E}{\partial q_1}	= \frac{\alpha}{1-q_1} \int D z_0 \frac{\int Dz_{1} \, H^{m}\left(\frac{\kappa + \sqrt{q_0} \, z_{0}+\sqrt{q_1-q_0}z_{1}}{\sqrt{1-q_1}}\right) GH^2\!\left(\frac{\kappa + \sqrt{q_0} \, z_{0}+\sqrt{q_1-q_0}z_{1}}{\sqrt{1-q_1}}\right) }{\int Dz_{1} \, H^{m}\left(\frac{\kappa + \sqrt{q_0} \, z_{0}+\sqrt{q_1-q_0}z_{1}}{\sqrt{1-q_1}}\right)}
		\end{align}
	\end{subequations}
	
	\subsubsection{Stability of the 1RSB ansatz}
	
	For completeness we report here the expression of the value of $\alpha$ above which the 1RSB ansatz is unstable. This is the so called \emph{Gardner transition} $\alpha_{\mathrm{G}}$; it plotted as a function of $\kappa$ in the phase diagram of Fig.~\ref{Fig::phase_diagram}). If $m$, $q_0$ and $q_1$ satisfy the 1RSB equations, then the value $\alpha_{\mathrm{G}}(\kappa)$ is found when the following equation is verified
	\begin{equation}
		\begin{split}
			\frac{1}{(1-q_1)^2} = \alpha \int D z_0 \, \frac{\int D z_1 \, H^m\left(\frac{\kappa + \sqrt{q_0} \, z_{0}+\sqrt{q_1-q_0}z_{1}}{\sqrt{1-q_1}}\right) \left[ \left.\frac{\partial^2}{\partial h^2} \ln H\left(\frac{h}{\sqrt{1-q_1}}\right) \right|_{h = \kappa + \sqrt{q_0} \, z_{0}+\sqrt{q_1-q_0}z_{1}} \right]^2 }{\int D z_1 \, H^m\left(\frac{\kappa + \sqrt{q_0} \, z_{0}+\sqrt{q_1-q_0}z_{1}}{\sqrt{1-q_1}}\right)} \\
			= \frac{\alpha}{(1-q_1)^2} \int D z_0 \, \frac{\int D z_1 \, H^m\left(\frac{\kappa + \sqrt{q_0} \, z_{0}+\sqrt{q_1-q_0}z_{1}}{\sqrt{1-q_1}}\right) \mathcal{W}^2 \left( \frac{\kappa + \sqrt{q_0} \, z_{0}+\sqrt{q_1-q_0}z_{1}}{\sqrt{1-q_1}}\right)}{\int D z_1 \, H^m\left(\frac{\kappa + \sqrt{q_0} \, z_{0}+\sqrt{q_1-q_0}z_{1}}{\sqrt{1-q_1}}\right)}
		\end{split}
	\end{equation} 
	where $\mathcal{W}$ was defined in~\eqref{eq::w(x)}.

	\subsubsection{The $\kappa_\text{max}$ limit at 1RSB level}
	In order to estimate the maximum margin at the 1RSB level, we need to perform the limit $q_1 \to 1$ and $m \to 0$ with fixed ratio $\tilde{m} \equiv \frac{m}{1-q_1}$~\cite{engel1992storage,relu_locent}. Therefore we express all the free entropy in terms of $m$ and $\tilde{m}$:
	\begin{equation}
		\phi = \frac{1}{2m}\left[ m \ln \left( \frac{m}{\tilde{m}} \right) + \ln\left( 1 - m + \tilde{m} (1-q_0) \right) + \frac{\tilde{m} q_0}{ 1 - m + \tilde{m} (1-q_0)} + 2m \alpha 	\mathcal{G}_{E} \right] \equiv \frac{\tilde{\phi}(m, \tilde m, q_0)}{2m}\,.
	\end{equation}
	In the limit $m \to 0$, we need to ensure that (as in the RS case) the entropy goes to $-\infty$,\footnote{In the spherical case this condition is the analog to the zero entropy condition for the binary weight cases. Indeed in the binary weight case the entropy corresponds to the log of the \emph{number} of solutions, whereas in the spherical case it is instead the log of their \emph{volume}.} so we need to impose that the coefficient of first order expansion of the free entropy (which is of order $1/m$) vanishes. This is equivalent to imposing that at the maximum possible margin
	\begin{equation}
		\tilde \phi(\tilde{m}, q_0) \equiv \lim\limits_{m \to 0} \tilde \phi(m, \tilde{m}, q_0) = \ln\left( 1 + \tilde{m} (1-q_0) \right) + \frac{\tilde{m} q_0}{ 1 + \tilde{m} (1-q_0)} + 2 \alpha f(\kappa_\text{max}; q_0, \tilde{m}) = 0
	\end{equation}
	where
	\begin{equation}
		f(\kappa_\text{max}; q_0, \tilde{m}) = \lim\limits_{m \to 0} m \mathcal{G}_E = \lim\limits_{m \to 0} \int Dz_{0}\,\ln\int Dz_{1} \, H^{m}\left(\frac{\kappa_\text{max} + \sqrt{q_0} \, z_{0} + \sqrt{1-q_0}\,z_{1}}{\sqrt{ \frac{m}{\tilde{m}}}}\right)
	\end{equation}
	In order to perform the vanishing $m$ limit in the previous expression, we use the following expansion
	\begin{equation}
		\begin{split}
			&\int Dz_{1} \, H^{m}\left(\frac{\kappa_\text{max} + \sqrt{q_0} \, z_{0}+\sqrt{1-q_0}z_{1}}{\sqrt{\frac{m}{\tilde{m}}}}\right) \\
			&= \int_{-\frac{\kappa_\text{max} + \sqrt{q_0} \, z_0}{\sqrt{1-q_0}}}^{+\infty} Dz_{1} \, H^{m}\left(\frac{\kappa_\text{max} + \sqrt{q_0} \, z_{0}+\sqrt{1-q_0} \, z_{1}}{\sqrt{\frac{m}{\tilde{m}}}}\right) +\int_{-\infty}^{-\frac{\kappa_\text{max} + \sqrt{q_0} z_0}{\sqrt{1-q_0}}} Dz_{1} \, \left[1 - H\left(-\frac{\kappa_\text{max} + \sqrt{q_0} \, z_{0}+\sqrt{1-q_0} \, z_{1}}{\sqrt{\frac{m}{\tilde{m}}}}\right) \right]^m \\
			&\simeq \int_{-\frac{\kappa_\text{max} + \sqrt{q_0} z_0}{\sqrt{1-q_0}}}^{+\infty} Dz_{1} \, e^{-\frac{\tilde{m}\left( \kappa_\text{max} + \sqrt{q_0} \, z_{0}+\sqrt{1-q_0} \, z_{1} \right)^2}{2}} + H\left(\frac{\kappa_\text{max} + \sqrt{q_0} z_0}{\sqrt{1-q_0}} \right)
		\end{split}
	\end{equation}
	where in the last line we have used the fact that $H(x) \simeq \frac{G(x)}{\sqrt{2\pi} x}$ for $x \to +\infty$ since $m\to 0$; therefore in the second integral of the second line the $H$ term can be safely neglected, since it is giving exponentially small corrections to the final result.
	Performing the integral over $z_1$ using the identity
	\begin{equation}
		\label{eq::general_identity}
		\int_\gamma^\infty Dz_1 \; e^{- \alpha \frac{z_1^2}{2} - \beta z_1 + \delta} = \frac{e^{\frac{\beta^2}{2(1+\alpha)} + \delta} H\left( \frac{\beta + \gamma (1+\alpha)}{\sqrt{1+\alpha}} \right)}{\sqrt{1+\alpha}}
	\end{equation}
	one finally gets
	\begin{equation}
		\begin{split}
			&f\left(\kappa_\text{max}; q_{0},\tilde{m}\right)=\int Dz_{0}\,\ln\left[\frac{e^{-\frac{\tilde{m} \left(\kappa_\text{max} + \sqrt{q_0} z_0\right)^{2}}{2\left( 1+\left(1-q_0\right)\tilde{m} \right)}}}{\sqrt{1+\left(1-q_0\right)\tilde{m}}}H\left(-\sqrt{\frac{1}{1-q_0}}\frac{\kappa_\text{max} + \sqrt{q_0} \,z_{0} }{\sqrt{1+\left(1-q_0\right)\tilde{m}}}\right)+H\left(\frac{\kappa_\text{max} +\sqrt{q_0} \, z_0}{\sqrt{1-q_0}} \right)\right]
		\end{split}
	\end{equation}
	The values of $\tilde m$ and $q_0$ are found by differentiation of $\tilde{\phi}(\tilde m, q_0)$ with respect to them:
	\begin{subequations}
		\begin{align}
			q_0 &= - 2\alpha \left( \frac{1 + \tilde{m} (1-q_0)}{\tilde{m}}\right)^2\frac{\partial f}{\partial q_0}\\
			\tilde{m} &= - 2 \alpha \left( \frac{1 + \tilde{m} (1-q_0)}{1-q_0}\right)^2 \frac{\partial f}{\partial \tilde{m}} - \frac{1}{(1-q_0)^2}
		\end{align}
	\end{subequations}
	
	\section{Replica computation of the Franz-Parisi free entropy: error-landscape around a sampled configuration from a generic loss function} \label{sec::FP}
	Here we review all the basic steps of the computation of the average local entropy around a given reference configuration, or briefly, the so called Franz-Parisi free entropy. As stated in the main text, it is defined as the typical log-volume of configurations $\boldsymbol{w}$, \emph{constrained} to be at a given distance $d = \frac{1-t_{\mathrm{d}}}{2}$ from a given reference solution $\tilde{\boldsymbol{w}}$. We suppose, in full generality, that the reference configuration $\tilde{\boldsymbol{w}}$ is extracted from the Boltzmann measure with a margin parameter $\tilde{\kappa}$ optimizing a loss $\ell$. The constrained configuration $\boldsymbol{w}$ will be extracted from the Boltzmann measure with margin $\kappa$ optimizing a loss $\overline{\ell}$. In formulas, the Franz-Parisi free entropy is
	\begin{equation}
		\label{eq::FP_entropy}
		\phi_{\text{FP}}(t_{\mathrm{d}}; \tilde \kappa, \kappa) = \frac{1}{N} \left\langle\frac{1}{Z} \int \! d\mu(\tilde{\boldsymbol{w}}) e^{-\beta \sum_\mu \ell(\Delta^\mu(\tilde{\boldsymbol{w}}; \tilde \kappa))}
		\ln \mathcal{N}(\tilde{\boldsymbol{w}}, t_{\mathrm{d}}; \kappa) \right\rangle_{\xi} \
	\end{equation}
	where
	\begin{equation*}
		\mathcal{N}(\tilde{\boldsymbol{w}}, t_{\mathrm{d}}; \kappa) \equiv \int \! d\mu(\boldsymbol{w}) e^{-\beta' \sum_\mu \overline{\ell}(\Delta^\mu(\boldsymbol{w}; \kappa))}\delta\left( Nt_{\mathrm{d}}- \sum_i w_i \tilde w_i \right)
	\end{equation*}
	is the constrained Gardner volume. In the following, we choose the loss of the constrained configuration to be the number of error loss $\overline{\ell}(x) = \Theta(-x)$. We also consider the limit $\beta' \to \infty$ for simplicity. The distance $d$ considered in the main text is related to the overlap by the relation
	\begin{equation}
		\label{eq::distance_overlap_connection}
		d = \frac{1 - t_{\mathrm{d}}}{2} \,. 
	\end{equation}
	
	\subsection{Upper bound to the Franz-Parisi entropy}~\label{sec::upper_bound_franz_parisi}
	As observed in the main text, when $\alpha = 0$ the Franz-Parisi entropy is maximal since it measures the log of the total volume of configurations at a certain overlap $t_{\mathrm{d}}$ from a given point on the sphere; since every point on the sphere is equivalent, this quantity does not depend at all on the reference $\tilde{\boldsymbol{w}}$ that is picked. Therefore we can safely choose $\tilde{w}_i = 1$ for all $i = 1, \dots, N$. In formulas:
	\begin{equation}
		\label{eq::max_local_entropy}
		\mathcal{S}_\text{max}(t_{\mathrm{d}}) \equiv \frac{1}{N}\int d\mu(\boldsymbol{w}) \, \delta\left( N t_{\mathrm{d}} - \sum_i w_i\right)
	\end{equation}
	Inserting the integral representations of the two delta function in~\eqref{eq::max_local_entropy} (one delta function is hided inside the spherical measure over the weights $d\mu(\boldsymbol{w})$) and integrating over the weights, we have 
	\begin{equation}
		\mathcal{S}_\text{max}(t_{\mathrm{d}}) = \frac{1}{N} \int \frac{d \hat t_{\mathrm{d}}}{2\pi} \frac{d \hat q}{2\pi} \, e^{N\left[ t_{\mathrm{d}} \hat t_{\mathrm{d}} + \frac{\hat q}{2} + \frac{1}{2} \ln\left( \frac{2\pi}{\hat q}\right) + \frac{\hat t_{\mathrm{d}}}{2\hat q}\right]}
	\end{equation}
	One can resort then to a saddle point approximation in the limit of large $N$; the corresponding equations can be exactly solved, leading to
	\begin{equation}
		\label{eq::upper_bound}
		\mathcal{S}_\text{max}(t_{\mathrm{d}}) = \frac{1}{2} \left[1 + \ln(2\pi) + \ln\left( 1-t_{\mathrm{d}}^2 \right)\right] \,.
	\end{equation}
	which, reminding the relation between overlap and distance in~\eqref{eq::distance_overlap_connection} is equation~\eqref{eq::upper_bound_FP_entropy} presented in the main text.

	\subsection{Sketch of the replica computation}
	For generic $\alpha$ we have to resort to the replica approach. We sketch here the main steps of the derivation. We introduce two sets of replicas, one for the log of $\mathcal{N}$ and one for the partition function $Z$ in the denominator of~\eqref{eq::FP_entropy}
	\begin{subequations}
		\begin{align}
			\ln\mathcal{N} & =\lim_{s\to0}\partial_{s}\mathcal{N}^{s}\\
			Z^{-1} & =\lim_{n\to0}Z^{n-1}
		\end{align}
	\end{subequations} 
	This variation of the replica trick is needed in order to have the average of the log of $\mathcal{N}$ over the same replicated measure as in Appendix~\ref{sec::Typical_case_scenario}; as a matter of fact, the saddle point equations for the reference will be the same, see also~\cite{Huang_2013}. The overlap $q_{ab}$, which concerns only the reference solution $\tilde{\boldsymbol{w}}$, will come out naturally and will satisfy the same saddle point equations for any anzatz~\cite{Huang_2013}.
	In the following we will use indices $a,b\in[n]$ and $c,d\in[s]$. The Franz-Parisi entropy is:
	
	\begin{equation}
		\begin{split}\phi_{FP}(t_{\mathrm{d}}) & = \frac{1}{N}\lim\limits _{\substack{n\to0\\
					s\to0}}\partial_{s} \int \! d\mu(\tilde{\boldsymbol{w}}^a) \prod_{a}\left\langle e^{-\beta \ell(\Delta^\mu(\tilde{\boldsymbol{w}}^a; \tilde{\kappa}))}\,\mathcal{N}^{s}(\tilde{\boldsymbol{w}}^{a=1},t_{\mathrm{d}}; \kappa)\right\rangle _{\xi} =\\
			&= \frac{1}{N}\lim\limits _{\substack{n\to0\\
					s\to0
				}
			}\partial_{s}\int\prod_{\mu a}\frac{dv_{\mu}^{a}d\hat{v}_{\mu}^{a}}{2\pi}\, \prod_{\mu a} e^{-\beta \ell\left(v_{\mu}^{a}-\tilde{\kappa}\right) -iv_{\mu}^{a}\hat{v}_{\mu}^{a}}\int\prod_{\mu c}\frac{du_{\mu}^{c}d\hat{u}_{\mu}^{c}}{2\pi}\,\Theta\left(u_{\mu}^{c}-\kappa\right)e^{-iu_{\mu}^{c}\hat{u}_{\mu}^{c}}\\
			& \quad \times\,\int \! \prod_{a}d\mu(\tilde{\boldsymbol{w}}^a) \prod_{c}d\mu(\boldsymbol{w}^c) \prod_{\mu}\left\langle e^{\frac{i}{\sqrt{N}}\sum_{i}\xi_{i}^{\mu}\left(\sum_{a}\tilde{w}_{i}^{a}\hat{v}_{\mu}^{a}+\sum_{c}w_{i}^{c}\hat{u}_{\mu}^{c}\right)}\right\rangle _{\boldsymbol{\xi}^{\mu}}\prod_{c}\delta\left(N t_{\mathrm{d}}-\sum_{i=1}^{N}\tilde{w}_{i}^{a=1}w_{i}^{c}\right)%
		\end{split}
		\label{eq::FP_replicas}
	\end{equation}
	where two auxiliary variables have been introduced 
	\begin{subequations}
		\label{eq::vu} 
		\begin{align}
			v_{\mu}^{a} & =\frac{1}{\sqrt{N}}\,\sum_{i}\tilde{w}_{i}^{a}\xi_{i}^{\mu}\\
			u_{\mu}^{c} & =\frac{1}{\sqrt{N}}\sum_{i}w_{i}^{c}\xi_{i}^{\mu}
		\end{align}
	\end{subequations} 
	and we have enforced those definitions by using
	delta functions. Next, the average
	over patterns can be performed
	\begin{equation}
		\begin{split}\left\langle e^{\frac{i}{\sqrt{N}}\sum_{i}\xi_{i}^{\mu}\left(\sum_{a}\tilde{w}_{i}^{a}\hat{v}_{\mu}^{a}+\sum_{c}w_{i}^{c}\hat{u}_{\mu}^{c}\right)}\right\rangle _{\boldsymbol{\xi}^{\mu}} & \simeq e^{-\frac{1}{2N}\sum_{i}\left(\sum_{a}\tilde{w}_{i}^{a}\hat{v}_{\mu}^{a}+\sum_{c}w_{i}^{c}\hat{u}_{\mu}^{c}\right)^{2}}\\
			& =e^{-\frac{1}{2}\sum_{ab}\left(\frac{1}{N}\sum_{i}\tilde{w}_{i}^{a}\tilde{w}_{i}^{b}\right)\hat{v}_{\mu}^{a}\hat{v}_{\mu}^{b}-\frac{1}{2}\sum_{cd}\left(\frac{1}{N}\sum_{i}w_{i}^{c}w_{i}^{d}\right)\hat{u}_{\mu}^{c}\hat{u}_{\mu}^{d}-\sum_{ac}\left(\frac{1}{N}\sum_{i}\tilde{w}_{i}^{a}w_{i}^{c}\right)\hat{v}_{\mu}^{a}\hat{u}_{\mu}^{c}}
		\end{split}
	\end{equation}
	The previous expression only depends on first
	two moments of the variables~\eqref{eq::vu}. We can therefore define
	the order parameters 
	\begin{subequations}
		\label{eq::overlapsFP}
		\begin{align}
			q_{ab} &\equiv \left\langle v_{\mu}^{a} v_{\mu}^{b} \right\rangle_{\boldsymbol{\xi}} 
			= \frac{1}{N} \sum_{i} \tilde{w}^{a}_{i} \tilde{w}^{b}_{i}\,, \\
			p_{cd} &\equiv  \left\langle u_{\mu}^c v_{\mu}^d \right\rangle_{\boldsymbol{\xi}} = \frac{1}{N} \sum_{i} w^{c}_{i} w^{d}_{i} \,, \\
			t_{ac} &\equiv \left\langle v_{\mu}^{a} u_{\mu}^{c} \right\rangle_{\boldsymbol{\xi}} 
			= \frac{1}{N} \sum_{i} \tilde{w}^{a}_{i} w^{c}_{i} \,.
		\end{align}
	\end{subequations}
	Notice that because	the reference $\tilde{\boldsymbol{w}}$ is constrained to be at an overlap $t_{\mathrm{d}}$ with $\boldsymbol{w}$, we
	have $t_{1c}=t_{\mathrm{d}}$ for every $c = 1, \dots, s$. We can enforce the definitions in equations~\eqref{eq::overlapsFP} by using delta functions and their integral representations. The
	Franz-Parisi entropy can be finally written as 
	\begin{equation}
		\begin{aligned}\phi_{FP}(S) & =\frac{1}{N}\lim\limits _{\substack{n\to0\\
					s\to0
				}
			}\partial_{s}\int\prod_{a<b}\frac{dq_{ab}d\hat{q}_{ab}}{2\pi}\,\prod_{c<d}\frac{dp_{cd}d\hat{p}_{cd}}{2\pi}\,\prod_{c,\,a\ne1}\frac{dt_{ac}d\hat{t}_{ac}}{2\pi}\prod_{c}\frac{d\hat{t}_{1c}}{2\pi}\,e^{N \left[ G_S + \alpha G_E\right]}\end{aligned}
		\label{S}
	\end{equation}
	where we have defined
	\begin{subequations}
		\begin{align}
			G_{S}  =& -\frac{1}{2}\sum_{ab}q_{ab}\hat{q}_{ab}-\frac{1}{2}\sum_{cd}p_{cd}\hat{p}_{cd}-\sum_{ac}t_{ac}\hat{t}_{ac}\\
			&+\ln\int\prod_{a}d\tilde{w}^{a}\prod_{c}dw^{c}\prod_{l}e^{\frac{1}{2}\sum_{ab}\hat{q}^{ab}\tilde{w}^{a}\tilde{w}^{b}+\frac{1}{2}\sum_{cd}\hat{p}^{cd}w^{c}w^{d}+\sum_{ac}\hat{t}^{ac}\tilde{w}^{a}w^{c}}\nonumber \\
			G_{E} & =\ln\int\prod_{a}\frac{dv_{a}d\hat{v}_{a}}{2\pi}e^{iv_{a}\hat{v}_{a}}\int\frac{du_{c}d\hat{u}_{c}}{2\pi}e^{iu_{c}\hat{u}_{c}}\prod_{a}e^{-\beta \ell(v_{a}-\tilde{\kappa})}\prod_{c}\Theta(u_{c}-\kappa)\,\\ 
			& \quad \times e^{-\frac{1}{2}\sum_{ab}q_{ab}\hat{v}_{a}\hat{v}_{b}-\frac{1}{2}\sum_{cd}p_{cd}\hat{u}_{c}\hat{u}_{d}-\sum_{ac}t_{ac}\hat{v}_{a}\hat{u}_{c}} \nonumber
		\end{align}
	\end{subequations}
	
	\subsection{RS ansatz}
	We impose an RS ansatz on the order parameters
	\begin{subequations} 
		\begin{align}
			p_{cd} & =\delta_{cd}+p(1-\delta_{cd})\\
			t_{ac} & =t_{\mathrm{d}}\,\delta_{a1}\,+\,t\,(1-\delta_{a1})\\
			\hat{p}_{cd} & = - \hat{P} \, \delta_{cd} + \hat{p}(1-\delta_{cd})\\
			\hat{t}_{ac} & =\hat{t}_{\mathrm{d}}\,\delta_{a1}\,+\,\hat{t}\,(1-\delta_{a1})
		\end{align}
	\end{subequations} 
	together with an RS ansatz on the order parameters characterizing the reference as in~\eqref{eq::RS_reference}. In this case the Franz-Parisi free entropy reads
	\begin{equation}
		\label{eq::FP_free_entropy}
		\phi_{\text{FP}}=\mathcal{G}_{S}+\alpha\,\mathcal{G}_{E}
	\end{equation}
	where the entropic term and the energetic terms are
	\begin{subequations}
		\begin{align}
			\mathcal{G}_{S}  &\equiv \lim\limits _{\substack{n\to0\\
					s\to0}} \partial_{s} G_S = \frac{1}{2}\hat{P}+\frac{1}{2}\hat{p}p+\hat{t}t-\hat{t}_{\mathrm{d}} t_{\mathrm{d}} +\frac{1}{2}\ln\frac{2\pi}{\hat{P}+\hat{p}}+\frac{1}{\hat{P}+\hat{p}}\left[\frac{\hat{p}}{2}+\frac{\left(\hat{t}_{\mathrm{d}}-\hat{t}\right)^{2}\left(\frac{\hat{Q}}{2}+\hat{q}\right)}{\left(\hat{Q}+\hat{q}\right)^{2}}+\frac{\hat{t}\left(\hat{t}_{\mathrm{d}}-\hat{t}\right)}{\hat{Q}+\hat{q}}\right] \,,\\
			\mathcal{G}_E &\equiv \lim\limits _{\substack{n\to0\\
					s\to0}} \partial_{s} G_E =   \int Dz_0 \, \frac{\int Dz_1 \, e^{-\beta \ell\left(\sqrt{q} z_0 + \sqrt{1-q} z_1 - \tilde \kappa \right) } \int D z_2 \, \ln H\left( \frac{\kappa - \frac{t}{\sqrt{q}} z_0 - \frac{t_{\mathrm{d}} - t}{\sqrt{1 - q}} z_1 - \sqrt{\gamma-\frac{(t_{\mathrm{d}} - t)^2}{1 - q}} z_2}{\sqrt{1-p}} \right)}{ \int Dz_1 \, e^{-\beta \ell\left(\sqrt{q} z_0 + \sqrt{1-q} z_1 - \tilde \kappa \right) } } \label{eq::FP_energetic_term}
		\end{align}
	\end{subequations}
	where 
	\begin{equation}
		\gamma \equiv p - \frac{t^2}{q} \,.
	\end{equation} 
	This expression must be optimized with respect to the 6 order parameters $p$, $\hat p$, $t$, $\hat t$, $\hat P$, $\hat{t}_{\mathrm{d}}$. Notice that order parameters $q$, $\hat q$ and $\hat Q$ satisfy the saddle point equations of a typical configuration of the problem with margin $\tilde \kappa$ i.e.~\eqref{eq::SPeq_RS}.
	
	As usual in spherical models, the conjugated parameter can be again integrated analytically, giving
	\begin{subequations}
		\label{eq::FPconjugated_solved}
		\begin{align}
			\hat P &= \frac{1 - 2 p (1 - q)^2 + q^2 + t_{\mathrm{d}}^2 - t^2 - 2 q (1 + t_{\mathrm{d}}^2 - t_{\mathrm{d}} t)}{(1 - p)^2 (1 - q)^2} \\
			\hat p &= \frac{p (1 - q)^2 - t_{\mathrm{d}}^2 + 2 q t_{\mathrm{d}} (t_{\mathrm{d}} - t) + t^2}{(1 - p)^2 (1 - q)^2} \\
			\hat t &= \frac{t-q t_{\mathrm{d}}}{(1 - p) (1 - q)^2} \\
			\hat{t}_{\mathrm{d}} &= \frac{t_{\mathrm{d}} - 2 q t_{\mathrm{d}} + q t}{(1 - p) (1 - q)^2} 
		\end{align}
	\end{subequations}
	Using also the relations~\eqref{eq::sol_eqRS_hat}, the spherical term becomes
	\begin{equation}
		\mathcal{G}_{S}  =
		\frac{(1 - t_{\mathrm{d}}^2) (1 - 2 q ) + q^2 - 2 q t_{\mathrm{d}} t + t^2}{2(1-p)(1-q)^2} +\frac{1}{2} \ln(2\pi) + \frac{1}{2}\ln\left(1-p\right) \,.
	\end{equation}
	This expression is actually useful when analyzing the $\tilde{\kappa} \to \kappa_{\text{max}}$ of the Franz-Parisi free entropy.
	
	Notice also how when $\alpha = 0$ we recover~\eqref{eq::upper_bound}. Indeed the saddle point for $p$ and $t$ can be readily solved, giving
	\begin{subequations}
		\begin{align}
			p &= t_{\mathrm{d}}^2\\
			t &= q t_{\mathrm{d}} \\
			\hat t &= \hat p = 0 \\
			\hat{P} &= 1			
		\end{align}
	\end{subequations}	
	which, inserted back into the Franz-Parisi entropy~\eqref{eq::FP_free_entropy} give~\eqref{eq::upper_bound}.

	\subsection{The case of a convex loss: the infinite $\beta$ limit}
	
	As in~\ref{app::infinite_beta_limit_reference}, here we consider the case in which the loss $\ell(x)$ is convex with a unique minimizer. In this case the large $\beta$ limit makes $q$ (the overlap between two different replica configuration) tend to 1, with the scaling in~\eqref{eq::q_large_beta}; correspondingly, the overlap $t$ must be scaled as
	\begin{equation}
		\label{eq::scaling_large_beta}
		t_{\mathrm{d}} - t = (1-q) \delta t = \frac{\delta q \delta t}{\beta}
	\end{equation}
	This can be obtained by inspecting the term $\frac{t_{\mathrm{d}} - t}{\sqrt{1 - q}} z_1$ in the argument of the $H$ function in the energetic term, see equation~\eqref{eq::FP_energetic_term}. Indeed since $1-q = O(1/\beta)$, $z_1$ should be scaled as $z_1 \to \sqrt{\beta} z_1$ (as it has been already done in section~\ref{app::infinite_beta_limit_reference}). Therefore $\frac{z_1}{\sqrt{1 - q}}$ scales as $\beta$ and this tells us that $t_{\mathrm{d}} - t = O(1/\beta)$. 
	With this scaling the entropic term remains finite as it should be, and it simplifies to
	\begin{equation}
		\mathcal{G}_{S}  =
		\frac{1-2 t_{\mathrm{d}} \delta t + \delta t^2}{2(1-p)} +\frac{1}{2} \ln(2\pi) + \frac{1}{2}\ln\left( 1-p\right) \,.
	\end{equation}
	Concerning the energetic term, we scale $z_1 \to \sqrt{\beta} z_1$ in~\eqref{eq::FP_energetic_term}, obtaining
	\begin{equation}
		\mathcal{G}_E =  \int Dz_0 \, D z_2 \, \ln H\left( \frac{\kappa - t_{\mathrm{d}} \, z_0 - \frac{ \delta t \, \delta q}{\sqrt{\delta q}} z_1^\star(z_0; \tilde{\kappa}) - \sqrt{\bar{\gamma}} z_2}{\sqrt{1 - p}} \right) 
	\end{equation}
	where $z_1^\star(z_0; \tilde{\kappa})$ satisfies equation~\eqref{eq::z1star} (with margin $\tilde{\kappa}$). We have also redefined $\gamma$ as
	\begin{equation}
		\bar\gamma = p - t_{\mathrm{d}}^2 \,.
	\end{equation}
	
	\subsection{The error-counting loss case} \label{app::FP_error_counting_loss}
	In the case we take a reference sampled from the error counting loss function the energetic term in the large $\beta$ limit reads
	\begin{equation}
		\label{eq::GEerr_RS}
		\mathcal{G}_{E} = \int \! Dz_{0} \frac{\int Dz_{1}H\text{\ensuremath{\left(\frac{\sqrt{\gamma} \left(\tilde \kappa + \sqrt{q } \, z_{0} \right) + (t_{\mathrm{d}}-t) z_{1}}{\sqrt{(1-q) \gamma - (t_{\mathrm{d}} - t)^{2}}}\right)} \ensuremath{\ln H\left(\frac{ \kappa + \frac{t}{\sqrt{q}} z_{0} + \sqrt{\gamma}z_{1}}{\sqrt{1-p}}\right)}}}{H\left( \frac{\tilde\kappa + \sqrt{q} z_0}{\sqrt{1-q}}\right)} \,.
	\end{equation}

	\subsubsection{Entropy landscape around the maximum margin solution}
	In this section we want to compute the Franz-Parisi free entropy of a configuration having the largest possible margin. This is a much more challenging task with respect to binary models, where the maximum margin solution can be found numerically by finding the value of the margin for which the typical entropy vanishes and then plugging this value (together with the corresponding order parameters of the reference) in the Franz-Parisi free entropy. As we remarked in section~\ref{sec::kmax_limit}, the volume of solutions space shrinks to a point in the $\kappa_{\text{max}}$ limit, i.e. $q \to 1$. Therefore, this limit must be performed \emph{analytically} in the Franz-Parisi free entropy directly.
	
	In the limit, $q\to 1$ the scaling for $t$, using a similar argument to the one exposed for~\eqref{eq::scaling_large_beta}, is
	\begin{equation}
		\label{eq::FP_RS_kmax_limit_scaling}
		t_{\mathrm{d}} - t = (1-q) \delta t = \delta t \, \delta q
	\end{equation}
	where $\delta q$ is small. Notice that imposing the previous scaling in~\eqref{eq::FPconjugated_solved} it can be seen that both $\hat{t}_{\mathrm{d}}$ and $\hat{t}$ diverge as
	\begin{subequations}
		\begin{align}
			\hat{t}_{\mathrm{d}} &= \frac{\delta q \delta t + t_{\mathrm{d}} - \delta t}{(1-p) \delta q} \\
			\hat{t} &= \frac{t_{\mathrm{d}} - \delta t}{(1-p) \delta q}
		\end{align}
	\end{subequations}
	but their difference 
	\begin{equation}
		\hat{t}_{\mathrm{d}} - \hat{t} = \frac{\delta t}{1-p}
	\end{equation}
	stays finite. The conjugated order parameters $\hat{P}$ and $\hat{p}$ remain, instead, finite in the limit. 
	
	The entropic term simplifies as
	\begin{equation}
		\mathcal{G}_{S}  =
		\frac{1-2 t_{\mathrm{d}} \delta t + \delta t^2}{2(1-p)} +\frac{1}{2} \ln(2\pi) + \frac{1}{2}\ln\left( 1-p\right) \,.
	\end{equation} 
	In doing the limit $q \to 1$ in the energetic term in~\eqref{eq::GEerr_RS} we need to pay attention, since the argument of the error function $H$ in the numerator and denominator diverges. The computation is very involved and it has been performed with Mathematica that is a well suited software for algebraic manipulations. Nevertheless, we sketch here the main steps:
	\begin{itemize}
		\item First of all we perform a rotation over $z_0$ and $z_1$ 
		\begin{equation}
			\begin{split}
				z_0 & \to \frac{1}{\sqrt{p}} \left( \frac{t}{\sqrt{q}} z_0 - \sqrt{\gamma} z_1 \right)\\
				z_1 & \to \frac{1}{\sqrt{p}} \left( \sqrt{\gamma} z_0 + \frac{t}{\sqrt{q}} z_1 \right) 
			\end{split}
		\end{equation}
		so that we can extract in the $\ln H$ term in~\eqref{eq::GEerr_RS} all the quantities that can in principle contribute with terms depending on $\delta q$. After the rotation all the terms behaving like that and the dependence on the $z_1$ variable are inside the other error function terms in the numerator or in the denominator.
		\item Secondly, we can use the identity
		\begin{equation}
			\label{eq::identity}
			\frac{H\left( \frac{A}{\sqrt{\delta q}} + \sqrt{\delta q} B \right)}{H\left( \frac{A}{\sqrt{\delta q}} + C \sqrt{\delta q} \right)} \simeq \Theta\left( A \right) e^{-A(B - C)} + \Theta\left(-A\right)
		\end{equation}
		which follows from the asymptotic expansion valid of the complementary error function around $x \to \infty$ 
		\begin{equation}
			\label{eq::Hexpansion}
			H(x) \simeq \frac{G(x)}{\sqrt{2\pi} x}
		\end{equation}
		The quantities $A$, $B$ and $C$ are 
		\begin{subequations}
			\begin{align}
				A &= \tilde{\kappa}_\text{max} + \frac{t_{\mathrm{d}}}{\sqrt{p}} z_0 -  \sqrt{1 - \frac{t_{\mathrm{d}}^2}{p}} z_1 \\
				B &= \frac{1}{2 \sqrt{1 - \frac{t_{\mathrm{d}}^2}{p}}} z_1 - \frac{\delta t^2 \left( \tilde \kappa_\text{max}  + \frac{t_{\mathrm{d}}}{\sqrt{p}} z_0 - \sqrt{1-\frac{t_{\mathrm{d}}^2}{p}} z_1\right)}{2(t_{\mathrm{d}}^2 - p)} \\
				C &= - \frac{\delta t}{\sqrt{p}} z_0 + \frac{p - 2 t_{\mathrm{d}} \delta t}{2 p \sqrt{1 - \frac{t_{\mathrm{d}}^2}{p}}} z_1
			\end{align}
		\end{subequations}
		The $\Theta(-A)$ term in~\eqref{eq::identity} is easy and gives 
		\begin{equation}
			\begin{split}
				\int Dz_{0} \, D z_1 \; \Theta(-A) \ln H\left(\frac{ \kappa + \sqrt{p}z_{1}}{\sqrt{1-p}}\right) = \int Dz_{0} \,  H\left( \frac{ \tilde{\kappa}_\text{max} + z_0\frac{t_{\mathrm{d}}}{\sqrt{p}}}{ \sqrt{1 - \frac{t_{\mathrm{d}}^2}{p}}}\right) \ln H\left(\frac{ \kappa + \sqrt{p} z_{0} }{\sqrt{1-p}}\right)
			\end{split}
		\end{equation}
		For the $\Theta(A)$ terms we can perform the $z_1$ integration by using the identity~\eqref{eq::general_identity} that we report here for convenience
		\begin{equation*}
			\int_\gamma^\infty Dz_1 \; e^{- \alpha \frac{z_1^2}{2} - \beta z_1 + \delta} = \frac{e^{\frac{\beta^2}{2(1+\alpha)} + \delta} H\left( \frac{\beta + \gamma (1+\alpha)}{\sqrt{1+\alpha}} \right)}{\sqrt{1+\alpha}}
		\end{equation*}
		It can be seen that $\beta$, $\alpha$ and $\delta$ are such that 
		\begin{equation}
			- \frac{z_0^2}{2} + \frac{\beta^2}{2(1+\alpha)} + \delta = - \frac{\left( \delta t  \tilde{\kappa} + \sqrt{p} z_0 \right)^2}{2 \eta}
		\end{equation}
		where 
		\begin{equation}
			\eta \equiv p + \delta t \left(\delta t  - 2 t_{\mathrm{d}}  \right)
		\end{equation}
		\item Finally we perform the change of variable 
		\begin{equation}
			z_0 \to \frac{\sqrt{\eta} z_0 - \delta t \tilde{\kappa}}{\sqrt{p}}
		\end{equation}
		It can be shown that the Jacobian of this transformation is \emph{exactly} equal to $\sqrt{1 + \alpha}$. We therefore get a standard Gaussian measure for the integration over $z_0$.
	\end{itemize}
	
	\noindent
	The final result is therefore
	\begin{equation}
		\begin{split}
			\mathcal{G}_{E} &= \int Dz_{0} \,  H\left( \frac{ \tilde{\kappa}_\text{max} + z_0\frac{t_{\mathrm{d}}}{\sqrt{p}}}{ \sqrt{1 - \frac{t_{\mathrm{d}}^2}{p}}}\right) \ln H\left(\frac{ \kappa + \sqrt{p} z_{0} }{\sqrt{1-p}}\right) + \int Dz_{0} \, H\left( -\frac{\tilde{\kappa}_\text{max} \sqrt{\eta} + \left( \delta t - t_{\mathrm{d}} \right) z_0}{\sqrt{p - t_{\mathrm{d}}^2}} \right) \ln H \left(  \frac{\kappa - \delta t \tilde{\kappa}_\text{max} - \sqrt{\eta} z_0 }{\sqrt{1-p}}\right)
		\end{split}
	\end{equation}	
	In our computation there can be two sources of replica symmetry breaking (RSB):
	\begin{itemize}
		\item Replica symmetry breaking for the overlaps involving the reference solution. Those effects are for sure relevant if at a given value of $\alpha$, $\tilde{\kappa} > \kappa_{\text{dAT}}(\alpha)$, where $\kappa_{\text{dAT}}(\alpha)$ is given in~\eqref{eq::dAT}. In the left panel of Figure~\ref{Fig::FP_1RSB} we plot the RS Franz-Parisi entropy for a value of $\kappa$ and $\alpha$ and several values of $\tilde \kappa$; we have $\kappa$ and $\alpha$ in such a way that even typical solutions are in the fRSB phase. One can see that the primary maximum of some of the curves is located at small distances and not anymore at large ones; this clearly is an indication of a problem on the RS ansatz. We therefore expect huge corrections to the curves plotted. In addition the curves are ordered in the opposite way as in the low $\alpha$ regime which was analyzed in the main text; namely here we see that (at least for not so small distances) the local entropy of high-margin configuration is lower that the ones corresponding to small margin solutions. We expect that including RSB effects on the reference will tend to mild this problem. 
		\item Replica symmetry breaking for the overlaps involving the constrained configuration.
	\end{itemize}
	Of course both of them in principle can contribute. Since taking into account both effects is very challenging, in this work we have studied them separately, leaving for future work a detailed analysis of the full-RSB equations of the Franz-Parisi free entropy.

	\subsubsection{Breaking the Replica-Symmetry on the reference}~\label{sec::app_FP_1RSB}
	We study here the impact of breaking the replica symmetry on the reference configuration. We therefore impose~\eqref{eq::1rsb_ansatz} on the overlaps of the reference configurations, we assume an RS ansatz on the overlaps corresponding to the constrained configuration (i.e. $p_{cd}$ and $\hat{p}_{cd}$) and we finally impose an 1RSB ansatz on the overlaps $t_{ac}$ and $\hat{t}_{ac}$ containing both $\boldsymbol{w}$ and $\tilde{\boldsymbol{w}}$:		
	\begin{subequations} 
		\begin{align}
			p_{cd} & =\delta_{cd}+p(1-\delta_{cd})\\
			t_{ac} & = t_0 + (t_1 - t_0)J_{ac}^m + (t_{\mathrm{d}}-t_1)\,\delta_{a1}\\
			\hat{p}_{cd} & =-\hat{P} \, \delta_{cd} + \hat{p}(1-\delta_{cd})\\
			\hat{t}_{ac} &= \hat{t}_0 + (\hat{t}_1 - \hat{t}_0)J_{ac}^m + (\hat{t}_{\mathrm{d}}-\hat{t}_{1})\,\delta_{a1} \,.
		\end{align}
	\end{subequations} 
	Here $J_{ac}^m$ is the $n\times s$ matrix that has elements equal to 1 in the first $m$ rows and 0 otherwise.

	Imposing this ansatz the Franz-Parisi free entropy can be still written as~\eqref{eq::FP_free_entropy} where the 1RSB entropic term is
	\begin{equation}
		\begin{aligned}
			\mathcal{G}_{S}   =& \;\frac{\hat{P}}{2}+ \frac{p \, \hat{p}}{2} - t_{\mathrm{d}} \hat{t}_{\mathrm{d}}-(m-1)\,t_1\hat{t}_1+ m\, t_0 \hat{t}_0 \\ &+\frac{1}{2}\frac{\hat{p}}{\hat{P}+\hat{p}} + \frac{1}{2} \ln \left( \frac{2\pi}{\hat{P}+\hat{p}}\right) + \frac{\hat{t}_0 (\hat{t}_{\mathrm{d}}-\hat{t}_1)}{(\hat{Q}+\hat{q}_1)(\hat{P}+\hat{p})} + m\frac{(\hat{t}_{\mathrm{d}}-\hat{t}_1)\left[ \hat{q}_0\,(\hat{t}_1-\hat{t}_0)+\hat{t}_0\,(\hat{q}_1-\hat{q}_0) \right]}{(\hat{Q}+\hat{q}_1)(\hat{P}+\hat{p})(\hat{Q}+\hat{q}_1-m(\hat{q}_1-\hat{q}_0))}\\
			&+ \frac{(\hat{t}_{\mathrm{d}}-\hat{t}_1)\,(\hat{t}_1-\hat{t}_0)}{(\hat{P}+\hat{p})\,(\hat{Q}+\hat{q}_1-m(\hat{q}_1-\hat{q}_0))} + m^2 \frac{\hat{q}_0\,(\hat{q}_1-\hat{q}_0)\,(\hat{t}_{\mathrm{d}}-\hat{t}_1)\,(\hat{t}_1-\hat{t}_0)}{(\hat{Q}+\hat{q}_1)\,(\hat{P}+\hat{p})\,(\hat{Q}+\hat{q}_1-m(\hat{q}_1-\hat{q}_0))^2}\\
			&+ \frac{(\hat{t}_{\mathrm{d}}-\hat{t}_1)^2}{2\,(\hat{P}+\hat{p})\,(\hat{Q}+\hat{q}_1)}\left[1+ \frac{\hat{q}_0}{(\hat{Q}+\hat{q}_1)} + \frac{\hat{q}_1-\hat{q}_0}{(\hat{Q}+\hat{q}_1-m(\hat{q}_1-\hat{q}_0))} \right] \\
			&+ \frac{(\hat{t}_{\mathrm{d}}-\hat{t}_1)^2}{2\,(\hat{P}+\hat{p})\,(\hat{Q}+\hat{q}_1)^2}\left[m^2 \frac{\hat{q}_0\, ( \hat{q}_1-\hat{q}_0)^2}{(\hat{Q}+\hat{q}_1-m(\hat{q}_1-\hat{q}_0))^2} + 2m \frac{\hat{q}_0 \,( \hat{q}_1-\hat{q}_0)}{(\hat{Q}+\hat{q}_1-m(\hat{q}_1-\hat{q}_0))}\right] \\
			&+m \, \frac{(\hat{t}_1-\hat{t}_0)^2}{2\,(\hat{P}+\hat{p})\,(\hat{Q}+\hat{q}_1 -m(\hat{q}_1-\hat{q}_0))}\\
			&+ m \, \frac{\hat{t}_0\,(\hat{t}_1-\hat{t}_0)}{(\hat{P}+\hat{p})\,(\hat{Q}+\hat{q}_1 -m(\hat{q}_1-\hat{q}_0))} + m^2 \, \frac{(\hat{t}_1-\hat{t}_0)^2\,\hat{q}_0}{2\,(\hat{P}+\hat{p})\,(\hat{Q}+\hat{q}_1 -m(\hat{q}_1-\hat{q}_0))^2}
		\end{aligned}
	\end{equation}
	and the energetic is
	\begin{subequations}
		\begin{align}
			&\mathcal{G}_E = \int Dx \, \frac{\int Dy Dz \, H\left( C_1(x, y, z)\right) H^{m-1}\left( C_m(x, z) \right) \ln 
				H \left( B(x,y,z) \right)
			}{\int Dz \, H^{m}\left( C_m(x, z)\right)} \\
			&C_1(x, y, z) = \frac{ \sqrt{\gamma}\left( \tilde{\kappa} + \sqrt{q_0} x + \sqrt{q_1-q_0} z \right) + (t_{\mathrm{d}}-t_1) y }{\sqrt{(1-q_1) \gamma- (t_{\mathrm{d}}-t_1)^2}} \\
			&C_m(x, z) = \frac{\tilde{\kappa} + \sqrt{q_0}x+\sqrt{q_1-q_0}z}{\sqrt{1-q_1}} \\
			&B(x,y,z) = \frac{\kappa + \frac{t_0}{\sqrt{q_0}}x+ \sqrt{\gamma} y  + \frac{t_1 - t_0}{\sqrt{q_1 - q_0}} z}{\sqrt{1 - p}} \\
			&\gamma = p - \frac{t_0^2}{q_0} - \frac{(t_1 - t_0)^2}{q_1 - q_0}
		\end{align}
	\end{subequations}
	
\begin{figure*}	
		\begin{centering}
			\includegraphics[width=0.49\textwidth]{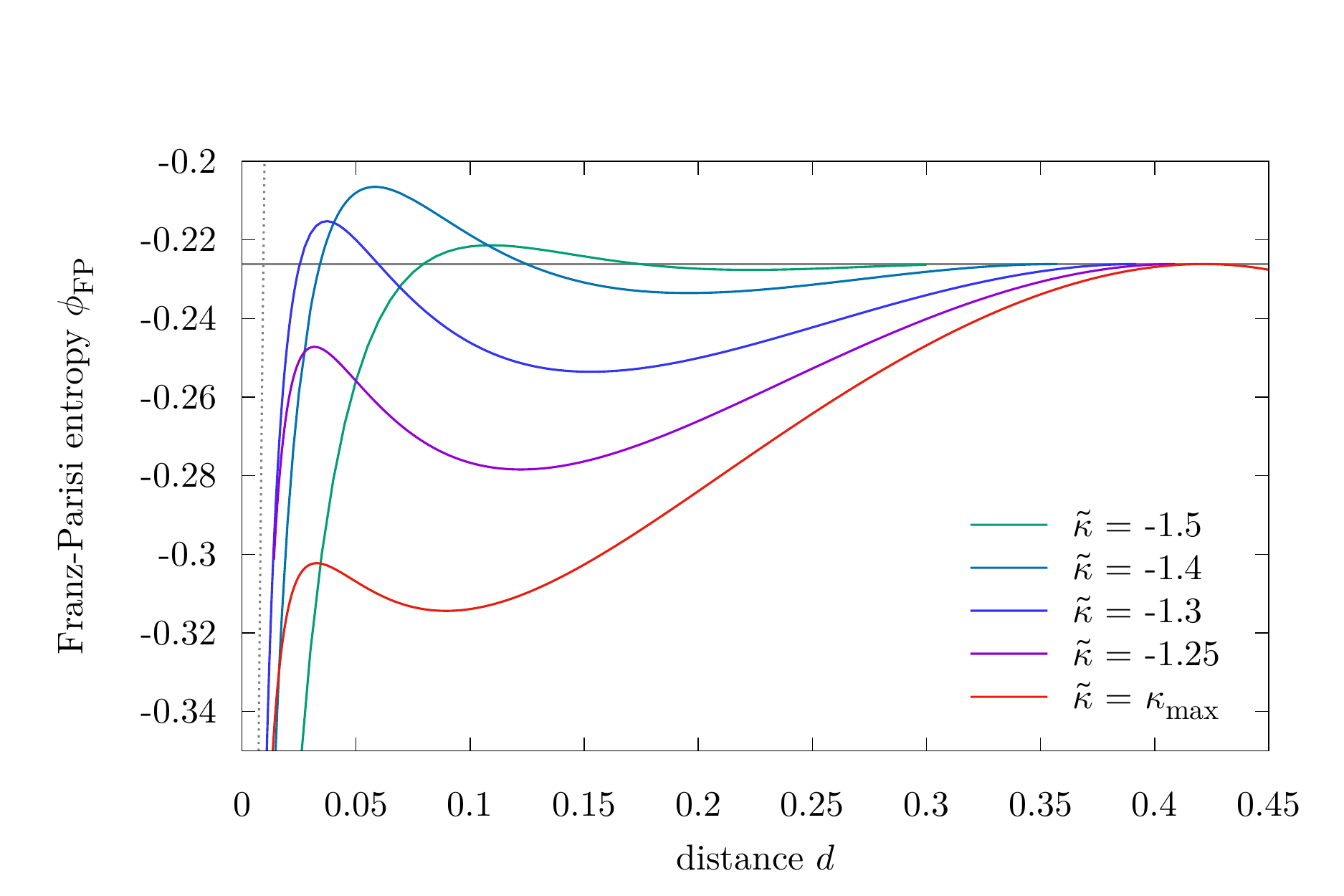}
			\includegraphics[width=0.49\textwidth]{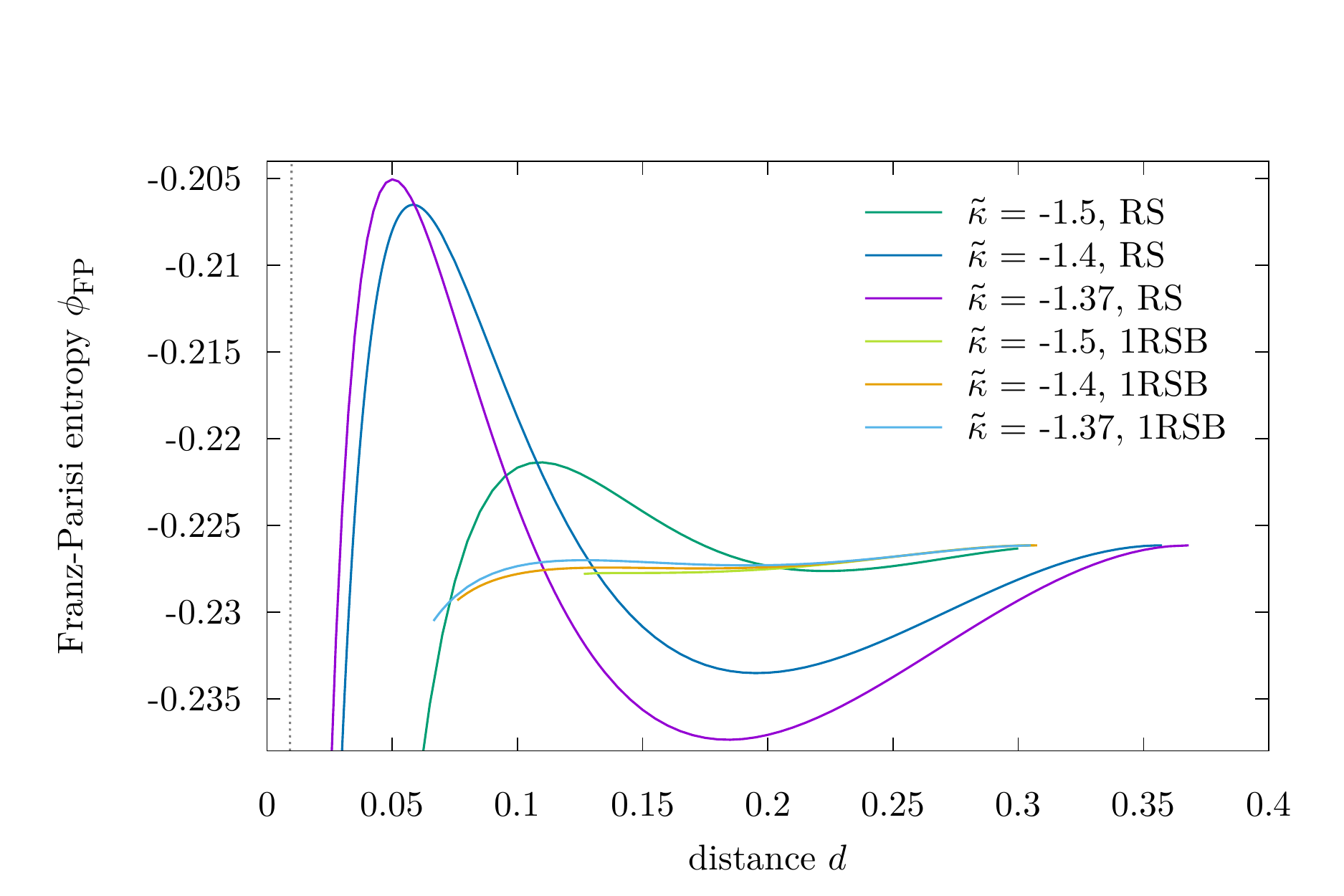}
		\end{centering}
		\caption{
			Left panel: Franz-Parisi entropy $\phi_{\text{FP}}$ as a function of distance $d=(1-t_{\mathrm{d}})/2$ for $\kappa = -1.5$, $\alpha = 22$ and several values of the reference margin $\tilde{\kappa}$. The curves are computed by using a RS approximation both on the reference and on the slaved solution. Notice that for this value of $\kappa$ and $\alpha$ the reference is in the fRSB phase, since $\alpha > \alpha_{\text{dAT}} \simeq 18.74$ for this value of $\kappa$. The grey line represents the value of the entropy of typical solutions as computed in the RS approximation. The dashed grey line represents the upper bound in equation~\eqref{eq::upper_bound}. 
			Right panel: same as left panel, but we also add the corresponding curves obtained by using a 1RSB ansatz on the reference configuration. We could not continue the curves at small distances because of the slowness with which the code converges to the saddle point. 
		}
		\label{Fig::FP_1RSB}
	\end{figure*}

	In the right panel of Figure~\ref{Fig::FP_1RSB} we compare the Franz-Parisi entropy in the 1RSB ansatz on the reference compared with the corresponding RS predictions, for several values of the margin and for $\alpha = 22$. Indeed we see that RSB effects are huge. The 1RSB approximation is enough to reorder the curves: high margin solutions possess a local entropy higher with respect to the ones having a smaller margin. Also, it seems that by increasing the margin the corresponding local entropy curve becomes non-monotonic. This could be an effect due to the fact that the 1RSB approximation is not yet accurate enough to reproduce the Franz-Parisi curve correctly. This also suggests that in this approximation the maximum margin solution could still be non-monotonic, therefore $\alpha_{\mathrm{LE}}$ in the 1RSB approximation is for sure lower than $\alpha = 22$ (the RS prediction is $\alpha_{\mathrm{LE}} \simeq 20.7$).
	
	In order to study precisely what is $\alpha_{\mathrm{LE}}$ in the 1RSB ansatz, one should do the limit $\tilde\kappa \to \kappa_{\mathrm{max}}$, which corresponds to do the limit $q_1 \to 1$ with $m\to 0$ and $\tilde m = \frac{m}{1-q_1}$ fixed, as done in~\ref{sec::1RSB_ansatz}. 
	The expressions are algebraically involved and we do not report them here; the numerics is even more challenging and we leave that to future work, together with a more refined analysis of the impact of higher levels of RSB.
	
	\subsubsection{Breaking the Replica-Symmetry on the constrained configuration}
	
	As reported in the main text we have tested the goodness of the RS approximation by plugging a more general ansatz for the order parameters. Since we have not changed the structure of the order parameters of the reference, we have imposed a one-step replica symmetry breaking ansatz (1RSB) for the order parameters that only involve the constrained configuration (i.e. $p_{cd}$ and $\hat p_{cd}$), leaving $q_{ab}$, $t_{ac}$ and their conjugated parameters unchanged. The new ansatz for $p_{cd}$ and $\hat p_{cd}$ is
	\begin{subequations} 
		\begin{align}
			p_{cd} &= p_0 + (p_1-p_0) I_{cd}^{(n,m)} + (1-p_1) I_{cd}^{(n,1)}\\
			\hat p_{cd} &= \hat p_0 + (\hat p_1 - \hat p_0) I_{cd}^{(n,m)} + (1-\hat p_1) I_{cd}^{(n,1)}
		\end{align}
	\end{subequations}
	where $I_{cd}^{(n,m)}$ is the $(c,d)$ element of a block matrix of size $n \times n$ whose diagonal blocks have size $m \times m$ and contain all ones and outside of them the matrix is composed of zeros.
	Therefore $p_1$ represents the overlap between different constrained solutions belonging to the same block while $p_0$ characterizes the overlap between two solutions belonging to different blocks of replicas.
	
	The entropic term is
	\begin{equation}
		\begin{split}
			\mathcal{G}_{S}  &= \frac{1}{2}\hat{P}+\frac{1}{2}\hat{p_1}p_1 - \frac{m}{2} \left( p_1 \hat p_1 - p_0 \hat p_0 \right) +\hat{t}t-\hat{t}_{\mathrm{d}}t_{\mathrm{d}}+\frac{1}{2}\ln\frac{2\pi}{\hat{P}+\hat{p}_1} - \frac{1}{2m} \ln\left( 1 - \frac{m(\hat{p_1} - 
				\hat{p_0})}{\hat P + \hat p_1}\right)\\
			&+\frac{1}{\hat{P}+\hat{p}_1 - m(\hat p_1 - \hat p_0)}\left[\frac{\hat{p}_0}{2}+\frac{\left(\hat{t}_{\mathrm{d}}-\hat{t}\right)^{2}\left(\frac{\hat{Q}}{2}+\hat{q}\right)}{\left(\hat{Q}+\hat{q}\right)^{2}}+\frac{\hat{t}\left(\hat{t}_{\mathrm{d}}-\hat{t}\right)}{\hat{Q}+\hat{q}}\right] \,.
		\end{split}
	\end{equation}
	Integrating the conjugated parameters we get
	\begin{equation}
		\begin{split}
			\mathcal{G}_{S}  &= \frac{(1 - t_{\mathrm{d}}^2) (1 - 2 q ) + q^2 - 2 q t_{\mathrm{d}} t + t^2 - (1-m)(p_1-p_0)(1-q)^2}{2(1-p_1+m(p_1-p_0))(1-q)^2} +\frac{1}{2} \ln(2\pi) + \frac{1}{2}\ln\left(1-p_1\right) \\
			&+\frac{1}{2m} \ln\left( \frac{1-p_1}{1-p_1+m(p_1-p_0)} \right)
		\end{split}
	\end{equation}
	The energetic term instead reads as
	\begin{equation}
		\label{eq::GEerr_1RSBslaved}
		\mathcal{G}_{E} = \frac{1}{m}\int Dz_{0}\frac{\int Dz_{1}H\text{\ensuremath{\left(\frac{\sqrt{\Gamma} \left(\tilde 	\kappa + \sqrt{q} \, z_{0} \right) + (t_{\mathrm{d}} - t) z_{1}}{\sqrt{(1 - q) \Gamma- (t_{\mathrm{d}} - t)^{2}}}\right)} \ensuremath{\ln \int D z_2 \, H^m\left(\frac{ \kappa + \frac{t}{\sqrt{q}} z_{0} + \sqrt{\gamma}z_{1} + \sqrt{p_1 - p_0} z_2 }{\sqrt{1 - p_1}}\right)}}}{H\left( \frac{\tilde\kappa + \sqrt{q} z_0}{\sqrt{1 - q}}\right)}
	\end{equation}
	where we have redefined in this case
	\begin{equation}
		\gamma = p_0 - \frac{t^2}{q} \,.
	\end{equation}	
	Now we are interested in the $\kappa_{\text{max}}$ limit; therefore we should send $q\to 1$. In this limit, the scaling for $t$ is, similarly for what we have done for the RS case~\eqref{eq::FP_RS_kmax_limit_scaling}
	\begin{equation}	
		t_{\mathrm{d}} - t = (1-q) \delta t = \delta t \, \delta q
	\end{equation}
	With this scaling the entropic term simplifies as
	\begin{equation}
		\mathcal{G}_{S}  =
		\frac{1-2 t_{\mathrm{d}} \delta t + \delta t^2 - (1-m)(p_1-p_0)}{2(1-p_1 + m(p_1-p_0))} +\frac{1}{2} \ln(2\pi) + \frac{1}{2}\ln\left( 1-p_1\right) - \frac{1}{2m} \ln\left( \frac{1-p_1}{1-p_1+m(p_1-p_0)} \right)\,.
	\end{equation}
	which is independent of $q$. 
	
	For the energetic term, notice that the argument of the first $H$ in the RS expression~\eqref{eq::GEerr_RS} and the 1RSB one~\eqref{eq::GEerr_1RSBslaved} are the same. Therefore we can repeat the same steps as before. We find
	\begin{equation}
		\begin{split}
			\mathcal{G}_{E} &= \frac{1}{m} \int Dz_{0} \,  H\left( \frac{ \tilde{\kappa} + \frac{t}{\sqrt{p_0}} z_0}{ \sqrt{1 - \frac{t^2}{p_0}}}\right) \, \ln \int \! Dz_2 \, H^m\left(\frac{ \kappa + \sqrt{p_0} \, z_{0} + \sqrt{p_1 - p_0} \, z_2}{\sqrt{1 - p_1}}\right) + \\
			&+ \frac{1}{m} \int Dz_{0} \, H\left( -\frac{\tilde{\kappa} \sqrt{\eta} + \left(  \delta t - t_{\mathrm{d}} \right) z_0}{\sqrt{p_0 - t_{\mathrm{d}}^2}} \right) \ln \int Dz_2 \, H^m \left(  \frac{ \kappa - \delta t \, \tilde{\kappa} - \sqrt{\eta} z_0 + \sqrt{p_1 - p_0} \, z_2}{\sqrt{1 - p_1}}\right)
		\end{split}
	\end{equation}
	where
	\begin{equation}
		\eta \equiv p_0 + \delta t \left( \delta t - 2 t_{\mathrm{d}} \right)
	\end{equation}
	We have solved numerically the corresponding saddle point equations; we didn't find any solutions other than that RS one $p_1 = p_0$.

	\section{Other approach: 1RSB formalism}
	\begin{figure*}	
		\begin{centering}
			\includegraphics[width=0.49\textwidth]{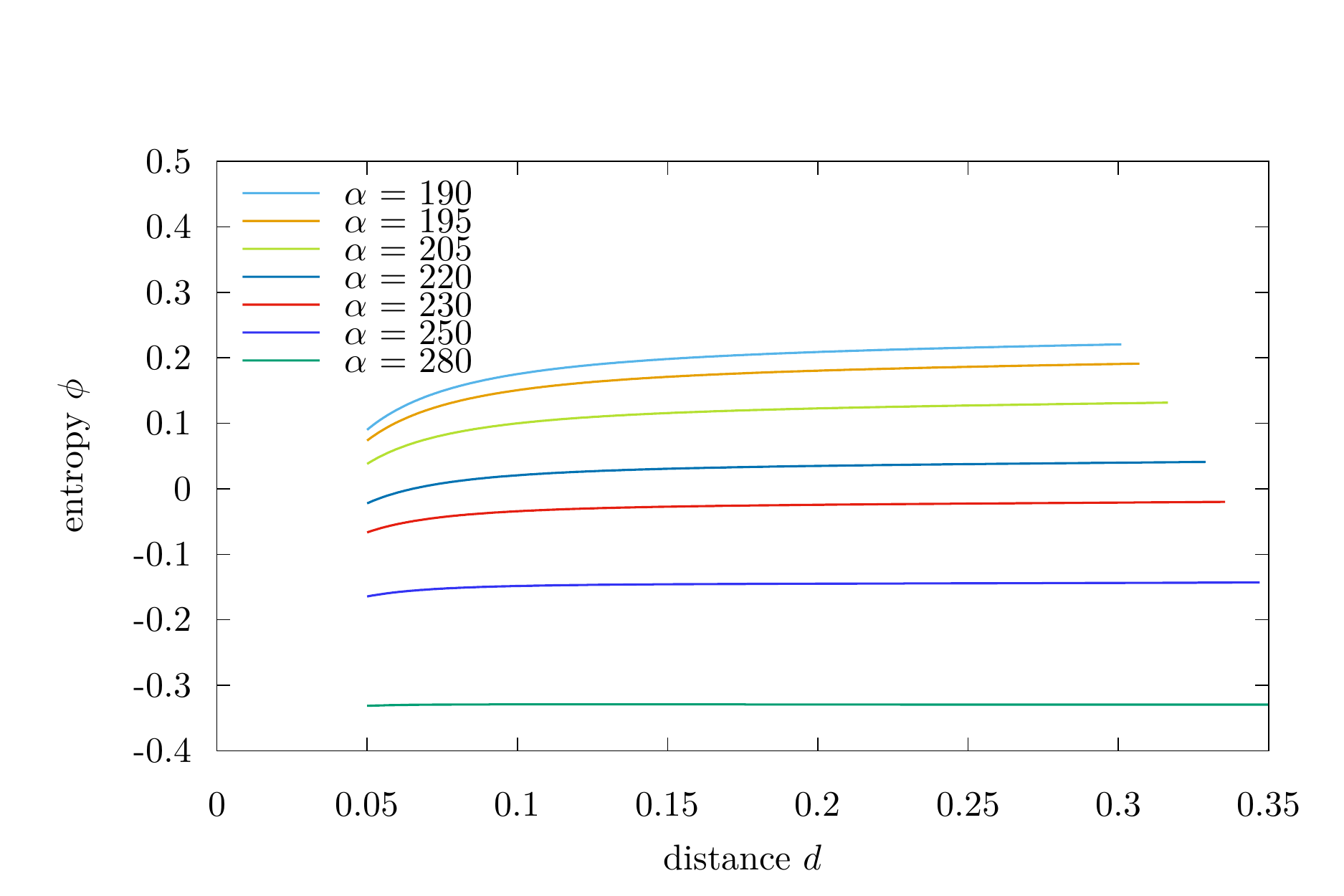}
			\includegraphics[width=0.49\textwidth]{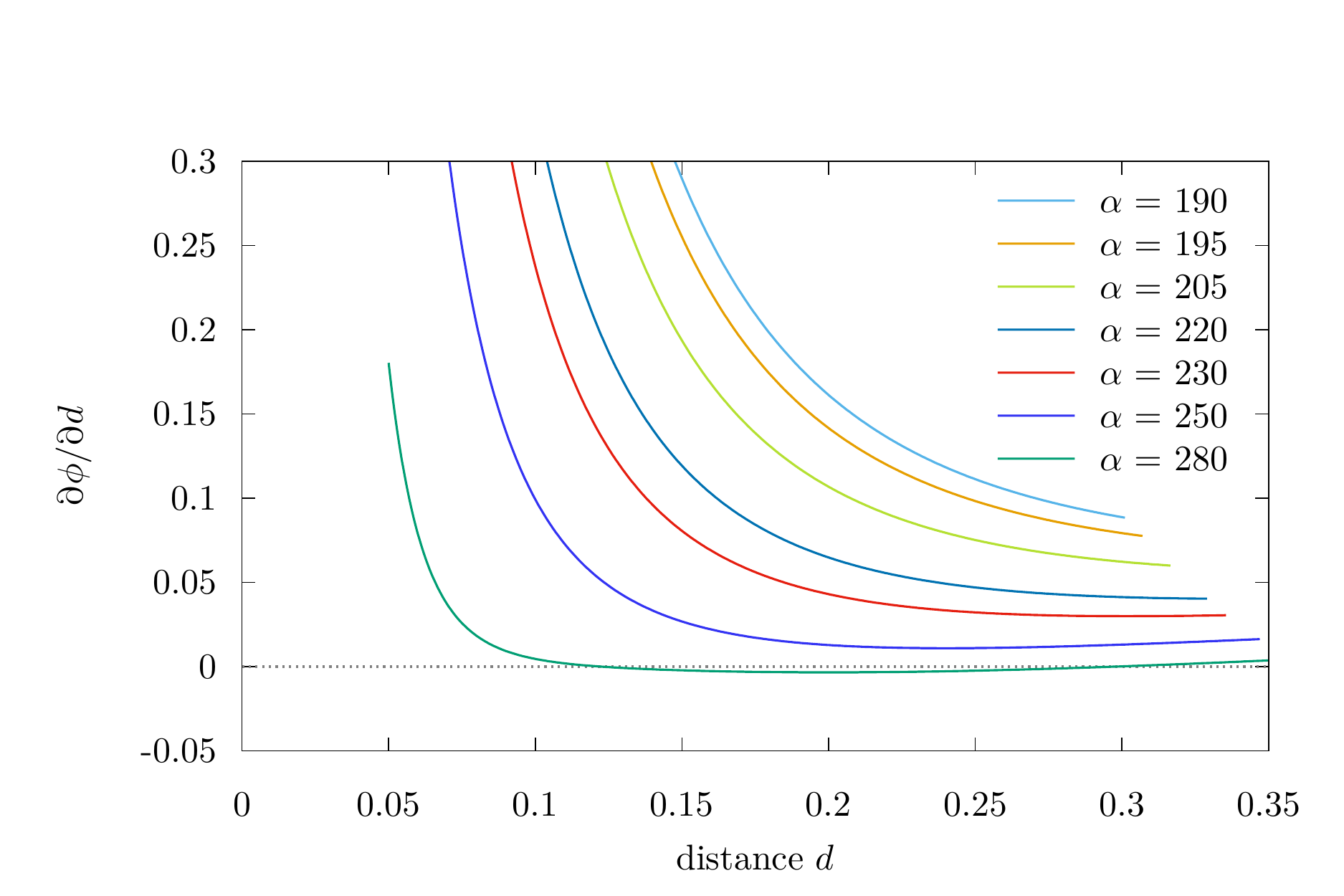}
		\end{centering}
		\caption{
			Left panel: Free entropy $\phi$ as a function of distance $d \equiv \frac{1-q_1}{2}$ for $\kappa = -2.5$ and several values of $\alpha$. As described in the main text the entropy $\phi$ has been obtained for every value of the distance by finding the value of $m$ for which the complexity in equation~\eqref{eq::complexity} vanishes. Right panel: same as left panel, but we plot the derivative of the entropy $\frac{\partial \phi}{\partial d}$.
		}
		\label{Fig::alphaLE_fake1rsb}
	\end{figure*}
	As cited in the main text, the monotonic/non-monotonic transition that is found in non-convex models can be also obtained using other methods not based on the Franz-Parisi approach. One of such methods is reminescent of the Monasson's method~\cite{monasson1995structural} that was firstly used in the context of the spherical $p$-spin model to locate the dynamical transition temperature. 
	
	The method was developed in~\cite{baldassi2020shaping}; we briefly summarize it here. The idea is to study the free entropy of a system of $m$ replicas constrained to be at a fixed overlap $q_1$
	\begin{equation}
		Z_m \equiv \int \prod_{a=1}^{m} d\mu(\boldsymbol{w}^a) \prod_{\mu=1}^{\alpha N} \Theta(\Delta^\mu(\boldsymbol{w};\kappa)) \prod_{a<b} \delta\left(N q_1 - \sum_{i=1}^{N} w_i^a w_i^b\right)
	\end{equation}
	We will call those $m$ replicas as ``\emph{real replicas}'' to distinguish them from the ``\emph{virtual}'' ones of the replica method. We want to compute the averaged free entropy of the system of $m$ real replicas we use the replica trick
	\begin{equation}
		\phi_m = \frac{1}{N}\left\langle \ln Z_m \right\rangle_{\xi} = \lim\limits_{s\to 0} \frac{1}{sN} \ln \left \langle Z_m^s \right \rangle_\xi \,.
	\end{equation}
	Now it is useful to note that if we consider $s = \frac{n}{m}$, we see that the free entropy \emph{per replica} is
	\begin{equation}
		\phi \equiv \frac{\phi_m}{m} = \lim\limits_{n\to 0} \frac{1}{nN} \ln \left \langle Z_m^{n/m} \right \rangle_\xi
	\end{equation}
	i.e. the total number of replicas (both virtual and real) $n = m s$ are grouped in $n/m$ groups containing $m$ replicas each; in each group in addition, the overlap between them is constrained to be $q_1$. Therefore the computation of $\phi_m/m$ is \emph{exactly} equivalent to impose a 1RSB ansatz on the usual typical free entropy of the model as in Section~\ref{sec::1RSB_ansatz}; the only difference being now that both $q_1$ and the blocks size $m$ are fixed as external parameters.
	
	As shown in~\cite{baldassi2015subdominant,baldassi_local_2016,baldassi2020shaping}, the value of $m$ is not restricted to the range $x \in [0,1]$ as is usually done in the equilibrium 1RSB ansatz; here instead one is instead interested to the regime $m>1$ in order to target configurations of large local entropy. 
	
\begin{figure*}	
		\begin{centering}
			\includegraphics[width=0.6\textwidth]{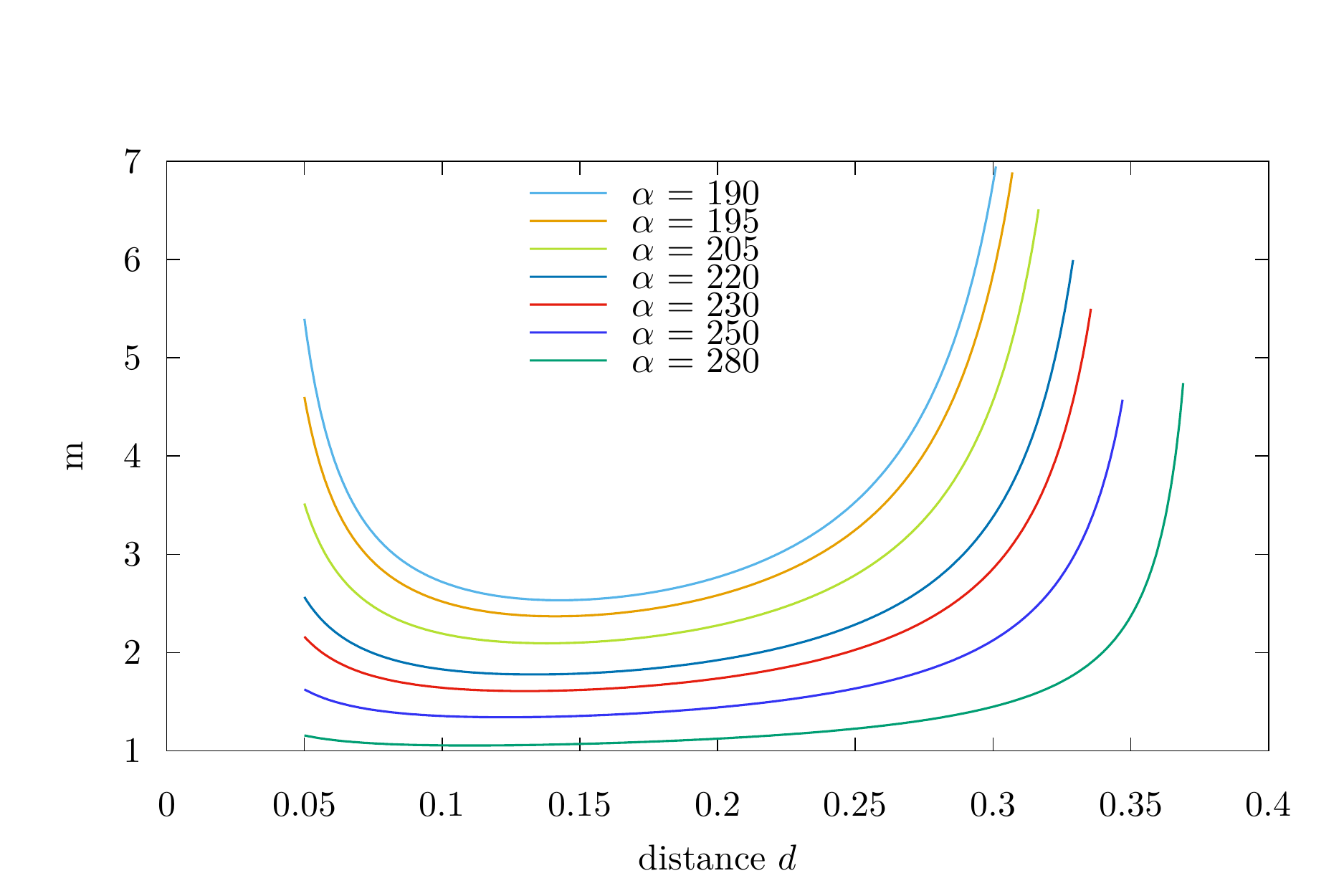}
		\end{centering}
		\caption{
			Plot of the value of $m$ for which the complexity in~\eqref{eq::complexity} vanishes as a function of distance and several values of $\alpha$. Here $\kappa = -2.5$.
		}
		\label{Fig::alphaLE_fake1rsb_m}
	\end{figure*}

	The so called \emph{complexity} can be written as
	\begin{equation}
		\label{eq::complexity}
		\Sigma = - m^2 \frac{\partial \phi}{\partial m}
	\end{equation}
	At a fixed value of $\alpha$ and $q_1$ we want to find the configuration having the maximal local entropy. This is achieved by finding the value of $m$ for which the complexity vanishes. This process is repeated for each value of $q_1$ and then for each value of $\alpha$. This method is expected to give more precise estimates of $\alpha_{\text{LE}}$ since there is no replica that is needed to be sampled in the deep RSB phase. For the behaviour of the value of $m$ for which the complexity vanishes as a function of $q_1$ for a fixed value of $\alpha$, see Figure~\ref{Fig::alphaLE_fake1rsb_m}.
	
	We plot in Figure~\ref{Fig::alphaLE_fake1rsb} the free entropy obtained in this way as a function of the distance $d \equiv (1-q_1)/2$, in the case $\kappa = -2.5$ and for several values of $\alpha$. One observes the same phenomenology obtained using the Franz-Parisi approach: as $\alpha$ increases, the free entropy $\phi$ goes from being monotonic to being non-monotonic when crossing a critical value of $\alpha$. This value of $\alpha$ can be considered as the estimate of the \emph{local entropy} transition
	$\alpha_{\text{LE}}$ obtained by using the Franz-Parisi methodology. Notice, however that the estimate of $\alpha_{\text{LE}}$ using this approach gives a different result with respect to the Franz-Parisi approach: for $\kappa = -2.5$ the local entropy transition is located around $\alpha_{\text{LE}} \simeq 280$ to be compared with the value $\simeq 195$ found using the Franz-Parisi maximum margin approach.

	\subsection{Large-$m$ limit}
	It is instructive to extract the large $m$ limit behaviour of the free entropy and the complexity.

	In the $m\to\infty$ case the order parameters $q_{0}$ and $\hat{q}_{0}$
	need to be rescaled with $m$ and reparametrized with two new quantities
	$\delta q_{0}$ and $\delta\hat{q}_{0}$, as follows:
	\begin{align}
		q_{0} & =q_{1}-\frac{\delta q}{m}\\
		\hat{q}_{0} & =\hat{q}_{1}-\frac{\delta\hat{q}}{m}
	\end{align}
	Integrating out the conjugated order parameters we have
	\begin{align}
		\mathcal{G}_{S} \left(q_{1}\right) & =\frac{1}{2}\frac{q_{1}}{1-q_{1}+\delta q}+\frac{1}{2}\text{\ensuremath{\ln}}\left(1-q_{1}\right)\\
		\mathcal{G}_{E}\left(q_{1}\right) & =\int Dz_{0} \, \underset{z_{1}}{\max}\left[-\frac{z_{1}^{2}}{2}+\log H\left(-\frac{\sqrt{q_1} \, z_{0}+z_{1}}{\sqrt{1-q_1}}\right)\right]
	\end{align}
	In order to derive the corresponding complexity we denote by $z_1^\star$ the quantity
	\begin{equation}
		z_1^\star = \argmax_{z_1}\left[-\frac{z_{1}^{2}}{2}+\log H\left(-\frac{\sqrt{q_1} \, z_{0}+z_{1}}{\sqrt{1-q_1}}\right)\right]
	\end{equation}
	which satisfies the following equation
	\begin{equation}
		z_1^\star = \sqrt{\frac{1}{1 - q_1}} GH\left(-\frac{\sqrt{q_1} \, z_{0}+z_{1}}{\sqrt{1-q_1}}\right)
	\end{equation}
	The saddle point equation for $q_0$ is
	\begin{equation}
		\label{eq:sp_eq_q_0}
		\frac{q_1}{\left( 1 - q_1 + \delta q\right)^2} = \frac{\alpha}{\delta q} \int Dz_0 \, \left(z_1^\star\right)^2
	\end{equation}
	The derivative of the energetic term with respect to $m$ in the large $m$ limit is
	\begin{equation}
		- m^2 \frac{\partial \mathcal{G}_E}{\partial m} = m \mathcal{G_E} - m \left\langle \ln H \left(-\frac{\sqrt{q_1} \, z_{0}+z_{1}}{\sqrt{1-q_1}}\right)  \right\rangle_{\mathcal{G}_E} \to -\frac{m}{2} \int Dz_0 \, (z_1^\star)^2
	\end{equation}
	and the entropic term is 
	\begin{equation}
		- m^2 \frac{\partial \mathcal{G}_S}{\partial m} \to \frac{1}{2} \left[ \frac{\delta q \left( (m+1)q_1 -2\delta q - 1\right)}{\left( 1+ \delta q - q_1\right)^2} - \ln(1-q_1) + \ln(1-q_1 + \delta q) \right]
	\end{equation}
	Using the saddle point equation~\eqref{eq:sp_eq_q_0} it turns out that the linear divergence in $m$ of the energetic and entropic term cancel out giving for the complexity the simple expression
	\begin{equation}
		\Sigma = - m^2 \frac{\partial \phi}{\partial m} = \frac{1}{2} \left[ \frac{\delta q \left( q_1 -2\delta q - 1\right)}{\left( 1+ \delta q - q_1\right)^2} - \ln(1-q_1) + \ln(1-q_1 + \delta q) \right]
	\end{equation}

	\section{Numerical Experiments details} ~\label{sec::APP_num_exp}
	
	\subsection{Sampling typical solutions}
	
	\begin{figure}	
		\begin{centering}
			\includegraphics[width=0.32\textwidth]{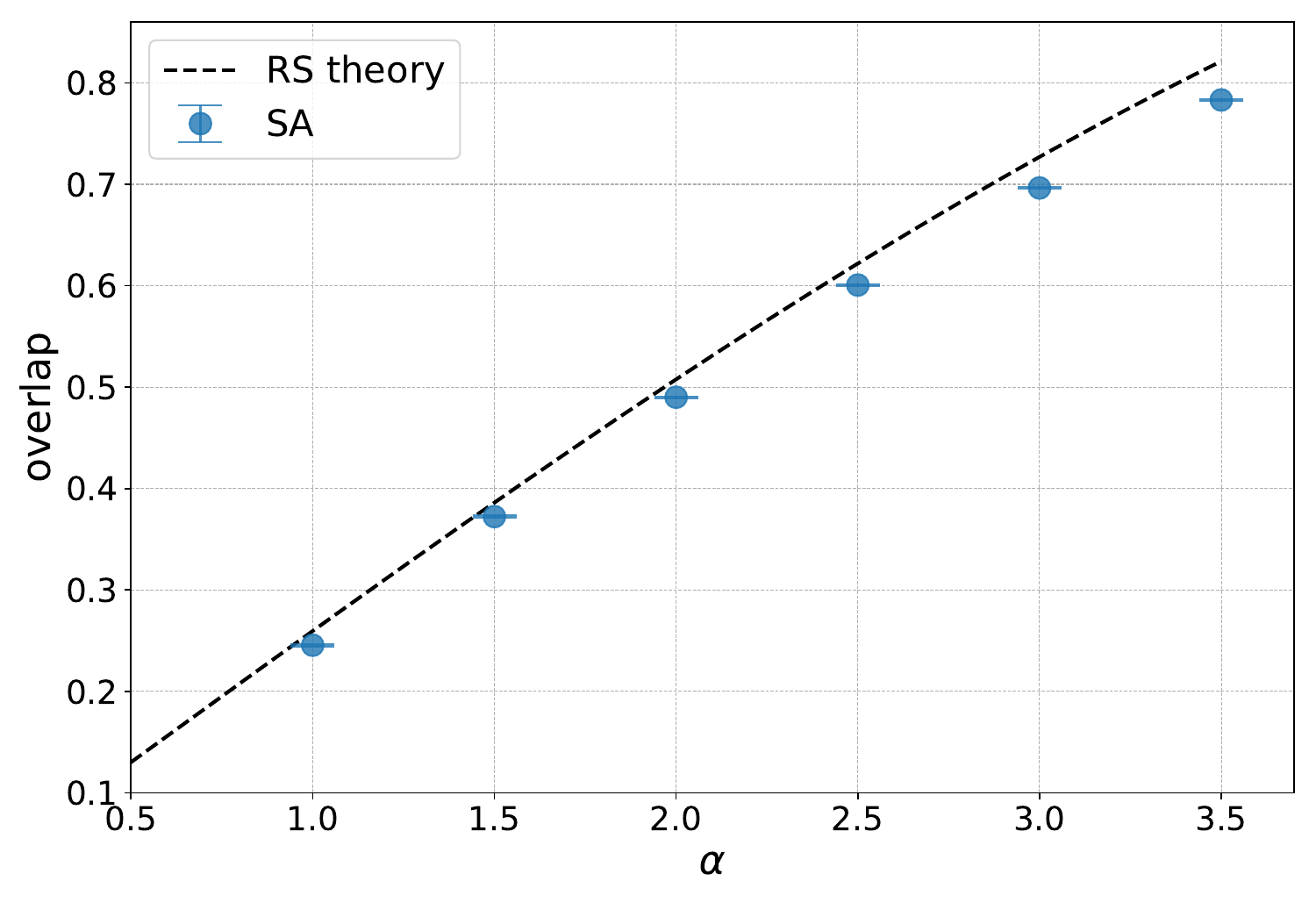}
			\includegraphics[width=0.32\textwidth]{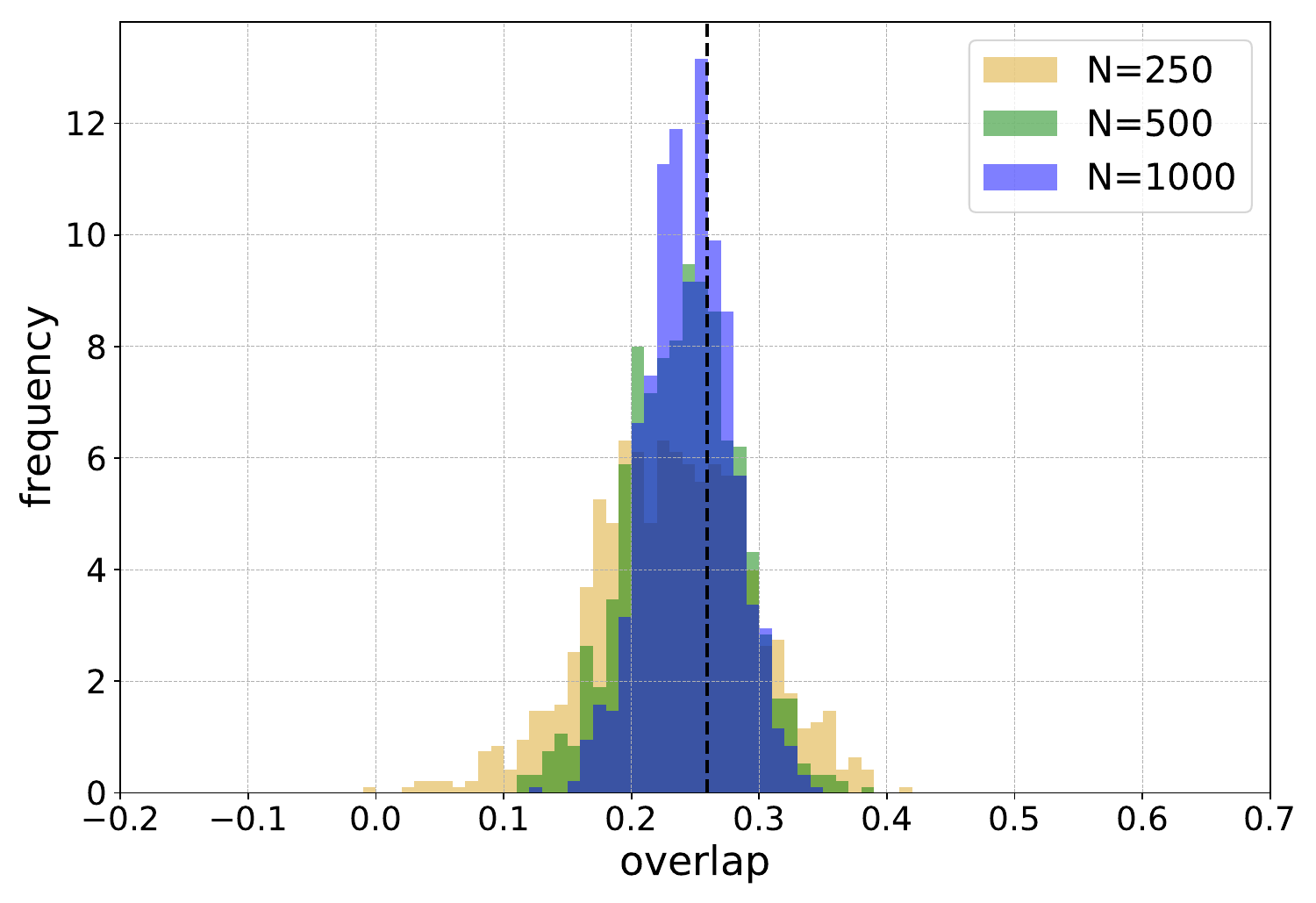}
			\includegraphics[width=0.32\textwidth]{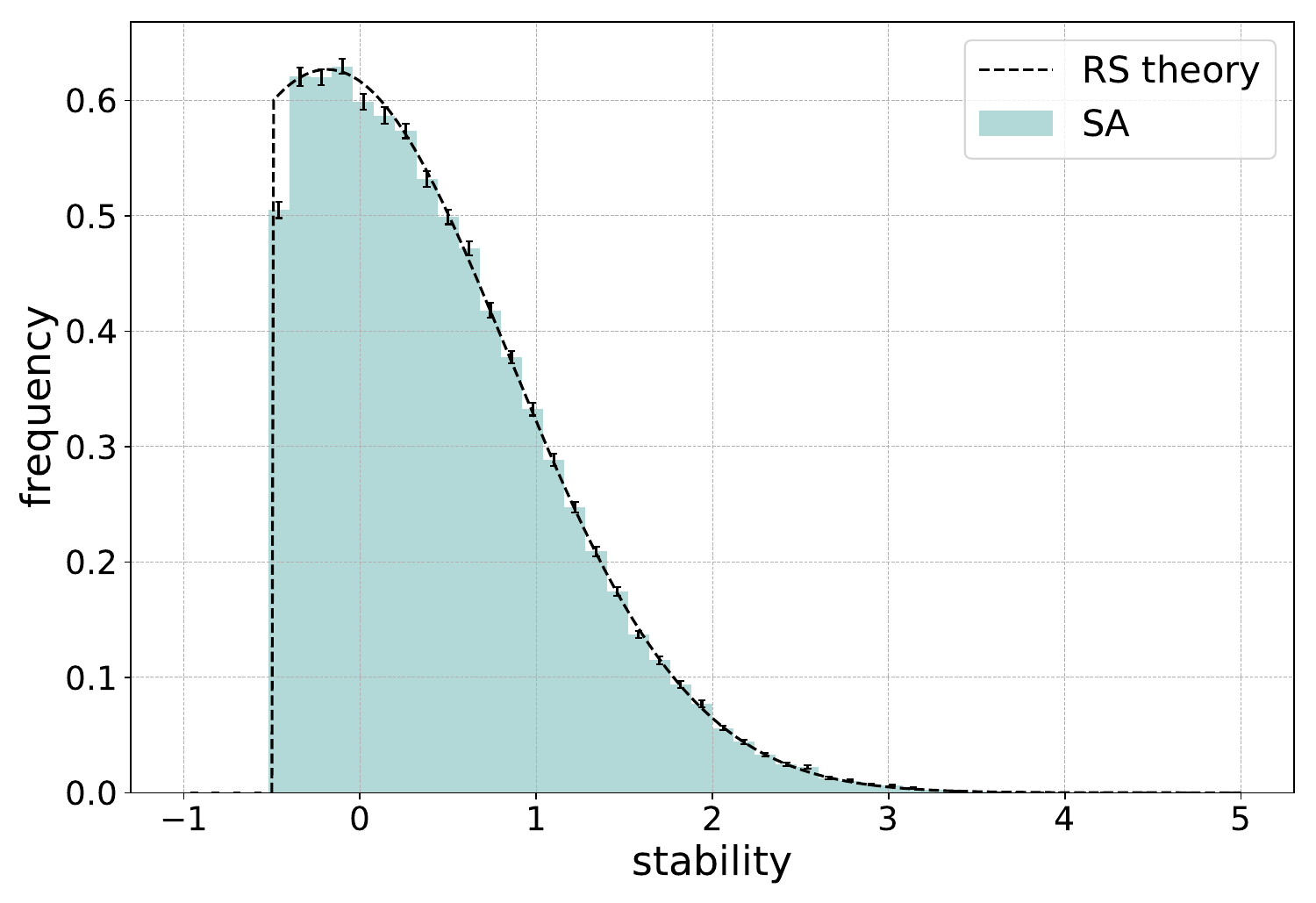}
		\end{centering}
		\caption{
			(Left) Overlap of typical solutions as a function of $\alpha$ for $\alpha < \alpha_{\mathrm{dat}}$ in the replica symmetric approximation (dashed line), for $\kappa=-0.5$. Points are numerical estimates obtained with solutions sampled with Simulated Annealing at $N=1000$. 
			(Center) Overlap distribution of solutions sampled with SA with different system sizes at $\alpha=1$ and $\kappa=-0.5$. The dashed vertical line is the RS prediction.
			(Right) Stability distribution for typical solutions at $\alpha=1$ and $\kappa=-0.5$. The dashed line represents the RS prediction while the histogram refers to the numerical estimate with SA solutions.
			All points are averages over $10$ datasets realization and $20$ solutions for each dataset.
		}
		\label{Fig::sa_typ}
	\end{figure}
	
	In order to sample typical solutions we use Simulated Annealing on a quadratic Hinge loss that is defined as follows
	\begin{equation}
		\ell(x) \equiv x^2 \Theta(-x)
	\end{equation} 
	By definition the Hinge loss has the same ground states as the error function loss; however the high energy solutions organization helps reaching the ground states in a much faster fashion than in the error function case.
	We start with a random vector on a unit sphere and at each step we propose a move in a random direction $w^{\prime} = w+\eta$, with $\eta = \epsilon \mathcal{N}(0,1)$. The move is accepted and the weights are updated with probability $p=\mathrm{min}\left(1, e^{-\beta\Delta\ell}\right)$, where $\Delta\ell$ is the loss difference between the current and proposed configurations. 
	After $N$ proposed moves we increase the inverse temperature using a linear scheduler $\beta \leftarrow \beta+d\beta$.
	In the simulations we used $\beta_0=2000$, $d\beta=2$ and $\epsilon=5\times 10^{-3}$.
	Using this scheme we are able to efficiently sample typical solutions in the whole RS region ($\alpha < \alpha_{\mathrm{dat}}$). Using these solutions we are able to make quantitatively good predictions even at moderate system sizes (see Fig.~\ref{Fig::sa_typ}).  
	
	\subsection{Sampling atypical solutions}
	We report here the details for the various algorithms used in text.
	\textit{Spherical Perceptron algorithms}. For SGD we initialized the weights uniformly on the unit sphere and optimized the cross entropy loss using batchsize $200$ and learning rate $1$, with a maximum number of allowed epochs equal to $20000$. In the fBP simulations (see~\cite{baldassi2020shaping} for details) we used $y=10$ replicas and an increasing coupling constant $\gamma$ with $15$ exponentially spaced values in the range $[0.1,10]$. For each value of $\gamma$ we initialized fBP with the messages obtained at the previous coupling value. We than made it run until the maximum absolute difference between messages was less than $10^{-5}$ or until it reached $500$ messages updates.  
	\textit{Binary Perceptron algorithms}. For gradinet based simulations we used the standard BinaryNet implementation (BNET) (see e.g.~\cite{hubara2016binarized}). We then optimized the cross entropy loss using full batch gradient descent with learing rate $1$ and a maximum number of $20000$ epochs. For Stochastic Belief Propagation Inspired (SBPI) simulations (see~\cite{baldassi2009generalization}) we set the maximum number of
	allowed iterations to $1000$ and used a threshold $\theta_{m}=2$ and a
	probability $p_m=0.3$ of updating the synapses with a stability $0\leq \Delta \leq \theta_{m}$. 
	For fBP simulations we used the same parameters as in the continuous weights case.
	
	\subsection{Maximally robust solutions beyond RS} 
\begin{figure}	
		\begin{centering}
			\includegraphics[width=0.5\textwidth]{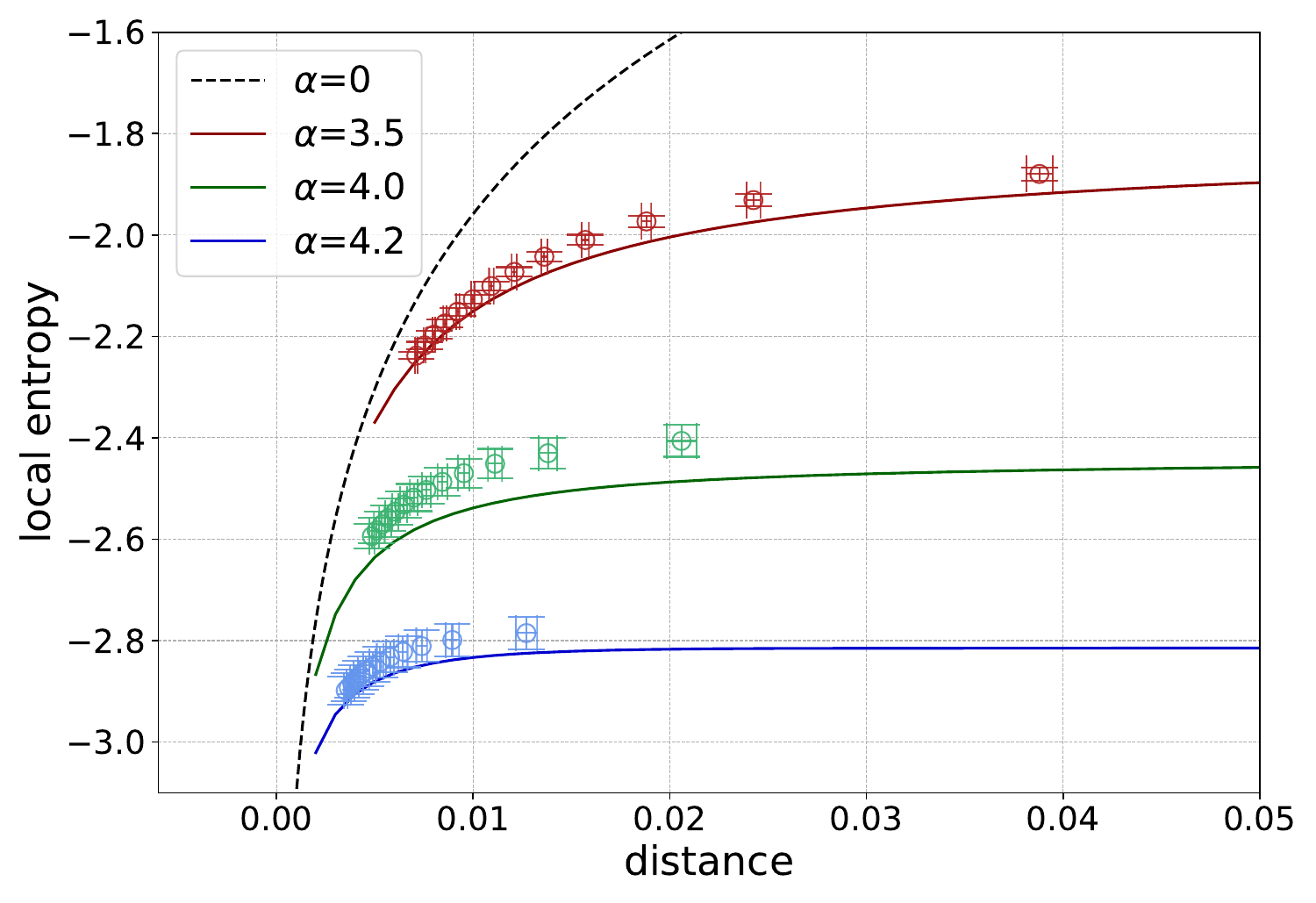}
		\end{centering}
		\caption{
			Local entropy of the $\kappa_{\mathrm{max}}(\kappa=-0.5)$ solutions at three different values of $\alpha$ at which the RS estimate of $\kappa_{\mathrm{max}}$ is known to be wrong due to RSB effects. Points are numerical estimates of the Local Entropy (using Belief Propagation) of solutions obtained with fBP at $N=1000$, averaged over $20$ independent data realizations. Even though the $\kappa_{\mathrm{max}}$ prediction is not correct in this range of $\alpha$, the RS theory is still able to qualitatively predict the Local Entropy of the maximally robust solutions.
		}
		\label{Fig::le_fbp}
	\end{figure}

	We show here the numerical estimate of the Local Entropy for maximally robust solutions in a range of $\alpha$ where the RS calculation does not give the correct prediction due to replica symmetry breaking effects. 
	We sample solutions with the fBP algorithm and estimate their local entropy with BP. Even in this region, the RS prediction for the local entropy of the $\kappa_{\mathrm{max}}$ solution is in a rather good agreement with simulations, see Fig.~\ref{Fig::le_fbp}. 
	
	\subsection{Generalization in the Binary Teacher-Student problem with Negative Margin}
	We report here the results for the teacher-student setting with a negative margin student, in the case of the binary perceptron (see discussion in Sec.~\ref{sec::numerical_experiments}). 
	The scenario is qualitatively the same as in the case of the spherical perceptron: depending on the algorithm bias toward flat minima, solutions with good generalization properties can be found also when increasing the negative margin modulus and the teacher signal is weak.
	
	\begin{figure}	
		\begin{centering}
			\includegraphics[width=0.5\textwidth]{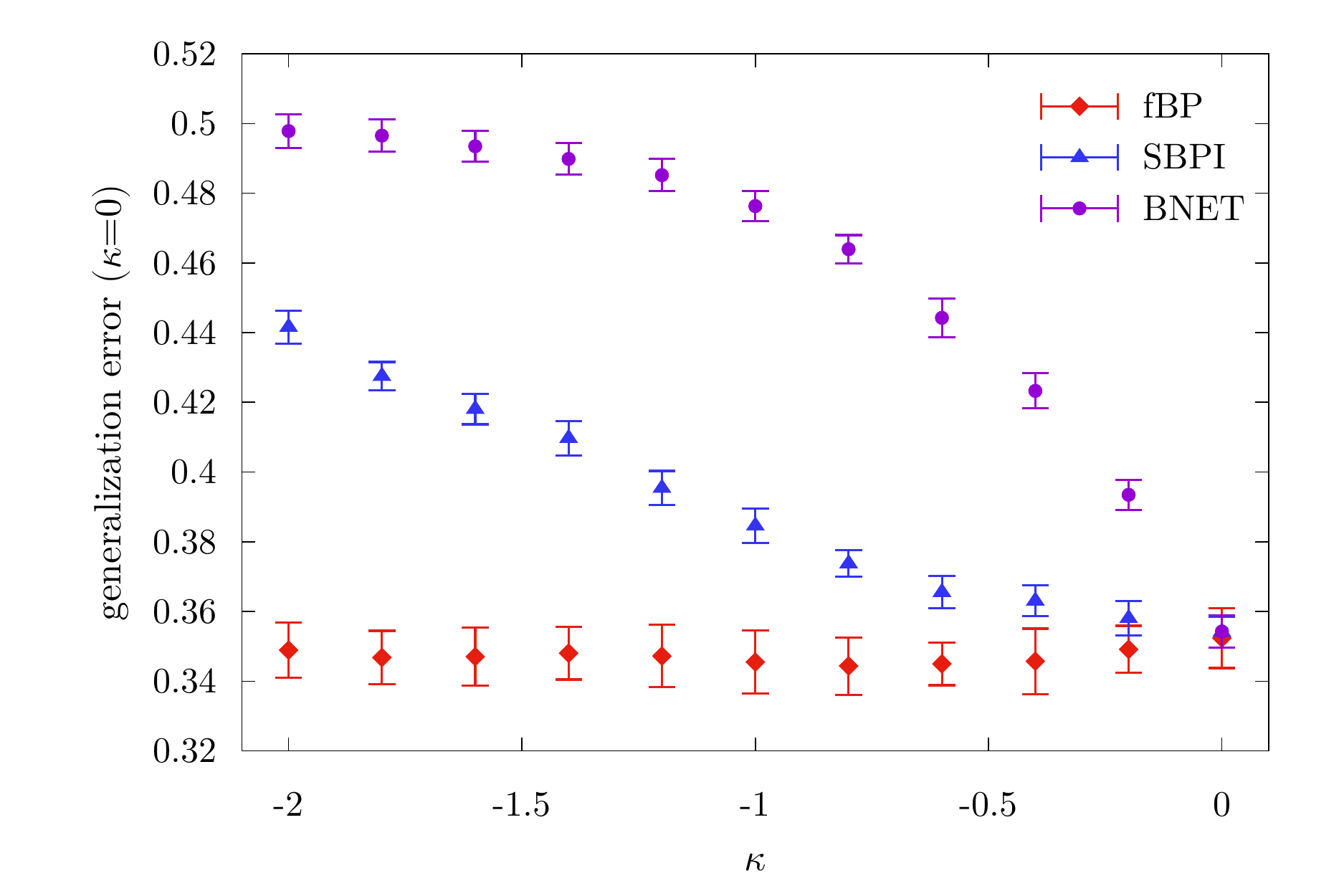}
		\end{centering}
		\caption{
			Same as Fig.~\ref{Fig::generalization} in the main text, but in the case of binary weights for $\alpha = 0.5$ and $N=1001$. The algorithms are fBP, BinaryNet (BNET), and SBPI. Results are averaged over $10$ independent teacher realizations and 5 random restarts for each dataset.
		}
		\label{Fig::ts_binary}
	\end{figure}
	
\end{document}